\documentclass[12pt]{article}
\usepackage{amsmath, amsthm, amsfonts}
\usepackage{amssymb}
\usepackage{graphicx,psfrag,epsf}
\usepackage{enumerate}
\usepackage{natbib}
\usepackage{url} 
\usepackage{booktabs}
\usepackage{multirow}
\usepackage{verbatim}
\usepackage{setspace}
\usepackage{authblk}

\RequirePackage[colorlinks,citecolor=blue,urlcolor=blue]{hyperref}

\newcommand\independent{\protect\mathpalette{\protect\independenT}{\perp}}
\def\independenT#1#2{\mathrel{\rlap{$#1#2$}\mkern2mu{#1#2}}}
\newcommand{\blind}{1}

\newcommand{\var}{\mbox{Var}}

\newcommand{\bmu}{\mbox{\boldmath $\mu$}}
\newcommand{\bbeta}{\mbox{\boldmath $\beta$}}

\newcommand{\hatbbeta}{\mbox{$\boldsymbol{\skew{2}\hat\beta}$}}

\newcommand{\boldeta}{\mbox{\boldmath $\eta$}}
\newcommand{\hatboldeta}{\mbox{$\boldsymbol{\skew{2}\hat\eta}$}}
\newcommand{\bSigma}{\mbox{\boldmath $\Sigma$}}

\newcommand{\btheta}{\mbox{\boldmath $\theta$}}
\newcommand{\blambda}{\mbox{\boldmath $\lambda$}}
\newcommand{\hatbtheta}{\mbox{$\boldsymbol{\skew{2}\hat\theta}$}}
\newcommand{\checkbtheta}{\mbox{$\boldsymbol{\skew{2}\check\theta}$}}
\newcommand{\checkboeta}{\mbox{$\boldsymbol{\skew{2}\check\eta}$}}

\DeclareMathOperator*{\argmin}{arg\,min}

\newtheorem{prop}{Proposition}
\newtheorem{theorem}{Theorem}
\newtheorem{lemma}{Lemma}

\begin{document}

\def\spacingset#1{\renewcommand{\baselinestretch}%
{#1}\small\normalsize} \spacingset{1}

\if1\blind
{
  \title{\bf Improved Small Domain Estimation via Compromise Regression Weights}
  
\author[1]{Nicholas C. Henderson}
\author[2,3]{Ravi Varadhan}
\author[3]{Thomas A. Louis}
\affil[1]{{\small Department of Biostatistics, University of Michigan, Ann Arbor}}
\affil[2]{ {\small Sidney Kimmel Comprehensive Cancer Center, Johns Hopkins University } }
\affil[3]{ {\small Department of Biostatistics, Bloomberg School of Public Health, Johns Hopkins University} }

\date{\vspace{-6ex}}
  \maketitle
} \fi

\if0\blind
{
  \bigskip
  \bigskip
  \bigskip
  \begin{center}
    {\LARGE\bf Improved Small Domain Estimation via Compromise Regression Weights}
\end{center}
  \medskip
} \fi

\bigskip

\begin{abstract}
Shrinkage estimates of small domain parameters typically utilize a combination of a noisy ``direct'' estimate that only uses data from a specific small domain and a more stable regression estimate.
When the regression model is misspecified, estimation performance
for the noisier domains can suffer due to substantial shrinkage towards a poorly estimated regression surface.
In this paper, we introduce a new class of robust, empirically-driven regression weights that target estimation
of the small domain means under potential misspecification of the global regression model. 
Our regression weights are a convex combination of the model-based weights associated with the best linear unbiased predictor (BLUP) and those associated with the observed best predictor (OBP). The compromise parameter in this convex combination is found by minimizing a novel, unbiased estimate of the 
mean-squared prediction error for the small domain means, and we label the associated small domain estimates the ``compromise best predictor", or CBP. Using a data-adaptive mixture for the regression weights enables the CBP to possess the robustness of the OBP while retaining the main advantages of the EBLUP whenever the regression model is correct. We demonstrate the use of the CBP in an application estimating gait speed in older adults.
\end{abstract}


\noindent%
\vfill

\spacingset{1.45}

\newpage

\section{Introduction} 
Analyzing clustered data where the targets of estimation are the cluster, area, or ``unit"-specific attributes is an important task that arises in a wide range of applied contexts. 
Common examples include estimating disease burden in specific geographic regions (e.g. \cite{Wakefield:2007}), estimating subgroup-specific treatment effects in clinical trials (e.g. \cite{Jones:2011}), quantifying hospital performance (e.g. \cite{Normand:2016}), and analyzing measures of gene expression (e.g. \cite{Smyth2004}).
A feature of many such applications is the availability of a ``direct" estimate for each unit, large standard errors for many of these direct estimates, and, consequently, considerable heterogeneity in estimation precision across units. When using hierarchical models to stabilize direct estimates and predictions for a collection of units, shrinkage estimates of unit-specific parameters often arise as a weighted combination of the direct, unit-specific estimate and a regression prediction for that unit. The direct estimates, while unbiased, typically have large variance, and while the regression estimates are biased, they are usually much more stable than the direct estimates. A version of shrinkage estimates are obtained by taking a weighted average of the direct estimates and the regression prediction with more influence coming from the regression model for units with larger variance. In standard practice (i.e., maximum likelihood estimation), the regression model itself is estimated by using regression weights which place more importance on units with smaller estimation variance. Consequently, while shrinkage estimates for high-variance units are more influenced by the regression estimate, they play a relatively minor role in determining the regression estimate itself. In other words, as noted in \cite{Jiang:2011} the units that really ``care about'' the regression model have relatively little impact on its estimation. 

When computing shrinkage estimates of unit-specific mean parameters, giving the relatively unstable units additional weight when estimating the regression model can substantially reduce overall bias while increasing variance. Overall, this may or may not reduce the mean-squared prediction error (MSPE) of the procedure. When the model is correctly specified, the MLE regression weights are optimal and cannot be improved upon. Under model misspecification however, regression weights targeting reduced prediction error can often result in substantial improvements in MSPE. Best predictive estimates (BPEs) of the regression coefficients (\cite{Jiang:2011}) target minimization of the MSPE for the resulting shrinkage estimates without relying on an assumption of correct model specification. In particular, the BPEs are found by minimizing an ``observed" MSPE associated with a particular choice of regression coefficients. 
In contrast to the MLE regression weights which minimize estimation variance when the model is correctly specified, the regression weights used in the BPE instead minimize a squared bias term which depends on the degree of model misspecification.

A natural way of building upon the strengths of the MLE and BPE weighting schemes is to allow, in a limited way, the form of the regression weights to depend on the observed responses. This enables the regression weights to adapt to the extent of regression function misspecification and the magnitude of the variance associated with a given set of shrinkage estimates. 
In this article, we consider regression weights that are an empirically-determined convex combination of the MLE and BPE regression weights. Such adaptive ``compromise" regression weights will automatically be closer to the MLE regression weights when the model is well-specified but will more closely resemble the BPE weights in scenarios with substantial model misspecification. Using such compromise regression weights to compute shrinkage estimates of small domain parameters can offer the robustness of the BPE while having MSPE performance which is close to the model-based estimates in cases where the model is well specified.
In addition, our compromise regression weights only depend on a single additional tuning parameter, namely the mixture term in the convex combination of the MLE and BPE regression weights, and hence estimating this additional tuning parameter will not introduce substantial additional estimation variance.

The foregoing discussion sets the context for our estimation approach that uses compromise regression weights constructed with the main goal of providing effective estimation of unit-specific parameters that can adapt to varying degrees of mean function misspecification. Our compromise regression weights induce a class of shrinkage estimates that depend on a mixing parameter and a variance component. To determine these terms empirically, we propose minimizing an unbiased estimate of the MSPE associated with these shrinkage estimates. This procedure resembles SURE-type shrinkage estimators (e.g., \cite{Xie:2012} or \cite{Donoho:1995}), all of which choose tuning parameters by minimizing Stein's unbiased risk estimate (\cite{Stein:1981}). We refer to the estimated regression coefficients associated with the variance component and mixing parameter that minimize our unbiased MSPE estimator as the ``compromise unbiased risk estimator", or CURE estimates of the regression coefficients, and we label the associated shrinkage estimates of the unit-specific parameters the ``compromise best predictor", or CBP. The CBP has the attractive property that its regression weights are not derived under an assumption of a correctly specified mean function. Rather, the CBP starts with an assumed class of shrinkage estimates and finds the best value of the shrinkage estimates within this class by minimizing a risk estimate whose unbiasedness does not rely on a correct mean model. 
We also examine the performance of a related predictor which we refer to as the ``plug-in" CBP where only the mixing parameter is computed and alternative values of the variance components are simply plugged in. This alternative form of the CBP often improves finite sample performance as the plugged in variance components frequently deliver better performance particularly when the model is correctly, or nearly correctly specified.

The remainder of this paper is organized in the following manner. Section \ref{sec:CBP_describe} describes the basic structure of the problem and discusses shrinkage estimation of unit-specific means using MLE or BPE regression weights. 
Section \ref{sec:CBP_describe} then describes our approach for improving the performance of the shrinkage estimates by using empirical combinations of the MLE and BPE regression weights. Section \ref{sec:pop_mean} details how such compromise weights can be used to estimate a single population-level attribute. Section \ref{sec:asymptotic_risk} discusses asymptotic properties of our proposed estimation scheme, and
Section \ref{sec:simulations} examines the performance of our method with several simulation studies. Section \ref{sec:application} demonstrates the application of our method to the estimation of ``normative'' gait speed in various subgroups of older adults, and we then conclude with a brief discussion.

\section{Regression Weights for Small Area Estimates} \label{sec:CBP_describe}
We let $Y_{1}, \ldots, Y_{K}$ denote measurements made from $K$ separate units with each $Y_{k}$ representing a direct estimate of a corresponding parameters of interest $\theta_{k}$. In addition to the direct estimate $Y_{k}$, each unit $k$ has an associated $p \times 1$ vector of covariates $\mathbf{x}_{k}$. The main goal here is to estimate each $\theta_{k}$ by combining the direct estimate $Y_{k}$ with a regression prediction that utilizes covariate information $\mathbf{x}_{k}$. 
Rather than assume a particular regression model to describe the variation in the $\theta_{k}$ we instead, as in \cite{Jiang:2011}, utilize a mixed model formulation which does not depend on an assumed regression structure for the unit-specific means. Specifically,
we consider the following mixed model representation
\begin{equation}
\begin{cases} 
Y_{k} = \mu_{k} + v_{k} + e_{k},\qquad 
k = 1, \ldots, K,  \\
E( e_{k} ) = E( v_{k} ) = 0, \\
\var(v_{k} ) = \tau_{0}^{2}, \quad \var(e_{k}) = \sigma_{k}^{2}, \quad v_{k} \independent e_{k}, 
\end{cases}
\label{eq:mixed_model_true}
\end{equation}
where the notation $v_{k} \independent e_{k}$ means that $v_{k}$ and $e_{k}$ are independent random variables. 
In addition to assuming that $v_{k}$ and $e_{k}$ are independent, we assume the values of $\sigma_{k}^{2}$ are known. 

Of primary interest is estimation/prediction of the mixed effects
\begin{equation}
\theta_{k} = \mu_{k} + v_{k},
\end{equation}
with mean-squared prediction error (MSPE) serving as the main measure of performance. 
For a given estimate/predictor $\hatbtheta$ of the vector $\btheta = (\theta_{1}, \ldots, \theta_{K})$ of mixed effects, the MSPE is defined as
\begin{equation}
\textrm{MSPE}( \hatbtheta )
= 
E\big\{ (\hatbtheta - \btheta)^{T}(\hatbtheta - \btheta) \big\}
= \sum_{k=1}^{K} E \big\{ (\hat{\theta}_{k} - \theta_{k})^{2} \big\}.
\nonumber
\end{equation}
Note that the $\hat{\theta}_{k}$ are often referred to as predictors in the context of mixed models, but we will use the terms estimates and predictors interchangeably when referring to the $\hat{\theta}_{k}$.

In estimating $\theta_{k}$, it is often assumed, as in the well-known Fay-Herriot model (\cite{Fay:1979}), that the means $\mu_{k}$ are related to the unit-specific covariates $\mathbf{x}_{k}$ via $\mu_{k} = \mathbf{x}_{k}^{T}\bbeta$. We will make such an assumption when deriving the form of particular estimating procedures, but we evaluate MSPE under the more general mixed model formulation (\ref{eq:mixed_model_true}).
For instance, if we add to model (\ref{eq:mixed_model_true}) the three additional working assumptions that $\mu_{k} = \mathbf{x}_{k}^{T}\bbeta$, $v_{k} \sim N(0, \tau^{2}), e_{k} \sim N(0, \sigma_{k}^{2})$, the ``estimate" of $\theta_{k}$ which minimizes MSPE for known values of $\bbeta$ and $\tau_{0}^{2} = \tau^{2}$ is 
\begin{equation}
\theta_{k}( \bbeta ) 
= E\big( \theta_{k} | Y_{k}, \bbeta, \tau^{2} \big)
= B_{k,\tau}\mathbf{x}_{k}^{T}\bbeta + (1 - B_{k,\tau}) Y_{k},
\label{eq:best_predictor}
\end{equation}
where $B_{k,\tau} = \sigma_{k}^{2}/(\sigma_{k}^{2} + \tau^{2})$. 
Additionally, under these three working assumptions, the maximum likelihood estimate of the regression coefficients $\bbeta$ (for an assumed value of $\tau$) is the following quantity
\begin{equation}
\hatbbeta_{MLE} = (\mathbf{X}^{T}\mathbf{V}^{-1}\mathbf{X})^{-1}\mathbf{X}^{T}\mathbf{V}^{-1}\mathbf{Y},
\label{eq:mle}
\end{equation}
where $\mathbf{V} = \tau^{2}\mathbf{I} + \mathbf{V}_{Y|\boldsymbol{\theta}}$. Here, $\mathbf{V}_{Y|\boldsymbol{\theta}} = \textrm{diag}\{\sigma_{1}^{2}, \ldots, \sigma_{K}^{2} \}$ denotes the covariance matrix of $\mathbf{Y}$ conditional on the vector of mixed effects $\btheta$.
If one plugs in $\hatbbeta_{MLE}$ into (\ref{eq:best_predictor}), the associated estimates of the $\theta_{k}$ are
\begin{equation}
\hat{\theta}_{k} = \theta_{k}(\hatbbeta_{MLE})
= B_{k,\tau}\mathbf{x}_{k}^{T}\hatbbeta_{MLE} + (1 - B_{k,\tau}) Y_{k}.
\label{eq:blup}
\end{equation}

The estimate in (\ref{eq:blup}) is commonly referred to as the best linear unbiased predictor (BLUP) of the mixed effect $\mathbf{x}_{k}^{T}\bbeta + v_{k}$ (see e.g., \cite{Henderson:1975}). The BLUPs $\hat{\theta}_{k}$ are optimal in the sense that, under the assumption that $\mu_{k} = \mathbf{x}_{k}^{T}\bbeta$, they achieve the smallest MSPE within the class of  linear unbiased estimators (see, e.g. \cite{Rao:2015} or \cite{Datta:2012}). Beyond the assumptions of model (\ref{eq:mixed_model_true}), the optimality of the BLUPs in (\ref{eq:blup}) only relies on the assumption that $\mu_{k} = \mathbf{x}_{k}^{T}\bbeta$ and that the variance of $v_{k}$ is correctly specified (i.e., $\tau^{2} = \tau_{0}^{2})$ and does not rely on any normality assumptions. In practice, $\tau$ is not usually known and is estimated from the data. When $\hatbbeta_{MLE}$ and the $B_{k,\tau}$ are computed using an estimated value of $\tau$, the resulting mixed-effects estimates in (\ref{eq:blup}) are usually referred to as the empirical best linear unbiased estimates (EBLUPs).

As an alternative to the BLUP estimates of the mixed effects, \cite{Jiang:2011} suggest plugging the following estimate of the regression coefficients into (\ref{eq:best_predictor})
\begin{equation}
\hatbbeta_{BPE} = (\mathbf{X}^{T}\mathbf{B}_{\tau}^{2}\mathbf{X})^{-1}\mathbf{X}^{T}\mathbf{B}_{\tau}^{2}\mathbf{Y}, \qquad \textrm{where} \quad
\mathbf{B}_{\tau} = \textrm{diag}\{ B_{1, \tau}, \ldots, B_{K, \tau} \}.
\label{eq:bpe}
\end{equation}
\cite{Jiang:2011} refer to $\hatbbeta_{BPE}$ as the best predictive estimator (BPE) of $\bbeta$, and they refer to the associated mixed effect estimates $\tilde{\theta}_{k} = \theta_{k}(\hatbbeta_{BPE})$ as the observed best predictor (OBP). The BPE is the best estimator of $\bbeta$ in the sense that it is the vector of regression coefficients minimizing the following estimate $\tilde{Q}(\bbeta)$ of the MSPE associated with any predictor $\theta_{k}(\bbeta)$ of the form (\ref{eq:best_predictor})
\begin{equation}
\tilde{Q}( \bbeta ) = C + \sum_{k=1}^{K} B_{k,\tau}^{2}(\mathbf{x}_{k}^{T}\bbeta)^{2} - 2\sum_{k=1}^{K} B_{k,\tau}^{2}( \mathbf{x}_{k}^{T}\bbeta ) Y_{k},
\end{equation}
where $C$ is a constant not depending on $\bbeta$. The quantity $\tilde{Q}( \bbeta )$ is an unbiased estimate of the MSPE associated with the best predictor (\ref{eq:best_predictor}) when both $\bbeta$ and $\tau$ are assumed to be fixed.

Upon inspection of (\ref{eq:mle}) and (\ref{eq:bpe}), both the MLE and the BPE of $\bbeta$ may be viewed as weighted least-squares estimates with \textit{regression weights} $w_{k}^{mle}(\tau) \propto 1/(\tau^{2} + \sigma_{k}^{2})$ 
and $w_{k}^{bpe}(\tau) \propto \{\sigma_{k}^{2}/(\tau^{2} + \sigma_{k}^{2})\}^{2} = B_{k,\tau}^{2}$ respectively.
Relative to the MLE, the BPE of $\bbeta$ uses regression weights which assign greater weight to units with larger sampling variances $\sigma_{k}^{2}$. Hence, the BPE enables the higher variance units to have more influence in determining the form of the estimate of $\bbeta$. In the context of prediction, this is sensible because it is the units with the largest sampling variances that are shrunken more closely to the fitted regression surface $\mathbf{x}_{k}^{T}\hatbbeta$ while the mixed-effects estimates for low-variance units are impacted much less from the fitted regression. In this sense, the fitted regression surface is more important for the highly variable units, and relative to the MLE, the BPE lets the more variable units play a larger role in fitting this regression surface.

\subsection{Estimating the MSPE for arbitrary regression weights} \label{subsec:estmspe}

While both the BLUP and OBP possess specific optimality properties, these procedures may be potentially improved upon by examining the MSPE associated with an arbitrary weighted least-squares estimate of the regression coefficients. To this end, we consider a vector of unit-specific weights $\mathbf{w} = (w_{1}, \ldots, w_{K})$ with $w_{k} \geq 0$, $\sum_{k} w_{k} = 1$ and where each $w_{k}$ will usually depend on an assumed value of $\tau$. For the choice of regression weights $\mathbf{w}$, the corresponding weighted least-squares estimate of $\bbeta$ is
\begin{equation}
\hatbbeta_{\mathbf{w}} = (\mathbf{X}^{T}\mathbf{W}\mathbf{X})^{-1}\mathbf{X}^{T}\mathbf{W}
\mathbf{Y},
\label{eq:general_fixed_effect}
\end{equation}
where $\mathbf{W} = \textrm{diag}\{ w_{1}, \ldots, w_{K} \}$.
By plugging $\hatbbeta_{\mathbf{w}}$ into (\ref{eq:best_predictor}),
one obtains that the mixed-effects estimates $\hat{\theta}_{k,\mathbf{w}, \tau}$ associated with these weights are $\hat{\theta}_{k,\mathbf{w},\tau} = \theta_{k}( \hatbbeta_{\mathbf{w}} )$. Note that the vector of mixed-effects estimates $\hatbtheta(\mathbf{w}, \tau) = (\hat{\theta}_{1,\mathbf{w},\tau}, \ldots, \hat{\theta}_{K,\mathbf{w},\tau})^{T}$ 
is a linear predictor that can be expressed as 
\begin{equation}
\hatbtheta(\mathbf{w}, \tau) 
= \big( \mathbf{U}_{\mathbf{w}, \tau} + \mathbf{I} \big)\mathbf{Y},
\label{eq:thetahat_w}
\end{equation}
where $\mathbf{U}_{\mathbf{w}, \tau}$ is the $K \times K$ matrix defined as
\begin{equation}
\mathbf{U}_{\mathbf{w},\tau} = \mathbf{B}_{\tau}\Big( \mathbf{X}(\mathbf{X}^{T}\mathbf{W}\mathbf{X})^{-1}\mathbf{X}^{T}\mathbf{W} - \mathbf{I} \Big),
\label{eq:umatrix}
\end{equation}
and where $\mathbf{B}_{\tau}$ is as defined in (\ref{eq:bpe}).
The mixed-effects estimates defined in (\ref{eq:thetahat_w}) can be thought of as defining a class of mixed-effects estimates indexed by both $\tau$ and the vector of unit-specific regression weights $\mathbf{w}$.

For fixed weights $\mathbf{w}$ and an assumed value of $\tau$, we let 
$\textrm{MSPE}(\mathbf{w}, \tau) = \textrm{MSPE}\{ \hatbtheta(\mathbf{w}, \tau) \}$ denote the MSPE associated with $\hatbtheta(\mathbf{w}, \tau)$. This is given by
\begin{equation}
\textrm{MSPE}(\mathbf{w}, \tau) = \bmu^{T}\mathbf{U}_{\mathbf{w}, \tau}^{T}\mathbf{U}_{\mathbf{w}, \tau}\bmu
+ \textrm{tr}\big\{ (\mathbf{U}_{\mathbf{w},\tau} + \mathbf{I})
\mathbf{V}_{Y|\boldsymbol{\theta}}
(\mathbf{U}_{\mathbf{w},\tau} + \mathbf{I})^{T} \big\} + \tau_{0}^{2}\textrm{tr}\big\{ \mathbf{U}_{\mathbf{w}, \tau}^{T}\mathbf{U}_{\mathbf{w}, \tau} \big\}. 
\label{eq:AMSE_population}
\end{equation}
While $\textrm{MSPE}(\mathbf{w}, \tau)$ is unobservable, one may use the fact that the vector $\mathbf{Y}$ has mean $\bmu$ and covariance matrix
$\tau_{0}^{2}\mathbf{I} + \mathbf{V}_{Y|\boldsymbol{\theta}}$
to show that, for fixed weights $\mathbf{w}$ and $\tau \geq 0$, the following quantity 
is an unbiased estimator of (\ref{eq:AMSE_population})
\begin{equation}
\hat{M}_{K}(\mathbf{w}, \tau)
= \mathbf{Y}^{T}\mathbf{U}_{\mathbf{w}, \tau}^{T}\mathbf{U}_{\mathbf{w}, \tau}\mathbf{Y} + 
2\textrm{tr}\big\{ \mathbf{U}_{\mathbf{w}, \tau} \mathbf{V}_{Y|\boldsymbol{\theta}} \big\}
 + 
\textrm{tr}\big\{ \mathbf{V}_{Y|\boldsymbol{\theta}} \big\}.
\label{eq:mspe-estimate}
\end{equation}

It is worth emphasizing that $\hat{M}_{K}(\mathbf{w}, \tau)$ is an unbiased estimator of $\textrm{MSPE}(\mathbf{w},\tau)$ under the assumption that the weight vector $\mathbf{w}$ and $\tau$ are fixed, and the unbiasedness does not hold when either $\mathbf{w}$ or $\tau$ are determined from the data. The estimator $\hat{M}_{K}(\mathbf{w}, \tau)$
is equivalent to Stein's unbiased risk estimate (SURE) (\cite{Stein:1981}) when $\sigma_{1}^{2} = \ldots = \sigma_{K}^{2}$, and hence, $\hat{M}_{K}( \mathbf{w}, \tau)$ may be viewed as a SURE-type estimator where heteroscedasticity is taken into consideration. In the special case of equal regression weights and an intercept-only model (i.e., $\mathbf{X}$ only has an intercept term), $\hat{M}_{K}(\mathbf{w}, \tau)$ is equivalent to the unbiased risk estimate of the shrinkage toward the grand mean estimator described in \cite{Xie:2012}.


It is worth noting that the unbiasedness of $\hat{M}_{K}(\mathbf{w}, \tau)$ only relies on the assumptions of model (\ref{eq:mixed_model_true}) and does not require any further assumptions about the distributions of $v_{k}$ or $e_{k}$.
Moreover, as stated in the following theorem, this unbiasedness holds even if one is interested evaluating the MSPE conditional on the unobserved $\btheta$ rather than marginally over $\btheta$.
\begin{theorem}
Under model (\ref{eq:mixed_model_true}), $\hat{M}_{K}(\mathbf{w}, \tau)$ is an unbiased estimate of $\textrm{MSPE}(\mathbf{w}, \tau)$ in the sense that 
\begin{equation}
E\{ \hat{M}_{K}(\mathbf{w}, \tau) \} = \textrm{MSPE}(\mathbf{w}, \tau).  \nonumber
\end{equation}
Moreover, the expectation of $\hat{M}_{K}(\mathbf{w}, \tau)$ conditional on $\btheta$ is equal to the conditional MSPE associated with $\hatbtheta( \mathbf{w}, \tau)$
\begin{equation}
E\big\{ \hat{M}_{K}( \mathbf{w}, \tau) \mid \btheta \big\} = 
E\Big[ \big\{ \hatbtheta(\mathbf{w}, \tau) - \btheta \big\}^{T}\big\{\hatbtheta(\mathbf{w}, \tau) - \btheta\big\} \Big| \btheta \Big]. \nonumber
\end{equation}
\label{thm:unbiased_estimate}
\end{theorem}

\subsection{Compromise Regression Weights and the CBP} \label{subsec:cbpdef}
The quantity $\hat{M}_{K}(\mathbf{w}, \tau)$ is only guaranteed to be an unbiased estimate of the MSPE of $\hatbtheta( \mathbf{w}, \tau )$ for fixed weights, and hence may be an inappropriate way of evaluating the MSPE associated with data-determined weights. Nevertheless, comparing $\hat{M}_{K}( \mathbf{w}, \tau)$ for different weights
can be a useful way for choosing among different weighting schemes when such weights are indexed by a small number of hyperparameters.
To allow the form of the weights to be partially driven by the observed value of $\hat{M}_{K}(\mathbf{w}, \tau)$ without spending many additional degrees of freedom, we consider a family of weights that are convex combinations of the MLE and BPE weights. This only requires that we estimate one additional hyperparameter (i.e., the mixing parameter) when compared to the EBLUP or the OBP.
Though one could use other weights to form the components of a convex combination, 
the choice of the MLE and BPE as the ``basis'' weights is motivated by a particular decomposition of
$\textrm{MSPE}(\mathbf{w}, \tau)$ described in \cite{Jiang:2011}. In the following proposition, we state a version of Theorem 1 in \cite{Jiang:2011} which specializes this theorem to our formulation of the mixed-effects prediction problem. 
\begin{prop}
(Due to \cite{Jiang:2011}) Consider the expression for the MSPE given in (\ref{eq:AMSE_population})
\begin{equation}
\textrm{MSPE}(\mathbf{w}, \tau) = \bmu^{T}\mathbf{U}_{\mathbf{w}, \tau}^{T}\mathbf{U}_{\mathbf{w}, \tau}\bmu
+ \textrm{tr}\big\{ (\mathbf{U}_{\mathbf{w},\tau} + \mathbf{I})
\mathbf{V}_{Y|\theta}
(\mathbf{U}_{\mathbf{w},\tau} + \mathbf{I})^{T} + \tau_{0}^{2} \mathbf{U}_{\mathbf{w}, \tau}^{T}\mathbf{U}_{\mathbf{w}, \tau} \big\}.
\nonumber
\end{equation}
For fixed $\tau$, $\bmu^{T}\mathbf{U}_{\mathbf{w}, \tau}^{T}\mathbf{U}_{\mathbf{w}, \tau}\bmu$ is minimized when $\mathbf{w}$ are the BPE weights, and, when $\tau$ is fixed at $\tau_{0}$, the second term is minimized by the MLE weights.
Moreover, $\bmu^{T}\mathbf{U}_{\mathbf{w},\tau}^{T}\mathbf{U}_{\mathbf{w},\tau}\bmu = \mathbf{0}$ 
whenever $\bmu = \mathbf{X}\bbeta$ for some $\bbeta \in \mathbb{R}^{p}$.
\label{thm:amse_decomp}
\end{prop}

Proposition \ref{thm:amse_decomp} states that the MSPE may be decomposed into a model misspecification term and a variance term which are minimized by the BPE weights and MLE weights respectively. This suggests that using weights which compromise between these two weighting schemes can potentially lead to meaningful reductions in MSPE. Moreover, allowing the degree of compromise to be data dependent enables the compromise weights to adapt to the extent of model misspecification and of estimation variance.

To compute the empirically-driven compromise weights, we adopt a direct approach which uses a convex combination of the MLE and BPE weights. Specifically, for $\alpha \in [0,1]$, we consider the family of compromise weights 
\begin{equation}
\mathbf{w}^{c}(\alpha, \tau) 
= (w_{1}^{c}(\alpha,\tau), \ldots, w_{K}^{c}(\alpha, \tau) )^{T},
\nonumber
\end{equation}
where the $k^{th}$ element of $\mathbf{w}^{c}(\alpha, \tau)$ is a convex combination of the $k^{th}$ MLE weight $w_{k}^{mle}(\tau)$ and the $k^{th}$ BPE weight $w_{k}^{bpe}(\tau)$
\begin{eqnarray}
w_{k}^{c}(\alpha, \tau) &=& \alpha w_{k}^{mle}(\tau) + (1 - \alpha)w_{k}^{bpe}(\tau)
\nonumber \\
&=& 
\frac{\alpha/(\sigma_{k}^{2} + \tau^{2})}{\sum_{k=1}^{K} (\sigma_{k}^{2} + \tau^{2})^{-1}}
+ \frac{(1 - \alpha)\{ \sigma_{k}^{2}/(\sigma_{k}^{2} + \tau^{2})\}^{2} }{
\sum_{k=1}^{K} \{ \sigma_{k}^{2}/(\sigma_{k}^{2} + \tau^{2})\}^{2} }.
\label{eq:compromise_weight}
\end{eqnarray}
To determine the optimal values of $\alpha$ and $\tau$ for the compromise weights, we minimize the estimate $\hat{M}_{K}\big( \mathbf{w}^{c}(\alpha, \tau), \tau \big)$ of the MSPE that is associated with this vector of regression weights. Because this estimate only depends on $(\alpha,\tau)$ for the family of weights (\ref{eq:compromise_weight}), we henceforth use $\hat{M}_{K}^{c}(\alpha, \tau) = \hat{M}_{K}\big( \mathbf{w}^{c}(\alpha, \tau), \tau \big)$ to denote the unbiased MSPE estimate when using compromise estimates $\mathbf{w}^{c}(\alpha, \tau)$. Using $\hat{M}_{K}^{c}(\alpha, \tau)$, the optimal values $(\alpha^{*}, \tau^{*})$ for the compromise weights are determined empirically as 
\begin{equation}
(\alpha^{*}, \tau^{*}) = \argmin_{\alpha \in [0,1], \tau \geq 0}\hat{M}_{K}^{c}(\alpha, \tau) = \argmin_{\alpha \in [0,1], \tau \geq 0} \hat{M}_{K}( \mathbf{w}^{c}(\alpha, \tau), \tau). 
\label{eq:optimal_cure}
\end{equation}
Recalling (\ref{eq:general_fixed_effect}), the weights $\mathbf{w}^{c}(\alpha^{*}, \tau^{*})$ will generate the following empirically-driven compromise estimate $\hatbbeta_{cure}$ of the fixed effects regression coefficients
\begin{equation}
\hatbbeta_{cure} =  
(\mathbf{X}^{T}\mathbf{W}_{\alpha^{*}, \tau^{*}}^{c}\mathbf{X})^{-1}
\mathbf{X}^{T}\mathbf{W}_{\alpha^{*}, \tau^{*}}^{c}\mathbf{Y},
\end{equation}
where $\mathbf{W}_{\alpha,\tau}^{c} = \textrm{diag}\{ w_{1}^{c}(\alpha,\tau), \ldots, 
w_{K}^{c}(\alpha, \tau) \}$. We label $\hatbbeta_{cure}$ the ``compromise unbiased risk estimator", or CURE of the regression coefficients.

The mixed-effects estimates $\check{\theta}_{k}$ associated with the optimal compromise regression weights are then defined as
\begin{equation}
\check{\theta}_{k} = 
B_{k,\tau^{*}}\mathbf{x}_{k}^{T}\hatbbeta_{cure} +  
(1 - B_{k, \tau^{*}})Y_{k}.
\label{eq:cbp-thetak}
\end{equation}
We refer to the vector of mixed-effects estimates $\checkbtheta^{CBP} = (\check{\theta}_{1}, \ldots, \check{\theta}_{K})^{T}$ 
as the ``compromise best predictor" or CBP of $\btheta$. Recalling (\ref{eq:thetahat_w}) and (\ref{eq:umatrix}), we can also express $\checkbtheta^{CBP}$ as
\begin{equation}
\checkbtheta^{CBP} = \hatbtheta\big( \mathbf{w}^{c}(\alpha^{*}, \tau^{*}), \tau^{*} \big) 
= \mathbf{Y} + \mathbf{B}_{\tau^{*}}\Big(\mathbf{X}(\mathbf{X}^{T}\mathbf{W}_{\alpha^{*}, \tau^{*}}^{c}\mathbf{X})^{-1}
\mathbf{X}^{T}\mathbf{W}_{\alpha^{*}, \tau^{*}}^{c} - \mathbf{I}\Big) \mathbf{Y}. \nonumber
\end{equation}


In practice, we compute the optimal values $(\alpha^{*}, \tau^{*})$ in (\ref{eq:optimal_cure})
using the constraints $(\alpha, \tau) \in [0, 1] \times [0, \tau_{max}]$. The maximal value
of $\tau$ is determined empirically and is set to $\tau_{max} = 10\sqrt{\tfrac{1}{K-1}\sum_{k=1}^{K} (Y_{k} - \bar{Y})^{2}}$.
Because $\tau^{2}$ represents the variance of the random effects $v_{k}$, the sample variance of the $Y_{k}$ is likely to be an overestimate of the best value of $\tau^{2}$ as $\tau^{2}$ only accounts for a fraction of the variation in the $Y_{k}$.
Hence, setting $\tau_{max}$ equal to 10 times the sample standard deviation can be interpreted as choosing an upper bound which is very likely to be a substantial overestimate of the optimal value of $\tau$.
Minimization of $\hat{M}_{K}^{c}(\alpha, \tau)$ with respect to these box constraints is performed using the limited-memory
BFGS algorithm (\cite{Byrd:1995}) which we have found to work quite well in this context.

It is also possible to construct compromise regression weights in the context of a nested-error regression model (\cite{Battese:1988})
using a very similar approach to the one outlined above.
Specifically, one would compute the optimal compromise regression weights by minimizing an unbiased estimator 
of the MSPE associated with a predictor of the mixed effects.
Section C of the appendix describes an unbiased estimator of the MSPE in the context of a nested-error regression model.

\subsection{Variations of the CBP} \label{subsec:variations}
We also consider two close variations of the CBP. The first of these, which we call the ``plug-in" CBP, uses a restricted maximum likelihood (REML) and an OBP-based estimate of $\tau$ as the starting point for defining the shrinkage and regression weights and then uses a convex combination of these two values of $\tau$ in both the regression and shrinkage weights. Another alternative which we explore is the ``multi-$\tau$" CBP. 
In a variety of numerical studies, we have observed that the plug-in version of the CBP often has better finite-sample performance than the CBP estimates defined in (\ref{eq:cbp-thetak}). For very large values of $K$, we typically see similar performance between the CBP and plug-in CBP. See Section \ref{sec:simulations} for comparisons of the performance of the CBP using both the plug-in and multi-$\tau$ approaches.

\medskip

\noindent
\textit{The plug-in CBP.} For the plug-in CBP, we consider regression weights $w_{k}^{c,1}(\alpha)$ and 
shrinkage weights $B_{k}^{1}(\alpha)$ of the form 
\begin{eqnarray}
w_{k}^{c,1}(\alpha) &=& \alpha w_{k}^{mle}(\hat{\tau}_{REML})
+ (1 - \alpha) w_{k}^{bpe}(\hat{\tau}_{OBP}) \nonumber \\
B_{k}^{1}(\alpha) &=& \frac{\sigma_{k}^{2}}{ \sigma_{k}^{2} + \alpha \hat{\tau}_{REML}^{2} + (1 - \alpha)\hat{\tau}_{OBP}^{2}}, \nonumber 
\end{eqnarray}
where $\hat{\tau}_{REML}$ denotes the restricted maximum likelihood (REML) estimate of $\tau$ while $\hat{\tau}_{OBP}$ denotes the OBP-based estimate of $\tau$. As described in \cite{Jiang:2011}, the OBP-based estimate $\hat{\tau}_{OBP}$ maximizes the following objective function
\begin{equation}
Q_{OBP}(\tau) = \mathbf{Y}^{T}(\mathbf{B}_{\tau}^{2} - \mathbf{B}_{\tau}^{2}(\mathbf{X}^{T}\mathbf{B}_{\tau}^{2}\mathbf{X})^{-1}\mathbf{X}^{T}\mathbf{B}_{\tau}^{2})\mathbf{Y} + 2\tau^{2}\textrm{tr}( \mathbf{B}_{\tau}). \nonumber
\end{equation}
Because $w_{k}^{c,1}(\alpha)$ and $B_{k}^{1}(\alpha)$ only depend on $\alpha$, the unbiased MSPE estimate $\hat{M}_{K}(\mathbf{w}, \tau)$ defined in (\ref{eq:mspe-estimate})
will only depend on $\alpha$. The plug-in CBP regression and shrinkage weights are then obtained by plugging in the value of $\alpha$ which minimizes $\hat{M}_{K}(\mathbf{w}, \tau)$ into both $w_{k}^{c,1}(\alpha)$ and $B_{k}^{1}(\alpha)$.

\medskip

\noindent
\textit{The multi-$\tau$ CBP.} For the multi-$\tau$ CBP, we consider weights similar to those for the plug-in CBP except that we do not restrict the values of the variance component terms
to equal specific values. In particular, the multi-$\tau$ CBP considers regression weights $w_{k}^{c,2}(\alpha, \tau_{0}, \tau_{1})$ and 
shrinkage weights $B_{k}^{2}(\alpha, \tau_{0}, \tau_{1})$ of the form 
\begin{eqnarray}
w_{k}^{c,2}(\alpha, \tau_{0}, \tau_{1}) &=& \alpha w_{k}^{mle}(\tau_{1} )
+ (1 - \alpha) w_{k}^{bpe}(\tau_{0} ) \nonumber \\
B_{k}^{2}(\alpha, \tau_{0}, \tau_{1}) &=& \frac{\sigma_{k}^{2}}{ \sigma_{k}^{2} + \alpha \tau_{1}^{2} + (1 - \alpha)\tau_{0}^{2}}, \nonumber 
\end{eqnarray}
with the values of $(\alpha, \tau_{0}, \tau_{1})$ being chosen to minimize (\ref{eq:mspe-estimate}).

\subsection{A More General CBP}
In this subsection, we consider the following more general formulation of the linear mixed model
\begin{equation}
\mathbf{Y} = \bmu + \mathbf{Z}\mathbf{v} + \mathbf{e},
\label{eq:mixmodel_general}
\end{equation}
where $\mathbf{Z}$ is a known $p_{z} \times K$ model matrix and where $E(\mathbf{v}) = E(\mathbf{e}) = \mathbf{0}$, 
$\var(\mathbf{v}) = \mathbf{G}_{\lambda}$, and $\var(\mathbf{e}) = \bSigma$. The covariance matrix $\bSigma$ of $\mathbf{e}$ is assumed to be known, and the entries of the covariance matrix $\mathbf{G}_{\lambda}$ of $\mathbf{v}$ are assumed to be determined by the $g \times 1$ parameter vector $\blambda = (\lambda_{1}, \ldots, \lambda_{g})^{T}$. The parameter vector $\blambda \in \Lambda$ could, for example, represent parameters modeling spatial dependence.
The marginal covariance matrix of $\mathbf{Y}$ under model (\ref{eq:mixmodel_general}) is given by $\var(\mathbf{Y}) = \mathbf{V}_{\lambda} = \mathbf{Z}\mathbf{G}_{\lambda}\mathbf{Z}^{T} + \bSigma$.

We now consider the situation where one is interested in predicting the following $p_{\eta} \times 1$ vector of mixed effects
\begin{equation}
\boldeta = \mathbf{A}^{T}\bmu + \mathbf{R}^{T}\mathbf{v},
\label{eq:general_target}
\end{equation}
where $\mathbf{A}$ and $\mathbf{R}$ are fixed $K \times p_{\eta}$ matrices. Under model (\ref{eq:mixmodel_general}) with the additional assumptions that $\bmu = \mathbf{X}\bbeta$, $\mathbf{v} \sim N(\mathbf{0}, \mathbf{G}_{\lambda})$, and $\mathbf{e} \sim N(\mathbf{0}, \bSigma)$, the best predictor of $\boldeta$ for known $\bbeta$ is given by $\boldeta(\bbeta) = \mathbf{A}^{T}\mathbf{X}\bbeta + \mathbf{R}^{T}\mathbf{G}_{\lambda}\mathbf{Z}^{T}\mathbf{V}_{\lambda}^{-1}(\mathbf{Y} - \mathbf{X}\bbeta)$. 
Because $\mathbf{G}_{\lambda}$ or $\bSigma$ may be non-diagonal, we now consider estimates of $\bbeta$ which depend on a potentially non-diagonal positive definite weight matrix $\mathbf{W}$ rather than the diagonal matrix $\mathbf{W}$ considered in Sections \ref{subsec:estmspe} and \ref{subsec:cbpdef}. By plugging in the weighted least-squares estimate $\hatbbeta_{\mathbf{W}} = (\mathbf{X}^{T}\mathbf{W}\mathbf{X})^{-1}\mathbf{X}^{T}\mathbf{W}\mathbf{Y}$ into $\boldeta(\bbeta)$ we obtain the class of estimates $\hatboldeta_{\mathbf{W},\lambda} = \boldeta(\hatbbeta_{\mathbf{W}})$ given by
\begin{equation}
\hatboldeta_{\mathbf{W}, \lambda} = (\mathbf{L}_{\mathbf{W}, \lambda} + \mathbf{A}^{T})\mathbf{Y}, \quad \textrm{ where } \quad
\mathbf{L}_{\mathbf{W}, \lambda} = \mathbf{R}^{T}\mathbf{G}_{\lambda}\mathbf{Z}^{T}\mathbf{V}_{\lambda}^{-1}
\mathbf{P}_{\mathbf{W}} - \mathbf{A}^{T}\mathbf{P}_{\mathbf{W}},
\label{eq:boldeta_class}
\end{equation}
and where $\mathbf{P}_{\mathbf{W}}$ is the matrix $\mathbf{P}_{\mathbf{W}} = \mathbf{I} - \mathbf{X}(\mathbf{X}^{T}\mathbf{W}\mathbf{X})^{-1}\mathbf{X}^{T}\mathbf{W}$. The MSPE associated with $\hatboldeta_{\mathbf{W},\lambda}$ is
$\textrm{MSPE}(\mathbf{W}, \blambda) = E\{ (\hatboldeta_{\mathbf{W},\lambda} - \boldeta)^{T}(\hatboldeta_{\mathbf{W},\lambda} - \boldeta) \}$ which can be shown to equal
\begin{eqnarray}
\textrm{MSPE}(\mathbf{W}, \blambda)
&=&   \bmu^{T}\mathbf{L}_{\mathbf{W}, \lambda}^{T}\mathbf{L}_{\mathbf{W}, \lambda}\bmu + 
\textrm{tr}\Big\{ \mathbf{L}_{\mathbf{W},\lambda} \mathbf{V}_{\lambda}\mathbf{L}_{\mathbf{W},\lambda}^{T}  \Big\} 
+ 2\textrm{tr}\Big\{ \mathbf{L}_{\mathbf{W}, \lambda}(\mathbf{V}_{\lambda}\mathbf{A} - \mathbf{Z}\mathbf{G}_{\lambda}\mathbf{R}) \Big\} \nonumber \\
&+& \textrm{tr}\Big\{ \mathbf{A}^{T}\mathbf{V}_{\lambda}\mathbf{A} \Big\} - 2\textrm{tr}\Big\{ \mathbf{R}^{T}\mathbf{G}_{\lambda}\mathbf{Z}^{T}\mathbf{A} \Big\} + \textrm{tr}\Big\{ \mathbf{R}^{T}\mathbf{G}_{\lambda}\mathbf{R} \Big\}.
\end{eqnarray}
It can also be shown (see Section C of the supplementary material) that, for fixed weight matrix $\mathbf{W}$ and $\blambda$, 
\begin{eqnarray}
\hat{M}_{K,g}( \mathbf{W}, \blambda) &=& \mathbf{Y}^{T}\mathbf{L}_{\mathbf{W}, \lambda}^{T}\mathbf{L}_{\mathbf{W}, \lambda}\mathbf{Y}
+ 2\textrm{tr}\Big\{ \mathbf{L}_{\mathbf{W}, \lambda}(\mathbf{V}_{\lambda}\mathbf{A} - \mathbf{Z}\mathbf{G}_{\lambda}\mathbf{R}) \Big\} + \textrm{tr}\Big\{ \mathbf{A}^{T}\mathbf{V}_{\lambda}\mathbf{A} \Big\} \nonumber \\
 &-& 2\textrm{tr}\Big\{ \mathbf{R}^{T}\mathbf{G}_{\lambda}\mathbf{Z}^{T}\mathbf{A} \Big\} 
 + \textrm{tr}\Big\{ \mathbf{R}^{T}\mathbf{G}_{\lambda}\mathbf{R} \Big\}
\label{eq:general_m}
\end{eqnarray}
is an unbiased estimator of $\textrm{MSPE}(\mathbf{W},\blambda)$, i.e.,
$E\{ \hat{M}_{K,g}(\mathbf{W}, \blambda) \} = \textrm{MSPE}( \mathbf{W}, \blambda)$.
In this setting, the empirically driven CURE of $\bbeta$ is given by
\begin{equation}
\hatbbeta_{CURE}^{g} =  
(\mathbf{X}^{T}\mathbf{W}_{\alpha^{*g}, \lambda^{*g}}^{c,g}\mathbf{X})^{-1}
\mathbf{X}^{T}\mathbf{W}_{\alpha^{*g}, \lambda^{*g}}^{c,g}\mathbf{Y},
\label{eq:cure_general}
\end{equation}
where $\mathbf{W}_{\alpha, \lambda}^{c,g}$ is the convex combination matrix 
\begin{equation}
\mathbf{W}_{\alpha,\lambda}^{c,g} = \alpha\mathbf{V}_{\lambda}^{-1}
+ (1 - \alpha)(\mathbf{A}^{T} - \mathbf{R}^{T}\mathbf{G}_{\lambda}\mathbf{Z}^{T}\mathbf{V}_{\lambda}^{-1})^{T}(\mathbf{A}^{T} - \mathbf{R}^{T}\mathbf{G}_{\lambda}\mathbf{Z}^{T}\mathbf{V}_{\lambda}^{-1}). \nonumber 
\end{equation}
The optimal compromise parameters $\alpha^{*g}$ and $\blambda^{*g}$ used in (\ref{eq:cure_general}) are found by minimizing 
the unbiased risk estimate (\ref{eq:general_m}). Specifically, $(\alpha^{*g}, \blambda^{*g})
= \argmin_{\alpha \in [0,1], \lambda \in \Lambda}\hat{M}_{K,g}(\mathbf{W}_{\alpha, \lambda}^{c,g}, \blambda)$.
Using $\alpha^{*g}$ and $\blambda^{*g}$, the CBP $\checkboeta^{CBP}$ of $\boldeta$ is then defined as 
\begin{equation}
\checkboeta^{CBP} = (\mathbf{L}_{\mathbf{W}_{\alpha^{*g}, \mathbf{\lambda}^{*g}}^{c,g}, \lambda^{*g}} + \mathbf{A}^{T})\mathbf{Y}. \nonumber
\end{equation}

\section{Estimating a Population Mean} \label{sec:pop_mean}
We now consider the case where the primary target of inference is an average of unit-level attributes rather than the unit-level attributes themselves. Such a target of interest may arise, for example, if one is primarily interested in assessing the average test performance of randomly selected schools in a particular region 
rather than estimating the test performance of individual schools. 
For this type of goal, we take as the population estimand the equally weighted average of the unit-specific means
\begin{equation} 
\mu_0 = \frac{1}{K}\sum_{k = 1}^{K} \theta_k,
\label{eq:theta.bar}
\end{equation} 
but more general weighted averages may be approached in a similar manner.
Note that the single-parameter target (\ref{eq:theta.bar}) is a special case of (\ref{eq:general_target}), where $\mu_{0} = \boldeta$, $\mathbf{A} = \mathbf{R}$, and $\mathbf{A}$ is the $K \times 1$ column vector $\mathbf{A} = (1/K, \ldots, 1/K)^{T}$.

An inferential target such as $\mu_{0} = K^{-1}\sum_{k} \theta_{k}$ arises often, for example, when one uses a stratified random sample to estimate a population mean. In such cases, the population is divided into $K$ separate strata/units, and for each stratum/unit $k$, $n_{k}$ responses $Z_{ik}, i = 1,\ldots, n_{k}$ are drawn from stratum $k$. The sample mean from stratum $k$, $Y_{k} = n_{k}^{-1}\sum_{i} Z_{ik}$, is an unbiased estimate of the stratum-specific mean $\theta_{k}$, and hence, it is sensible to focus on the  ``mean model" wherein
$E(Y_{k}|\theta_{k}) = \theta_{k}$, $\var(Y_{k}|\theta_{k}) = \sigma_{k}^{2} = \sigma^{2}/n_{k}$, and where there are no unit-specific covariates used in the analysis. In other words, the working model is that $Y_{k} = \beta_{0} + v_{k} + e_{k}$ with $E(v_{k}) = E(e_{k}) = 0$, $\var(v_{k}) = \tau^{2}$, and $\var(e_{k}) = \sigma^{2}/n_{k}$.
In this context, estimating $\mu_{0}$ using compromise weights can be especially useful when the sample sizes are informative in the sense that the $n_{k}$ have some association
with the values of $\theta_{k}$.

In this setting, specializing (\ref{eq:boldeta_class}) to the estimation of $\mu_{0}$ leads to the following family of estimates 
\begin{eqnarray}
\hat{\mu}_{0}(\mathbf{w}, \tau)
&=& \frac{1}{K} \sum_{k=1}^{K} B_{k,\tau}\Big( \frac{\sum_{k=1}^{K}Y_{k}w_{k} }{K\sum_{k=1}^{K} w_{k}} \Big)  + \frac{1}{K}\sum_{k=1}^{K} (1 - B_{k,\tau})Y_{k} \nonumber \\
&=& B_{.\tau} \Big(\frac{\sum_{k=1}^{K}Y_{k}w_{k} }{K\sum_{k=1}^{K} w_{k}} \Big)
+ \frac{1}{K}\sum_{k=1}^{K} (1 - B_{k,\tau})Y_{k},
\label{eq:mu_naught}
\end{eqnarray}
where $B_{k, \tau} = \sigma_{k}^{2}/(\sigma_{k}^{2} + \tau^{2}) = \sigma^{2}/(\sigma^{2} + n_{k}\tau^{2})$ and $B_{.\tau} = \sum_{k=1}^{K} B_{k,\tau}$. Following (\ref{eq:general_m}), an unbiased estimate of $E\{ (\hat{\mu}_{0}(\mathbf{w}, \tau) - \mu_{0})^{2}\}$ is given by
\begin{eqnarray}
\hat{M}_{K,g,0}(\mathbf{w}, \tau)
&=& \Big(\frac{1}{K}\sum_{k=1}^{K}B_{k,\tau}Y_{k} - \frac{B_{.\tau}}{K}\sum_{j=1}^{K} w_{j}Y_{j}  \Big)^{2} + \frac{2B_{.\tau}\sigma^{2}}{K^{2}}\sum_{k=1}^{K} \frac{ w_{k}}{n_{k}} \nonumber \\
&-& \frac{2}{K^{2}}\sum_{k=1}^{K} \frac{\sigma^{4}}{n_{k}(\sigma^{2} + n_{k}\tau^{2})}  + \frac{\sigma^{2}}{K \ddot n},
\label{eq:hat_Mnaught}
\end{eqnarray}
where $\ddot n = (\frac{1}{K} \sum_{k} 1/n_{k} )^{-1}$ represents the harmonic mean of the unit-specific sample sizes.  Note that when the $\theta_{k}$ are treated deterministically (i.e., $\tau = 0)$ $B_{k,\tau} = 1$ and the estimator in (\ref{eq:mu_naught}) reduces to 
$\hat{\mu}(\mathbf{w}, 0) = (\sum_{k}Y_{k}w_{k})/ \sum_{k} w_{k}$.

There are a variety of approaches for selecting weights from which to plug into $\hat{\mu}_{0}(\mathbf{w}, \tau)$.
Regardless of the value of $\tau$, weights chosen so that $w_{k} \propto B_{k,\tau}$ lead to $\hat{\mu}_{0}(\mathbf{w}, \tau)$ becoming the direct unbiased estimator of $\mu_{0}$ given by 
\begin{equation}
\hat{\mu}_{0}^{direct} = \frac{1}{K}\sum_{k=1}^{K} Y_{k}. 
\end{equation}
It is also worth noting that when the $\theta_{k}$ are treated deterministically (i.e., $\tau = 0$) the BPE weights are uniform (i.e., $w_{k}^{BPE} \propto 1$), and the corresponding estimate of $\mu_{0}$ is the same as the direct estimate. In other words, $\hat{\mu}_{0}^{direct} = \hat{\mu}_{0}(\mathbf{w}^{BPE}, 0)$ when the vector of BPE weights $\mathbf{w}^{BPE}$ are formed under the assumption that $\tau = 0$.

Weights $w_{k}^{MV}$ minimizing the variance (conditional on the $\theta_{k})$ of $\hat{\mu}_{0}(\mathbf{w}, \tau)$ are given by $w_{k}^{MV} \propto \{ (B_{k,\tau} - 1)\bar{n} + n_{k} \}$, and these weights lead to the following estimator of $\mu_{0}$
\begin{equation}
\hat{\mu}_{0}^{MV} = \frac{1}{K \bar{n}}\sum_{k=1}^{K} Y_{k}n_{k}, 
\end{equation}
where $\bar{n} = K^{-1}\sum_{k} n_{k}$. When the $\theta_{k}$ are treated deterministically so that $\tau = 0$ and $B_{k,\tau} = 1$, the weights $w_{k}^{MV}$ and $w_{k}^{MLE}$ are equivalent. Hence, $\hat{\mu}_{0}^{MV} = \hat{\mu}(\mathbf{w}^{MLE}, 0)$ when the vector of MLE weights $\mathbf{w}^{MLE}$ are formed under the assumption that $\tau = 0$.

As in Section \ref{sec:CBP_describe}, we consider an arbitrary convex combination of two weighting schemes in order to improve estimation of $\mu_{0}$. Given two vectors of weights $\mathbf{w}_{0}(\tau)$ and $\mathbf{w}_{1}( \tau )$, compromise parameters $(\alpha, \tau)$ are obtained by minimizing $\hat{M}_{K,g,0}\big( \mathbf{w}^{c}(\alpha, \tau), \tau \big)$ where $\mathbf{w}^{c}(\alpha, \tau) = \alpha \mathbf{w}_{1}(\tau) + (1 - \alpha)\mathbf{w}_{0}(\alpha, \tau)$. As stated in the following theorem, for fixed $\tau$ the optimal value of the mixing proportion $\alpha$ actually has a closed-form expression.

\begin{theorem}
Let $w_{k}^{1}(\tau)$ and $w_{k}^{0}(\tau)$ be two weighting schemes such that $\sum_{k} w_{k}^{1}(\tau) = \sum_{k} w_{k}^{0}(\tau) = 1$, for all $\tau \geq 0$. Consider compromise weights $\mathbf{w}^{c}(\alpha, \tau) = \big( w_{1}^{c}(\alpha, \tau), \ldots, w_{K}^{c}(\alpha,\tau) \big)$ defined as $w_{k}^{c}(\alpha, \tau) = \alpha w_{k}^{1}(\tau) + (1 -  \alpha) w_{k}^{0}(\tau)$. If we consider $\hat{M}_{g,0}(\mathbf{w}^{c}(\alpha, \tau), \tau)$ as a function of $\alpha$ and $\tau$ where $\hat{M}_{K,g,0}$ is as defined in (\ref{eq:hat_Mnaught}), then, for a fixed $\tau$ such that $\sum_{k} w_{k}^{0}(\tau) Y_{k} \neq \sum_{k} w_{k}^{1}(\tau) Y_{k}$, the value of $\alpha \in [0,1]$ which minimizes $\hat{M}_{K,g,0}(\mathbf{w}^{c}(\alpha, \tau), \tau)$ is given by
\begin{equation}
    \alpha_{opt}(\tau) = 
\begin{cases}
    0,& \text{ if } C_{2}(\tau) \leq 0 \\
    1,              & \text{ if } C_{1}(\tau) \leq C_{2}(\tau)  \\
    C_{2}(\tau)/C_{1}(\tau), & \text{ otherwise}
\end{cases}
\end{equation}
where $C_{1}(\tau)$ and $C_{2}(\tau)$ are defined as:
\begin{eqnarray}
C_{1}(\tau) &=& \frac{B_{.\tau}^{2} }{K^{2}} \Big( \sum_{j=1}^{K} \{ w_{j}^{1}(\tau) -  w_{j}^{0}(\tau) \} Y_{j} \Big)^{2} \nonumber \\
C_{2}(\tau) &=& \frac{B_{.\tau}}{K^{2}}\Big( \sum_{j=1}^{K} \{ w_{j}^{1}(\tau) -  w_{j}^{0}(\tau) \} Y_{j} \Big)\Bigg( \sum_{j=1}^{K} B_{k,\tau} \Big[ Y_{k} - \sum_{j=1}^{K}w_{j}^{0}(\tau)Y_{j} \Big] \Bigg)  - \frac{B_{.\tau}\sigma^{2}}{K^{2}}\sum_{k=1}^{K} \frac{ w_{k}^{1}(\tau) - w_{k}^{0}(\tau) }{n_{k}}  \nonumber 
\end{eqnarray}
\label{thm:single_mean}
\end{theorem}

\medskip

\noindent
\textit{Example: Combining the Minimum Variance and Direct Estimates.} Suppose we want a compromise estimate based on the minimum variance $w_{k}^{1}(0) = n_{k}/K\bar{n}$ and the direct estimate weights $w_{k}^{0}(0) = 1/K$ while assuming that $\tau$ is fixed at zero. In this case, $w_{k}^{1}(0) - w_{k}^{0}(0) = (n_{k} - \bar{n})/K\bar{n}$, and a direct computation (see Section A of the appendix) shows that
\begin{equation}
\alpha_{opt}(0) = 
\min\Bigg\{ \max\Big\{ \frac{\sigma^{2}(\bar{n} - \ddot{n})}{K\bar{n}\ddot{n}(\hat{\mu}^{MV} - \hat{\mu}^{direct})^{2}}, 0 \Big\}, 1\Bigg\}. \label{eq:alpha_popmean_direct}
\end{equation}

\bigskip

An alternative approach for estimating a single population quantity such as $\mu_{0}$ is to construct a flexible regression model relating the unit-specific means and unit-specific sample sizes. As described in \cite{Zheng:2005} and \cite{Little:2004} in the survey context, if sample inclusion is informative, bias can be reduced or removed by building a regression model that adjusts for the association and then uses weighted least squares. Though our concern in this context is informative sample size, the structure of the problem is essentially the same. Similar to \cite{Zheng:2005}, we consider the following model for the direct estimates $Y_{k}$
\begin{eqnarray}
Y_{k}|\theta_{k} &\sim& \textrm{Normal}\big( \theta_{k}, \sigma^{2}/n_{k} \big)  \nonumber \\
\theta_{k} &=& h(n_{k}; \bbeta^{h}), \label{eq:spline-estimate}
\end{eqnarray}
where the function $h$ is a spline with coefficients $\bbeta^{h}$ though more general models for $h$ could be considered. After using weighted least squares to estimate the spline coefficients $\hatbbeta^{h}$, the estimate $\hat{\mu}_{0}^{sr}$ of $\mu_{0}$ is then obtained by taking the average of the fitted values. That is, $\hat{\mu}_{0}^{sr} = K^{-1}\sum_{k=1}^{K} h(n_{k}, \hatbbeta^{h})$. The estimator $\hat{\mu}^{sr}$ of the population average is also very similar to the approach described in \cite{Matloff:1981} where an estimator is developed to utilize additional covariate information when the goal is to estimate the unconditional mean of an outcome. Section \ref{subsec:popavgsim} describes a simulation study comparing the performance of $\hat{\mu}_{0}^{sr}, \hat{\mu}_{0}^{direct}, \hat{\mu}_{0}^{MV}$, and a direct compromise estimator which uses $\alpha_{opt}(0)$ defined in (\ref{eq:alpha_popmean_direct}) as the compromise parameter.

\section{Asymptotic Risk of the CBP} \label{sec:asymptotic_risk}
In this section, we compare the mean-squared prediction error of the CBP with the MSPE obtained by an oracle who knew the true values of $\theta_{1}, \ldots, \theta_{K}$ but was restricted to use an estimate of the form (\ref{eq:thetahat_w}) with compromise weights of the form (\ref{eq:compromise_weight}). To be more precise, first consider the following loss function
\begin{equation}
\mathcal{L}_{K}( \btheta, \hatbtheta) =
\frac{1}{K} \sum_{k=1}^{K}\{ \theta_{k} - \hat{\theta}_{k} \}^{2}.
\end{equation}
The (pre-posterior) risk $\mathcal{R}_{K}( \hatbtheta ) = E\{ \mathcal{L}_{K}(\btheta, \hatbtheta ) \}$ associated with $\mathcal{L}_{K}$ is just the MSPE scaled by the number of units, i.e., $\mathcal{R}_{K}( \hatbtheta ) = K^{-1} \times \textrm{MSPE}(\hatbtheta)$.

With respect to this loss function, the oracle ``estimate" $\hatbtheta^{OR}$ of $\btheta$ is defined as $\hatbtheta^{OR} = \hatbtheta\{ \mathbf{w}^{c}(\alpha^{OR}, \tau^{OR}), \tau^{OR}\}$ where $\hatbtheta(\mathbf{w}, \tau)$ is as defined in (\ref{eq:thetahat_w}) and (\ref{eq:umatrix}).
The oracle hyperparameters $(\alpha^{OR}, \tau^{OR})$ are found by minimizing the (unobservable) loss 
\begin{equation}
(\alpha^{OR}, \tau^{OR}) = \argmin_{\alpha \in [0,1], \tau \geq 0} \frac{1}{K} \sum_{k=1}^{K}\{ \theta_{k} - \hat{\theta}_{k}^{c}(\alpha,\tau) \}^{2}, 
\label{eq:alpha_loss}
\end{equation}
where $\hat{\theta}_{k}^{c}(\alpha, \tau)$ denotes the $k^{th}$ component of $\hatbtheta( \mathbf{w}^{c}(\alpha, \tau), \tau)$.
By definition, the oracle risk $\mathcal{R}_{K}( \hatbtheta^{OR} )$ is, for any $K$, less than or equal to the risk associated with either the CBP, OBP, or EBLUP. Despite this, we show that, under appropriate conditions, the risk obtained by the CBP is asymptotically the same as the oracle risk. Specifically, the difference between the oracle and CBP risk goes to zero as the number of units $K$ goes to infinity.
To show this asymptotic equivalence, we will assume that the following conditions hold. 
\begin{itemize}
\item[(A1)] There is a $\delta \in (0, 1/2)$ such that
\begin{equation}
\lim_{K \longrightarrow \infty} \frac{\sigma_{max,K}^{2}}{K^{1/2 - \delta}\sigma_{min,K}^{2}} = 0, \nonumber
\end{equation}
where $\sigma_{max,K}^{2} = \max\{ \sigma_{1}^{2}, \ldots, \sigma_{K}^{2} \}$ and $\sigma_{min,K}^{2} = \min\{ \sigma_{1}^{2}, \ldots, \sigma_{K}^{2} \}$.
\item[(A2)] For the same $\delta \in (0, 1/2)$ used in condition (A1),
\begin{equation}
\lim_{K \longrightarrow \infty} \frac{1}{K^{1 + \delta}}\sum_{k=1}^{K} \mu_{k}^{2} = 0
\qquad \textrm{ and } \qquad
\lim_{K \longrightarrow \infty}\frac{1}{K^{1 + \delta/2}}\sum_{k=1}^{K} \sigma_{k}^{2} = 0. \nonumber
\end{equation}
\item[(A3)]
For each $k$, $E( e_{k}^{4}) < \infty$, and
\begin{equation}
\lim_{K \longrightarrow \infty} \frac{1}{K^{2}}\sum_{k=1}^{K} \mu_{k}^{4} = 0, \qquad 
\lim_{K \longrightarrow \infty} \frac{1}{K^{2}}\sum_{k=1}^{K} \sigma_{k}^{4} = 0, \qquad
\lim_{K \longrightarrow \infty} \frac{1}{K^{2}}\sum_{k=1}^{K} E(e_{k}^{4}) = 0, \nonumber
\end{equation}
\item[(A4)]
For each $K$, the design matrix $\mathbf{X}$ has full rank. Moreover, for the same $\delta \in (0, 1/2)$ used in conditions (A1) and (A2)
\begin{equation}
\lim_{K \longrightarrow \infty} K^{1 - \delta/2}D_{max}( \mathbf{P}_{X} ) = 0, \nonumber
\end{equation}
where $D_{max}( \mathbf{P}_{X} )$ denotes the maximum value of the diagonal elements of the matrix $\mathbf{P}_{X} = \mathbf{X}( \mathbf{X}^{T}\mathbf{X} )^{-1}\mathbf{X}^{T}$.
\end{itemize}

Condition (A1) places a fairly moderate restriction on the spread of the $\sigma_{k}^{2}$. For instance, condition (A1) would be satisfied if the $\sigma_{k}^{2}$ had the form $\sigma_{k}^{2} = \sigma^{2}/n_{k}$ for positive integers $n_{k}$ that were bounded by some number $M$. Condition (A2) is a moderately unrestrictive condition and is one that would be automatically satisfied if both the partial averages $K^{-1} \sum_{k=1}^{K}\mu_{k}^{2}$ and $K^{-1}\sum_{k=1}^{K} \sigma_{k}^{2}$ converged to some finite limit.  
Condition (A3) is a very weak assumption about the convergence of the sum of fourth moments. Condition (A4) requires that, for each $K$, the design matrix has full rank. Additionally, condition (A4) places a restriction on the maximal ``leverage'' $D_{max}( \mathbf{P}_{X} )$ that any one unit may have. In a classic regression setting, one often expects that $D_{max}( \mathbf{P}_{X} ) = O(1/K)$ to ensure no single observation has undue impact on the fitted regression line. Compared to this, condition (A4) makes the slightly weaker assumption that $D_{max}( \mathbf{P}_{K} ) = o(K^{-1 + \delta/2})$.

As in \cite{Li:1985}, the key to the asymptotic risk optimality of $\checkbtheta^{CBP}$ lies in the quality of $\hat{M}_{K}^{c}(\alpha,\tau)$ as an approximation of the loss function (not the risk). Specifically, the difference between $\hat{M}_{K}^{c}(\alpha, \tau)/K$ and the associated loss function approaches zero uniformly as $K$ goes to infinity. This result is stated by the following theorem. 
\begin{theorem}
Consider the family of estimates $\hatbtheta\{\mathbf{w}^{c}(\alpha, \tau), \tau\}$ with $\hatbtheta(\mathbf{w}, \tau)$ as defined in (\ref{eq:thetahat_w}). If conditions (A1)-(A4) hold, then $\mathcal{L}\big(\btheta, \hatbtheta\{\mathbf{w}^{c}(\alpha, \tau), \tau\} \big) - \hat{M}_{K}^{c}( \alpha, \tau)/K$ converges uniformly in $L^{1}$ to zero. That is,
\begin{equation}
\lim_{K \longrightarrow \infty} E \Bigg( \sup_{\alpha \in [0,1], \tau \geq 0}\Big| \mathcal{L}_{K}\Big( \btheta, \hatbtheta\{ \mathbf{w}^{c}(\alpha, \tau), \tau \}  \Big) - \hat{M}_{K}^{c}(\alpha, \tau)/K \Big| \Bigg) = 0. \nonumber
\end{equation}
\label{thm:uniform_conv}
\end{theorem}

Due to the close proximity of the loss function and $\hat{M}_{K}^{c}(\alpha,\tau)/K$, we should expect the performance of the oracle procedure and the CBP to be quite close as $\hatbtheta^{OR}$ and $\checkbtheta^{CBP}$ use values of $(\alpha, \tau)$ which minimize the loss  (\ref{eq:alpha_loss}) and $\hat{M}_{K}^{c}(\alpha, \tau)/K$ respectively. Indeed, as stated in the following theorem, the difference  between these risks goes to zero as the number of units goes to infinity.
\begin{theorem}
If conditions (A1)-(A4) hold, then
\begin{equation}
\lim_{K \longrightarrow \infty}\big[ \mathcal{R}_{K}( \checkbtheta^{CBP} ) - \mathcal{R}_{K}( \hatbtheta^{OR} ) \big] = 0. \nonumber
\end{equation}
\label{thm:asymp_risk}
\end{theorem}

\vspace{-.7cm}

Theorem \ref{thm:asymp_risk} establishes that the risk associated with the CBP is as good (asymptotically) as the oracle risk. Additionally, it follows from Theorem \ref{thm:asymp_risk} that the CBP risk is asymptotically at least as good as any other procedure using an estimate of the form $\hatbtheta\{ \mathbf{w}^{c}(\alpha,\tau), \tau\}$ - a class of estimates which includes the OBP, different versions of the EBLUP, or other procedures which might use alternative combinations of the OBP and EBLUP regression weights.

\section{Simulation Studies} \label{sec:simulations}
To evaluate the performance of the CBP and compare it with other methods, we conducted three main simulation studies. In the first two of these simulation studies, we compared the CBP and plug-in CBP with the following four approaches for predicting the mixed effects $\theta_{k}$: the OBP and three different versions of the EBLUP which vary according to how $\tau$ is estimated.  For these three versions of the EBLUP, we consider the following approaches for estimating $\tau$: marginal maximum likelihood (MLE), restricted maximum likelihood (REML), and unbiased risk estimation (URE). For the EBLUP with the URE of $\tau$, the unbiased risk estimate $\hat{\tau}_{URE}$ of $\tau$ is found by minimizing $\hat{M}_{K}\big( \mathbf{w}^{MLE}(\tau), \tau \big)$ with respect to $\tau$ when the regression weights are assumed to be the MLE weights $w_{k}^{MLE}(\tau) \propto 1/(\tau^{2} + \sigma_{k}^{2})$. To our knowledge, the use of such an unbiased risk estimate has not received substantial attention in the context of mixed models though, for example, \cite{Kou:2017} considers unbiased risk estimates of both the shrinkage weights and the target regression surface. We include $\hat{\tau}_{URE}$ in our simulations for two main reasons: to explore its use as an alternative approach approach to variance component estimation and to more clearly examine the benefits of combining the two weighting schemes. Because both EBLUP (URE) and the CBP are based on minimizing an unbiased risk criterion, comparing EBLUP (URE) and CBP provides a more direct way of examining the benefits of using compromise weights, since estimation of hyperparameters for both procedures is more closely related. The third simulation study concerns estimation of the population-average parameter discussed in Section \ref{sec:pop_mean}. Here, we also compare our compromise estimators with the direct, minimum-variance, and regression-based estimators described in Section \ref{sec:pop_mean}.

For each simulation setting, we estimate the MSPE with $\frac{1}{n_{rep}}\sum_{j=1}^{n_{rep}}\sum_{k=1}^{K} \{\hat{\theta}_{k}^{(j)} - \theta_{k} \}^{2}$, where 
$\hat{\theta}_{k}^{(j)}$ denotes the estimated value of $\theta_{k}$ in the $j^{th}$ simulation replication and $n_{reps}$ denotes the total number of replications used for that simulation setting. For every setting, we use $n_{rep} = 5,000$. 

\subsection{Two Unmodeled Latent Groups}
\textit{Simulation Description}

\noindent
We consider a scenario where units belong to two distinct clusters but such cluster membership is unmodeled in the analysis.
Specifically, we consider responses generated as
\begin{equation}
Y_{k} = \beta_{0} + \beta_{1} Z_{k} + v_{k} + \sigma_{k}e_{k}, \quad \textrm{ for } k = 1, \ldots, K, 
\label{eq:latent_clusters}
\end{equation}
where $K$ is a positive integer and the $Z_{k} \in \{0, 1\}$ are independent Bernoulli random variables with $P( Z_{k} = 1 ) = 1/2$. For these simulations, we assume that both $v_{k} \sim \textrm{Normal}(0, 1)$ and $e_{k} \sim \textrm{Normal}(0, 1)$. The residual variances $\sigma_{k}^{2}$ are assumed to take the form $\sigma_{k}^{2} = 1/n_{k}$ with the $n_{k}$ being determined by
\begin{equation}
n_{k} = 
10Z_{k} + 2(1 - Z_{k}).  \nonumber 
\end{equation}
Note that while the random effects $v_{k}$ are simulated from a known distribution, each of the estimation procedures considered in this simulation study (i.e., EBLUP, OBP, and CBP) does not use this known value of $\tau$ and uses a value of $\tau$ which is estimated from the data.

Model (\ref{eq:latent_clusters}) is meant to represent a situation involving two latent groups where the units in one group (i.e., the group where $Z_{k} = 1$) tend to have larger means than the other group, and moreover, the estimation precision in the $Z_{k} = 1$ group is much greater than in the $Z_{k} = 0$ group. Specifically, the residual standard deviation is $\sigma_{k} = 1/\sqrt{10}$ for those in the $Z_{k} = 1$ group and $\sigma_{k} = 1/\sqrt{2}$ in the $Z_{k} = 0$ group.

\bigskip

\noindent
\textit{Performance under an assumed intercept-only model.}

\noindent
While (\ref{eq:latent_clusters}) is the true data-generating model, we consider estimates of $\theta_{k}$ which assume an intercept-only model. That is, the assumed design matrix $\mathbf{X}$ when computing the shrinkage estimates of $\theta_{k}$ consists of a single $K \times 1$ column vector whose entries are all equal to $1$. Hence, we have misspecification of the mean model whenever $\beta_{1} \neq 0$ because the true $\mu_{k}$ can take one of two values. 

\begin{figure}
\centering
     \includegraphics[width=6.5in,height=4.5in]{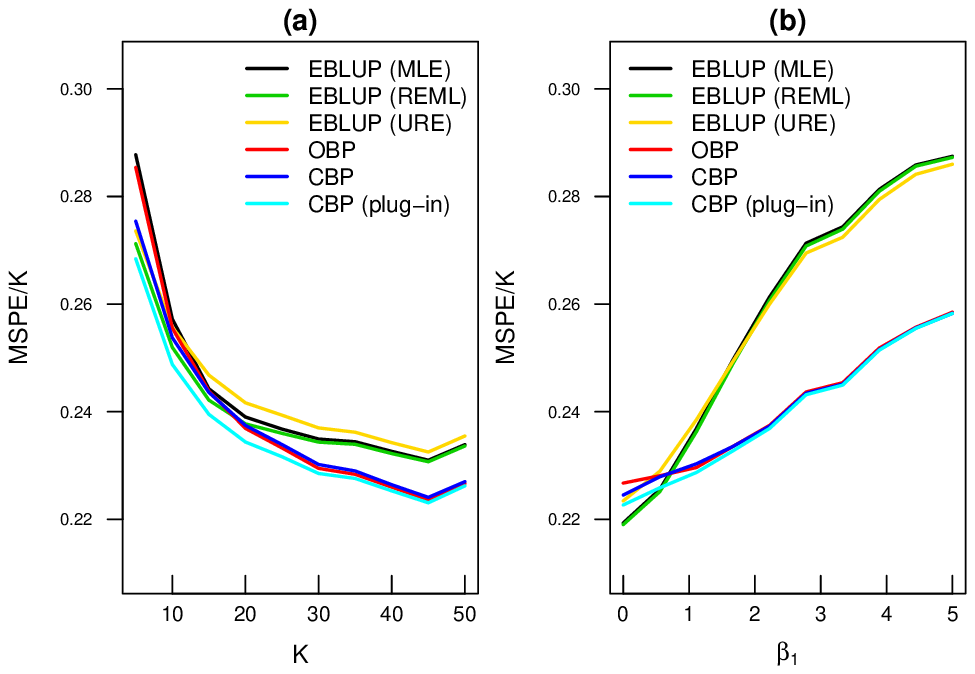}
\caption{Estimated MSPE for responses generated from model (\ref{eq:latent_clusters}) and with an intercept-only model used for each method. In (a), $\beta_{0} = 0$ and $\beta_{1} = 1$, and the number of units $K$ varies from $5$ to $50$. In (b), $\beta_{0} = 0$ and $K = 30$, and $\beta_{1}$ varies from $0$ to $5$. $5,000$ replications were used for each simulation setting to estimate the MSPE.}
\label{fig:latent2clustf1}
\end{figure}

Figure \ref{fig:latent2clustf1} shows results for the MSPE in this simulation setting. The left-hand panel of Figure \ref{fig:latent2clustf1} compares the MSPE across different methods where $\beta_{1}$ is fixed at one and the number of units $K$ varies from $5$ to $50$. When the number of units is very small (i.e., $K \leq 10$), the EBLUPs generally perform very well due to the greater role of variance in driving estimation performance. However, for settings with more units, the systematic bias in the assumed mean structure becomes much more important, and hence, the OBP tends to clearly outperform every version of the EBLUP. Though never quite the top performer, the CBP is quite robust here in the sense that it always has MSPE near the best performer for all values of $K$. Interestingly, the plug-in CBP is the top performer for all values of $K$. Though not much better than the CBP and OBP for $K \geq 30$, for the ranges $5 \leq K \leq 25$, the plug-in CBP provides a noticeable improvement in MSPE over the CBP.

The right-hand panel of Figure \ref{fig:latent2clustf1} shows how the MSPE changes when we consider a fixed number of units and vary the severity of mean model misspecification. When there is no model misspecficiation (i.e., when $\beta_{1} = 0$), the EBLUPs, as expected, have lower MSPE than the OBP with the two versions of the CBP falling in between the OBP and the EBLUPs. As $\beta_{1}$ increases however, the bias term in the MSPE grows substantially while the variance remains mostly unchanged. Hence, as $\beta_{1}$ increases, the OBP quickly dominates the EBLUPs due to the greater role that the model misspecification plays in impacting MSPE performance. Notably, the CBP never has poor MSPE performance regardless of the value of $\beta_{1}$. When $\beta_{1} = 0$, the MSPE of the CBP is marginally worse than the MLE and REML versions of the EBLUP and is just as good as EBLUP (URE), and for larger values of $\beta_{1}$, the weights of the CBP adapt in such a way that its performance is very similar to that of the OBP.

\medskip

\noindent
\textit{Performance when also including irrelevant covariates.}

\noindent
The OBP generally has very good performance under an intercept-only assumption especially when the number of units is large and $\beta_{1} > 0$. This is because, in these scenarios, the bias is the dominating factor in determining the MSPE. 
For scenarios having prominent roles for both bias and variance, the CBP can often clearly outperform both the OBP and the EBLUPs. We demonstrate this here by comparing MSPE when one also includes irrelevant covariates in the analysis of data simulated from model (\ref{eq:latent_clusters}). More specifically, for these simulations the data are simulated from model (\ref{eq:latent_clusters}), but when estimating the $\theta_{k}$, the $k^{th}$ row of the design matrix $\mathbf{X}$ is assumed to take the form $\mathbf{x}_{k}^{T} = (1, x_{k1}, \ldots, x_{kq})$ where the $x_{kj}$ are standard normal random variables generated independently from the $v_{k}$ and $e_{k}$ in (\ref{eq:latent_clusters}). Including such irrelevant covariates substantially increases the variance of each method while having a minimal impact on bias.

\begin{figure}
\centering
     \includegraphics[width=6.5in,height=4.5in]{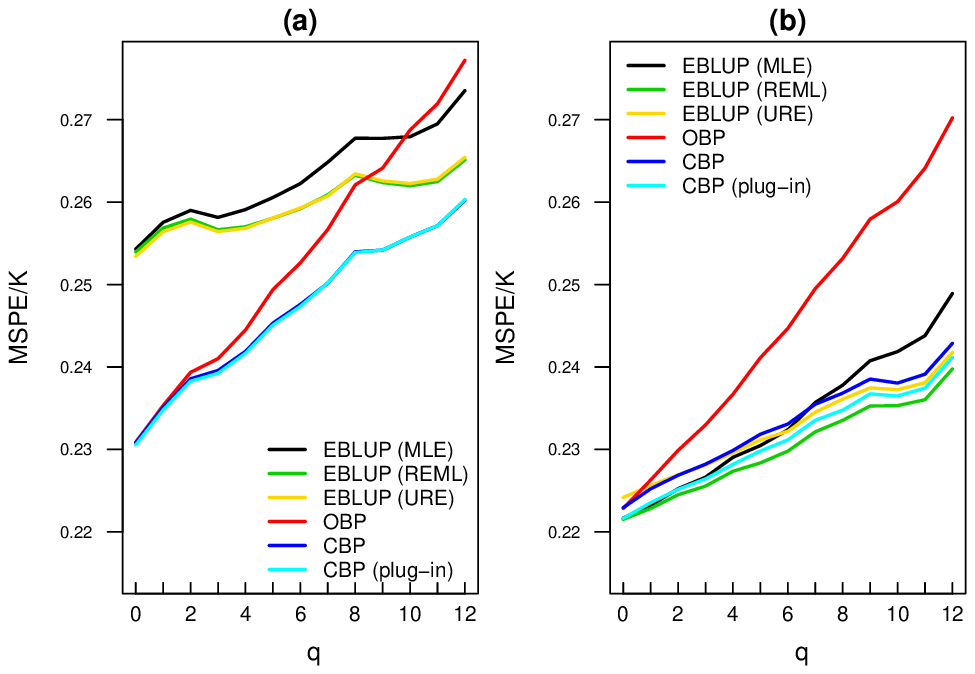}
\caption{Estimated MSPE for responses generated from model (\ref{eq:latent_clusters}) with $q$ irrelevant covariates used for each method. In (a), $\beta_{0} = 0$, $\beta_{1} = 2$, and $K = 50$ while $q$ ranges from $0$ to $12$. In (b), $\beta_{0} = 0$, $\beta_{1} = 1/2$, and $K = 50$ while $q$ ranges from $0$ to $12$. $5,000$ replications were used for each simulation setting to estimate the MSPE.}
\label{fig:latent2clustf2}
\end{figure}

Figure \ref{fig:latent2clustf2} displays the estimated MSPE for different methods with the number of irrelevant covariates $q$ ranging from $0$ to $12$ and with the number of units fixed at $K = 50$. Here, $q = 0$ corresponds to estimating the $\theta_{k}$ assuming an intercept-only model. In the left-hand panel where $\beta_{1} = 2$, the misspecification in the mean model is substantial which leads to strong performance of the OBP for $q = 0$. However, as more irrelevant covariates are added, the variance contribution to the MSPE grows which leads the OBP to perform even worse than all EBLUP methods for $q > 10$. Using weights which can adapt to different levels of bias and variance enables both versions of the CBP to perform very well. As shown in the left-hand panel of Figure \ref{fig:latent2clustf2}, when $\beta_{1} = 2$ the CBP has nearly identical MSPE to the OBP for $q = 0$, and it clearly dominates all other methods for $q \geq 4$. In the right-hand panel of Figure \ref{fig:latent2clustf2}, $\beta_{1}$ is set to $1/2$ so that model misspecification is much less severe than the scenario depicted in the left-hand panel. In this scenario, the EBLUPs generally have the best performance due the strong role of estimation variance in these scenarios. Despite this, the CBP has nearly identical performance to the EBLUP (URE) and is very competitive with the other EBLUP methods for all values of $q$ considered. Moreover, the plug-in CBP has consistently better performance than the CBP, and it has lower MSPE than both EBLUP(MLE) and EBLUP(URE) for most values of $q$ considered.

\subsection{Sample Size as an Ignored Covariate}
\textit{Simulation Description}

\noindent
We begin by examining the performance of the CBP when the mixed effects are related linearly to the unit-specific sample sizes and this dependency is not properly modeled. Specifically, in this simulation study we generate unit-specific responses $Y_{k}$, $k = 1,\ldots,K$
\begin{equation}
Y_{k} = \mathbf{x}_{k}^{T}\bbeta + \rho\tau n_{k}/sd(n) + v_{k}\tau\sqrt{1 - \rho^{2}} + \sigma e_{k}/\sqrt{n_{k}}, \quad k=1,\ldots,K,
\label{eq:iss_simulation}
\end{equation}
where $sd(n) = \{ \tfrac{1}{K-1}\sum_{k=1}^{K}(n_{k} - \bar{n})\}^{1/2}$ is the standard deviation of the $n_{k}$ and $\mathbf{x}_{k}$ is a $p \times 1$ vector of regression coefficients.
For each $k$, we draw $v_{k}$ independently from the $n_{k}$, and the distribution of $v_{k}$ is chosen so that $E(v_{k}) = 0$ and $\var(v_{k}) = 1$. Consequently, the sample correlation between the $\theta_{k}$ and $n_{k}$ will be approximately equal to $\rho$ in each simulation replication. In these simulations, the dependence of the mean of $Y_{k}$ on $n_{k}$ is not modeled as the sample sizes were not used as regression covariates in any of the methods used.

In each of our simulations, we use $\tau = 1/2$ and the following sequence for the unit-specific sample sizes: $\log n_{k} = 3(k-1)/(K-1)$ to generate the data.
Moreover, for each simulation setting, we use $3 \times 1$ covariate vectors of the form $\mathbf{x}_{k} = (1, x_{k1}, x_{k2})^{T}$ where $x_{k1}$ and $x_{k2}$ were generated independently as $x_{kj} \sim \textrm{Normal}(0, 1)$.

\begin{figure}
\centering
     \includegraphics[width=6in,height=4.5in]{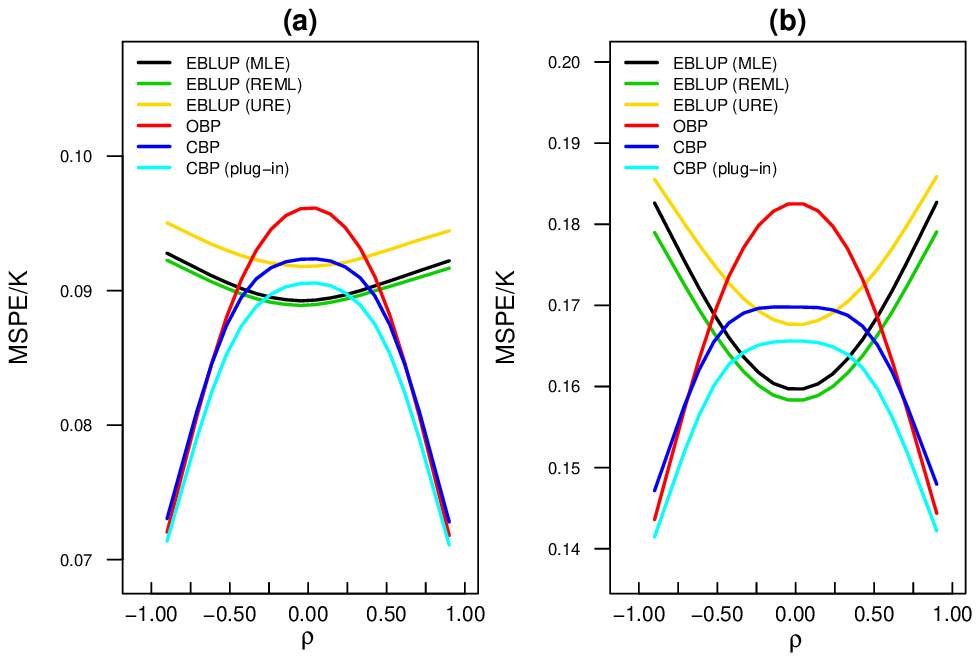}
\caption{Estimated MSPE for various methods using data generated from model (\ref{eq:iss_simulation}). In these simulations, $K=50$ and $v_{k} \sim \textrm{Normal}(0, 1)$. In (a), $\sigma^{2} = 0.5$ while in panel (b), $\sigma^{2} = 1.5$. Both figures show the role that the correlation parameter $\rho$ plays in determining the relative performance of the methods considered. In both cases, the plug-in CBP performs very well as it has the lowest MSPE for larger values of $|\rho|$ and is not much worse than EBLUP(MLE) and EBLUP(REML) for small values of $|\rho|$ where the model is nearly correctly specified.}
\label{fig:linearsampsize1}
\end{figure}

\medskip

\noindent
\textit{Performance when varying $\sigma^{2}$.}

\noindent
Figure \ref{fig:linearsampsize1} shows the results for simulations from model (\ref{eq:iss_simulation}) where $K = 50$ and $v_{k} \sim \textrm{Normal}(0,1)$, and the correlation parameter $\rho$ varies from $-0.9$ to $0.9$. The left-hand panel of Figure \ref{fig:linearsampsize1} corresponds to a simulation setting with $\sigma^{2} = 0.5$ while the right-hand panel of Figure \ref{fig:linearsampsize1} corresponds to a simulation setting with $\sigma^{2} = 1.5$.  When $\rho = 0$, model (\ref{eq:iss_simulation}) is correctly specified, and thus, in these cases, we should expect the EBLUP procedures to generally perform the best. As shown in Figure \ref{fig:linearsampsize1}, this is indeed the case. For each plot shown, the EBLUP (MLE) and EBLUP (REML) procedures have the lowest MSPE whenever $\rho = 0$. For values of $\rho$ which are larger in absolute value however, both the CBP and the plug-in CBP can provide substantial improvements over the EBLUPs in terms of MSPE. For these simulations, both the CBP and the plug-in CBP exhibit the same downward facing parabola as $\rho$ varies from $-0.9$ to $0.9$, but the plug-in CBP clearly has better performance for both the $\sigma^{2} = 0.5$ and $\sigma^{2} = 1.5$ settings. For both settings of $\sigma^{2}$, the plug-in CBP dominates the OBP for all values of $\rho$ while the CBP dominates the OBP except for very large values of $|\rho|$ where the model is highly misspecified.
 
\begin{figure}
\centering
     \includegraphics[width=6.5in,height=4.5in]{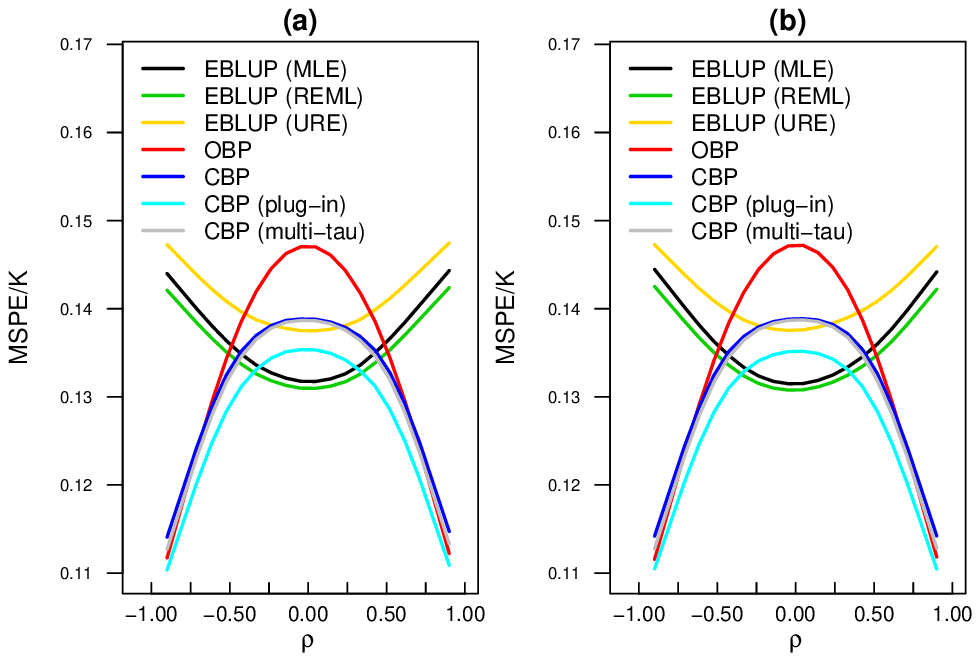}
\caption{Estimated MSPE for various methods using data generated from model (\ref{eq:iss_simulation}). In these simulations, $K=50$ and $\sigma^{2} = 1$ while the distribution of $v_{k}$ is non-Gaussian in both panels. In (a), $v_{k}$ is drawn from a mixture of two Gaussian distributions, and in panel (b), $v_{k} \sim \textrm{Uniform}(-\sqrt{3}, \sqrt{3})$. Both figures again show the role the correlation parameter $\rho$ plays in determining the relative performance of the methods considered.}
\label{fig:linearsampsize2}
\end{figure}

\noindent
\textit{Performance for different distributions of $v_{k}$.}

\noindent
Figure \ref{fig:linearsampsize2} shows the results for simulations corresponding to model (\ref{eq:iss_simulation}) where $K = 50$ and $\sigma^{2}=1$. Again, the correlation parameter $\rho = \textrm{Corr}(\theta_{k}, n_{k})$ is varied from $-0.9$ to $0.9$. We consider two choices for the distribution of $v_{k}$: a Gaussian mixture distribution with two components and a uniform distribution. For the Gaussian mixture model, the $v_{k}$ are generated under the assumption that $v_{k}|Z_{k} = 0 \sim \textrm{Normal}(\tfrac{-1}{\sqrt{2}}, \tfrac{1}{2})$ and $v_{k}|Z_{k} = 1 \sim \textrm{Normal}(\tfrac{1}{\sqrt{2}}, \tfrac{1}{2})$ with $P(Z_{k} = 0) = 1/2$. For the uniform distribution, we use $v_{k} \sim \textrm{Uniform}(-\sqrt{3}, \sqrt{3})$ so that $\textrm{Var}(v_{k}) = 1$.

The left-hand panel of Figure \ref{fig:linearsampsize2} shows the estimated values of the MSPE when $v_{k}$ is generated from the Gaussian mixture distribution while the right-hand panel corresponds to the cases when $v_{k}$ is generated from a uniform distribution. In both panels of Figure \ref{fig:linearsampsize2}, we see an overall pattern which is similar to that in Figure \ref{fig:linearsampsize1}. Namely, the EBLUPs dominate for values of $\rho$ near $0$ while the OBP and the different versions of the CBP dominate for more extreme values of the correlation parameter $\rho$. Similar to the results shown in Figure \ref{fig:linearsampsize1}, the plug-in CBP performs better here than the CBP across all simulation settings. Indeed, the plug-in CBP is the clear winner with respect to MSPE for values of $\rho$ such that $0.4 \leq |\rho| \leq 0.7$. Figure \ref{fig:linearsampsize2} also shows results for the ``multi-$\tau$" CBP approach that was described in Section \ref{subsec:variations}. As shown in this figure, the differences between the CBP and the multi-$\tau$ CBP were very minimal in these simulation scenarios.

\subsection{Estimating a Population Average} \label{subsec:popavgsim}
This simulation study concerns estimation
of the population-average parameter $\mu_{0} = K^{-1}\sum_{k=1}^{K} \theta_{k}$ discussed in Section \ref{sec:pop_mean}.
For this simulation study, we simulate the direct estimates $Y_{k}$ under the assumption that
\begin{equation}
Y_{k}|\theta_{k} \sim \textrm{Normal}( \theta_{k}, \sigma^{2}/n_{k}), \nonumber
\end{equation}
where the $n_{k}$ can be thought of as unit-specific sample sizes though we do not constrain
the $n_{k}$ to be integers in our simulations. The sample sizes $n_{1}, \ldots, n_{K}$ and the unit-specific parameters $\theta_{1}, \ldots, \theta_{K}$ are generated from the following scheme
\begin{eqnarray}
n_{k} &=& \frac{ K\bar{n} \exp\Big\{ \frac{a(2k - K - 1)}{K-1} \Big\} }{  \sum_{k=1}^{K} \exp\Big\{ \frac{a(2k - K - 1)}{K-1} \Big\}  } , \quad \textrm{ for } k = 1, \ldots, K \label{eq:setting1_setupa} \\
\theta_{k} &=& c_{1}(\rho, \xi)f_{1}( n_{k} ) + c_{2}(\rho, \xi)v_{k}, \quad \textrm{ for } k = 1, \ldots, K,
\label{eq:setting1_setup}
\end{eqnarray}
where $v_{k} \sim \textrm{Normal}(0, 1)$, $\mu_{ln} = \tfrac{1}{K}\sum_{k=1}^{K} \log(n_{k})$, and $\sigma_{ln} = [\tfrac{1}{K-1}\sum_{k=1}^{K} \{ \log(n_{k}) - \mu_{ln} \}^{2} ]^{1/2}$, and $f_{1}(n_{k})$ is defined as
\begin{equation}
f_{1}( n_{k} ) = \Phi\Big( \frac{2[\log(n_{k}) - \mu_{ln}] }{ \sigma_{ln} }  \Big) - \frac{1}{2}. \nonumber
\end{equation}
The sample sizes $n_{k}$ in this setting are equally spaced on the log scale, and the constant $a$ in (\ref{eq:setting1_setupa}) can be chosen to achieve a desired value for the standard deviation of the $n_{k}$, namely, $sd(n) = \{ \tfrac{1}{K-1}\sum_{k=1}^{K}(n_{k} - \bar{n})\}^{1/2}$. The constants $c_{1}(\rho, \xi)$, $c_{2}(\rho, \xi)$ in (\ref{eq:setting1_setup}) are defined as $c_{1}(\rho, \xi) = sd(n)\xi\rho/\kappa_{n}$ and $c_{2}(\rho, \xi) = \sqrt{\xi^{2} - c_{1}^{2}(\rho,\xi)\sigma_{f}^{2}}$, where $\kappa_{n} = \tfrac{1}{K}\sum_{k=1}^{K} f_{1}(n_{k})n_{k}$ and $\sigma_{f}^{2} = \tfrac{1}{K-1}\sum_{k=1}^{K} f_{1}^{2}(n_{k})$.
Defining the constants this way ensures that 
\begin{equation}
\frac{1}{K}\sum_{k=1}^{K} \textrm{Var}(\theta_{k}) = \xi^{2} \quad \textrm{ and } \quad 
\frac{ E(\frac{1}{K}\sum_{k=1}^{K} \theta_{k}n_{k})}{ sd(n)\sqrt{ \frac{1}{K}\sum_{k=1}^{K} \textrm{Var}(\theta_{k}) }    } = \rho.  \nonumber
\end{equation}
In other words, $\xi$ measures the standard deviation of the unit-specific means $\theta_{k}$ while $\rho$ measures the correlation between the $\theta_{k}$ and the unit-specific samples sizes $n_{k}$. Hence, larger values of $\rho$ correspond to settings where sample size is more informative for the magnitude of $\theta_{k}$.

Table \ref{tab:popavg_simulations} shows simulation-based estimates of the MSPE for $8$ estimation methods and different choices of $(K,\sigma^{2},\rho)$. In each row of Table \ref{tab:popavg_simulations}, the ratio between the MSPE and the minimum MSPE for that row is shown. For these simulations, we included the estimators $\hat{\mu}^{direct}$ and $\hat{\mu}^{mv}$ described in Section \ref{sec:pop_mean} along with the ``direct compromise" estimator $\hat{\mu}^{compr} = \alpha_{opt}(0)\hat{\mu}^{mv} + \{1 - \alpha_{opt}(0)\}\hat{\mu}^{direct}$, where $\alpha_{opt}(0)$ is as defined in (\ref{eq:alpha_popmean_direct}). In our comparisons, we also included the nonparametric regression-based estimator $\hat{\mu}_{0} = \tfrac{1}{K}\sum_{k=1}^{K} h(n_{k}, \hatbbeta^{h})$ described in Section \ref{sec:pop_mean}. For the function $h(n_{k}, \bbeta^{h})$ in (\ref{eq:spline-estimate}), we used a cubic smoothing spline with the smoothing parameter selected using generalized cross-validation (\cite{Craven1978}). 

The results in Table \ref{tab:popavg_simulations} show that either the EBLUP(REML) estimate of $\mu_{0}$ or $\hat{\mu}^{mv}$ generally perform the best whenever $\rho = 0$. This is expected as $\rho = 0$ corresponds to a correctly specified regression model. The EBLUP estimator generally performs better than $\hat{\mu}^{mv}$ in lower noise settings, i.e., when $\sigma^{2} = 1$. Moreover, for settings with high variance (i.e., $\sigma^{2}=4$), the EBLUP and $\hat{\mu}^{mv}$ both perform well even for larger values of $\rho$. For large values of $\rho$, the direct estimator $\hat{\mu}^{direct}$ generally does quite well though the estimator $\hat{\mu}^{sr}$ does somewhat better in many of these settings, particularly when $\sigma^{2}=4$. The compromise estimators (CBP, CBP(plug-in), Direct-Compr) are quite robust across different settings in the sense that their MSPE performance is never especially poor when compared with the best method. For each setting, they lie somewhere between the best and worst performer. Indeed, the worst relative performance of $1.82$ for the CBP occurs when $\rho = 0.0$ and $\sigma^{2} = 4$. The regression-based estimator $\hat{\mu}^{sr}$ is also quite robust in this sense. For settings with $\sigma^{2} = 1$, the compromise estimators are very competitive with $\hat{\mu}^{sr}$ and usually have lower MSPE when $\rho \leq 0.1$. This is despite the fact that
no modeling is involved in implementing the compromise approaches whereas the regression approach requires modeling the dependence of $\theta_{k}$ on the unit-specific sample sizes.

\begin{table}[ht]
\centering
\begin{tabular}{lllcccccccc}
  \toprule
K & $\sigma^{2}$ & $\rho$ & {\small EBLUP(REML)} & {\small OBP} & {\small CBP} & {\small CBP(plug-in}) & {\small Direct} & {\small MinVar} & {\small Direct-Compr    } & {\small SR} \\ 
  \midrule
\multirow{6}{*}{10} & \multirow{6}{*}{1} & 0.0 & 1.00 & 1.67 & 1.12 & 1.07 & 1.27 & 2.54 & 1.21 & 1.27 \\ 
 & & 0.1 & 1.00 & 1.62 & 1.11 & 1.05 & 1.23 & 2.61 & 1.18 & 1.24 \\ 
 & & 0.2 & 1.00 & 1.53 & 1.09 & 1.03 & 1.15 & 3.11 & 1.15 & 1.16 \\ 
 & & 0.3 & 1.01 & 1.46 & 1.06 & 1.00 & 1.08 & 3.70 & 1.11 & 1.07 \\ 
 & & 0.4 & 1.16 & 1.35 & 1.09 & 1.03 & 1.00 & 4.79 & 1.14 & 1.00 \\ 
 & & 0.5 & 1.41 & 1.36 & 1.15 & 1.12 & 1.00 & 6.08 & 1.18 & 1.00 \\ 
\midrule
\multirow{6}{*}{10} & \multirow{6}{*}{4} & 0.0 &  1.49 & 2.47 & 1.77 & 1.83 & 2.10 & 1.00 & 1.78 & 1.85 \\ 
 & & 0.1 & 1.47 & 2.47 & 1.74 & 1.81 & 2.09 & 1.00 & 1.75 & 1.85 \\ 
 & & 0.2 & 1.36 & 2.25 & 1.62 & 1.66 & 1.90 & 1.00 & 1.63 & 1.68 \\ 
 & & 0.3 & 1.28 & 2.12 & 1.53 & 1.57 & 1.79 & 1.00 & 1.54 & 1.58 \\ 
 & & 0.4 & 1.16 & 1.91 & 1.40 & 1.42 & 1.61 & 1.00 & 1.41 & 1.41 \\ 
 & & 0.5 & 1.02 & 1.64 & 1.22 & 1.23 & 1.38 & 1.00 & 1.24 & 1.24 \\
\midrule
\multirow{6}{*}{50} & \multirow{6}{*}{1} & 0.0 & 1.00 & 1.80 & 1.19 & 1.11 & 1.34 & 2.72 & 1.29 & 1.33 \\ 
 & & 0.1 & 1.00 & 1.64 & 1.14 & 1.07 & 1.22 & 3.41 & 1.21 & 1.22 \\ 
 & & 0.2 & 1.14 & 1.37 & 1.09 & 1.03 & 1.01 & 5.47 & 1.13 & 1.00 \\ 
 & & 0.3 & 1.56 & 1.48 & 1.15 & 1.12 & 1.00 & 9.54 & 1.17 & 1.00 \\ 
 & & 0.4 & 2.22 & 1.56 & 1.14 & 1.17 & 1.00 & 14.95 & 1.12 & 1.01 \\ 
 & & 0.5 & 3.25 & 1.63 & 1.14 & 1.15 & 1.00 & 22.32 & 1.09 & 1.01 \\ 
\midrule
\multirow{6}{*}{50} & \multirow{6}{*}{4} & 0.0 &  1.43 & 2.53 & 1.82 & 1.89 & 2.19 & 1.00 & 1.84 & 1.88 \\ 
 & & 0.1 & 1.33 & 2.32 & 1.69 & 1.74 & 2.02 & 1.00 & 1.71 & 1.70 \\ 
 & & 0.2 & 1.04 & 1.81 & 1.35 & 1.37 & 1.57 & 1.00 & 1.36 & 1.34 \\ 
 & & 0.3 & 1.00 & 1.64 & 1.28 & 1.27 & 1.42 & 1.32 & 1.31 & 1.24 \\ 
 & & 0.4 & 1.00 & 1.59 & 1.30 & 1.27 & 1.37 & 1.71 & 1.34 & 1.21 \\ 
 & & 0.5 & 1.00 & 1.54 & 1.29 & 1.24 & 1.32 & 2.31 & 1.34 & 1.17 \\ 
   \bottomrule
\end{tabular}
\caption{Estimates of the MSPE for estimates of the population-average parameter $\mu_{0}$. In each row of the table, the ratio between the MSPE and the minimum MSPE for that row is shown. The ``Direct" and ``MinVar" estimates correspond to the estimates $\hat{\mu}^{direct}$ and $\hat{\mu}^{mv}$ respectively. The ``Direct-Compr" estimate corresponds to the direct compromise estimate $\hat{\mu}^{compr} = \alpha_{opt}(0)\hat{\mu}^{mv} + \{1 - \alpha_{opt}(0)\}\hat{\mu}^{direct}$. The ``SR" estimate refers to the spline-based estimator $\hat{\mu}^{sr}$ of the population-average parameter described in Section \ref{sec:pop_mean}.} 
\label{tab:popavg_simulations}
\end{table}

\section{Estimation of Normative Gait Speed in Older Adults} \label{sec:application}
In this section, we apply the CBP and plug-in CBP approaches to estimate gait speed within a collection of demographically defined strata of older adults. The data analyzed for this purpose come from round 8 of the National Health and Aging Trends Study (NHATS) public use data. NHATS is a nationally representative survey of adults from the United States aged $65$ and older that is designed to track key measures of well-being related to the aging process. One such measure recorded by NHATS is gait speed.  The ability to walk is essential for independent living, and gait speed is a simple measure of the ability to walk.  It is a valid measure of the overall functional health of older adults. It is typically measured as the speed at which a person walks a specified, short distance at usual pace. Typically, two measurements are taken and averaged.  Slower gait speed has been shown to be a powerful predictor of mortality in older adults (\cite{Studenski:2011}) and is sometimes referred to as the ``sixth vital sign" (\cite{Middleton:2015}). In NHATS, gait speed was measured by instructing participants to ``walk at their usual pace'' over a 3-meter course (distance measured using a 5 meter colored chain). Participants started from a standing position and time was marked when the last foot crossed over the 3-meter mark on the link-chain. This was done twice and the average of the two trials was taken.

Our aim is to estimate the average gait speed within key demographic strata recorded by NHATS. Being derived from a nationally representative sample, these estimates may be considered as ``normative'' values of gait speed in older adults. Specifically, we look at 48 strata created from the following demographic characteristics: sex (male and female), race (white non-hispanic, black, hispanic, and other), and age (65-69, 70-74, 75-79, 80-84, 85-89, 90+). In this context, we define $Y_{k}$ to be the sample mean of gait speed within the $k^{th}$ demographically defined stratum, and $\sigma_{k}^{2} = s_{k}^{2}/n_{k}$, where $s_{k}$ is the sample standard deviation of gait speed within the $k^{th}$ stratum. It has also been recognized that height can play an important role in gait speed (\cite{Bohannon:1997}) and hence including height in our regression model can potentially improve the stratum-specific estimates of gait speed. We incorporate this into our analysis by defining $x_{k} = u_{k} - 65$ where $u_{k}$ is the mean height (in inches) within stratum $k$. 

In addition to an analysis involving all 48 strata of interest, we also
performed an analysis which only utilized data from a subset of 32 strata. This subset of 32 strata
was created by excluding the 12 strata that contained the ``other" race category and the 4 strata where the race was ``hispanic" and the age category was either 85-89 or 90+.  
This analysis of the subset of 32 strata was conducted to better highlight differences
between the CBP, OBP, and EBLUP that can often arise in practice.

\begin{figure}
\centering
     \includegraphics[width=6in,height=4.75in]{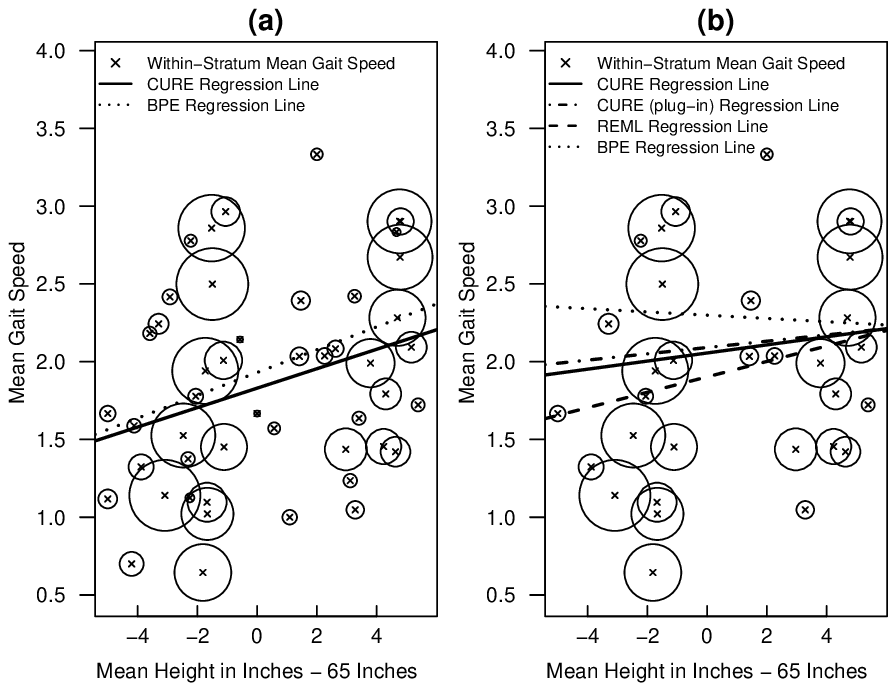}
\caption{Sample means $Y_{k}$ of gait speed  within demographically-defined strata and fitted regression lines $\hat{\beta}_{0} + \hat{\beta}_{1}(u_{k} - 65)$ estimated via CURE, CURE (plug-in), REML, or BPE, where $u_{k}$ represents the sample mean of height within stratum $k$. The size of the circles surrounding the within-stratum means are inversely proportional to the stratum-specific standard errors $\sigma_{k} = s_{k}/\sqrt{n_{k}}$, where $s_{k}$ is the sample standard deviation of gait speed within stratum $k$. In (a), sample means and fitted regression lines are displayed for all 48 of the strata defined by all combinations of the demographic categories of sex, race, and age. In (b), sample means and fitted regression lines are shown for a selected subset of 32 strata. The CURE (plug-in) and REML estimates are not shown in panel (a) because these are identical to the CURE estimates.
}
\label{fig:walkscatter}
\end{figure}

Figure \ref{fig:walkscatter} shows the direct estimates $Y_{k}$ of mean gait speed for each of the $48$ strata of interest in the left-hand panel and the subset of $32$ strata in the right-hand panel. The sizes of the circles surrounding each direct estimate are inversely proportional to the standard error of the direct estimate. 
For the small area model $Y_{k} = \beta_{0} + \beta_{1}x_{k} + v_{k} + e_{k}$ using data from all 48 strata, the REML estimates of $\beta_{0}$, $\beta_{1}$, and $\tau$ were $\hat{\beta}_{0} = 1.833$, $\hat{\beta}_{1} = 0.062$, and $\hat{\tau} = 0.60$ respectively, and the BPE estimates of these parameters were $(\hat{\beta}_{0}, \hat{\beta}_{1}, \hat{\tau}) = (1.930, 0.073, 0.45)$. The CURE estimates were $(\hat{\beta}_{0}, \hat{\beta}_{1}, \hat{\tau}) = (1.830, 0.062, 0.46)$ while the plug-in CURE estimates were $(\hat{\beta}_{0}, \hat{\beta}_{1}, \hat{\tau}) = (1.833, 0.062, 0.60)$. 
The values of $(\alpha^{*}, \tau^{*})$ used in the CURE estimates were $\alpha^{*} = 1.0$ and $\tau^{*} = 0.46$, and the value $\alpha_{plug}^{*}$ of the compromise parameter used for the plug-in CURE estimates was $0.999$. Because $\alpha_{plug}^{*}$ is very close to $1$, the plug-in CBP regression weights are essentially the same as the REML regression weights and hence the corresponding estimates of $\beta_{0}$ and $\beta_{1}$ were very similar.

As shown in the fitted regression lines of Figure \ref{fig:walkscatter}, the CURE estimates of $(\beta_{0}, \beta_{1}, \tau)$ were substantially 
different from both the REML and BPE estimates when only looking at the 32-strata subset. Specifically, we obtained
$(\hat{\beta}_{0}, \hat{\beta}_{1}, \hat{\tau}) = (2.056, 0.026, 0.73)$ for the CURE estimates 
while the REML and BPE estimates were $(1.902, 0.50, 0.64)$ and 
$(2.299, -0.10, 0.63)$ respectively. The optimal values of the mixing parameter $\alpha$ were $\alpha^{*} = 0.635$ 
and $\alpha_{plug}^{*} = 0.531$ 
for the CURE and plug-in CURE estimates respectively. 
When comparing these results with the analysis of all 48 strata, you may note that the fitted regression 
lines associated with the CURE and BPE estimates have changed more dramatically than the fitted regression line associated
with the REML estimates. This is mainly due to the fact that all of the larger strata present in the group of 48
are also present in the group of 32, and the removal of a number of smaller strata does not substantially change the 
values of the REML estimates. In contrast, both the CURE and BPE regression lines are much more sensitive to the presence/absence
of the smaller strata.

\begin{figure}
\centering
     \includegraphics[width=6in,height=4.75in]{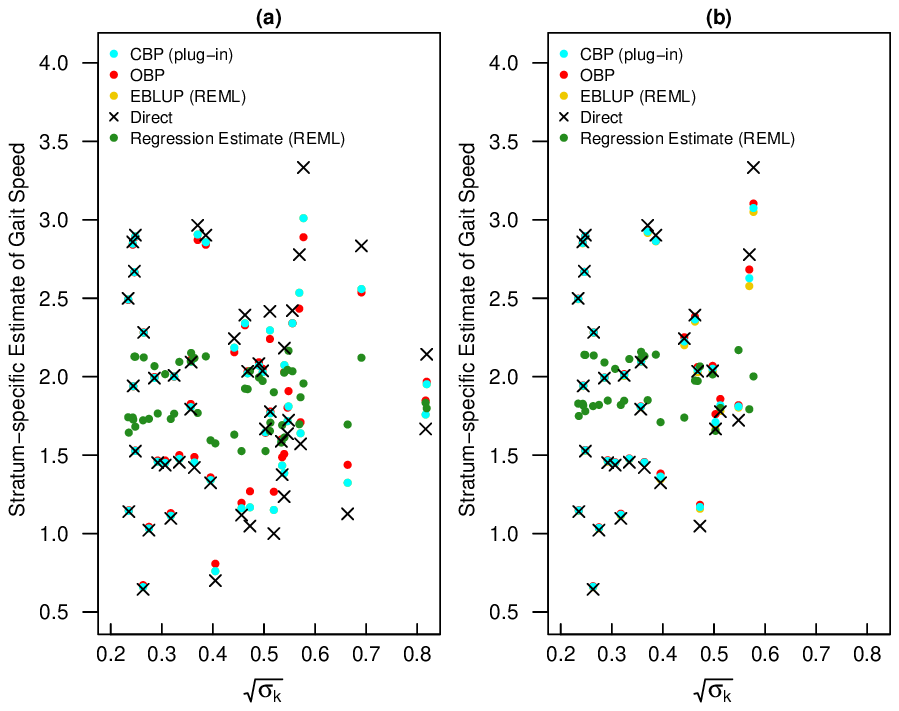}
\caption{Estimates of mean gait spped using the CBP (plug-in), OBP, and EBLUP (REML) methods within (a)48 demographically-defined strata, (b)a subset of 32 of these demographically-defined strata. The direct estimates are the within-stratum sample means $Y_{k}$. The regression estimates are given by $\hat{\beta}_{0}^{reml} + \hat{\beta}_{1}^{reml}(u_{k} - 65)$, where $\hat{\beta}_{0}^{reml}$, $\hat{\beta}_{1}^{reml}$ are the REML estimates and $u_{k}$ represents the sample mean of height within stratum $k$. Stratum-specific estimates are ordered according to the stratum-specific standard errors $\sigma_{k}$.  }
\label{fig:walktheta}
\end{figure}

The variation in the circle sizes in Figure \ref{fig:walkscatter} demonstrates the considerable variability in the stratum-specific sample sizes, and hence we should expect that many of the stratum-specific estimates of gait speed will have little impact from the overall regression fit $\hat{\beta}_{0} + \hat{\beta}_{1}x_{k}$ while others will have considerably more shrinkage towards the regression target. Figure \ref{fig:walktheta} shows that this is indeed the case for both the analysis of the 48 strata and the 32-strata subset. This figure shows the plug-in CBP, OBP, and EBLUP estimates of stratum-specific gait speed along with the fitted regression and direct estimates for each stratum. The fitted regression points are based on the REML estimates of $\beta_{0}$ and $\beta_{1}$. EBLUP (REML) estimates of gait speed are not shown in the 48 strata graph since the EBLUP estimates are essentially identical to the plug-in CBP estimates in this case. 
As shown in Figure \ref{fig:walktheta}(a), when $\sqrt{\sigma_{k}} \leq 0.4$ both the plug-in CBP and OBP provide almost no shrinkage of the direct estimates towards their corresponding regression targets. For $\sqrt{\sigma_{k}} \geq 0.5$, noticeable differences between the plug-in CBP and the OBP become more apparent. Specifically, the OBP appears to consistently shrink the direct estimates towards the REML-based estimates of the regression line more than the plug-in CBP though this is largely due to the fact that the OBP is shrinking the direct estimates towards a different regression target. 

As in the 48 strata analysis, the CBP, EBLUP, and OBP all apply very little 
shrinkage to the direct estimates for $\sqrt{\sigma_{k}} \leq 0.4$ in the analysis of the 32 strata subset shown in Figure \ref{fig:walktheta}(b). Only for smaller strata where $\sqrt{\sigma_{k}} \geq 0.5$ do the differences between the OBP and CBP become more apparent, and there are only noticeable differences between the EBLUP and CBP for the very small strata. The greater similarity between the CBP and EBLUP estimates than the EBLUP and OBP estimates mostly reflects the greater similarity between the CURE and REML estimates of the regression coefficients. The two strata with the smallest values of $\sigma_{k}$ shown in Figure \ref{fig:walktheta}(b) are the hispanic aged 65-69 strata. For these strata,
the EBLUP and the plug-in CBP estimates were $3.08$ and $3.05$ respectively for the hispanic/65-69/male subgroup where $\sqrt{\sigma_{k}} = 0.58$ and $n_{k} = 6$, and the EBLUP and CBP estimates were $2.63$ and $2.58$ respectively 
for the female subgroup where $\sqrt{\sigma_{k}} = 0.57$ and $n_{k} = 9$. 

\section{Discussion} \label{sec:discussion}

In this article, we have introduced a new approach for choosing regression weights in contexts where a regression
model and direct estimates are combined to estimate a collection of unit-specific mean parameters. In our approach, regression
weights are expressed as a convex combination of the MLE and BPE regression weights with the values of the regression weights depending on a mixing parameter $\alpha \in [0,1]$ and a variance component parameter $\tau \geq 0$. The terms $(\alpha, \tau)$ are determined empirically so that the corresponding estimates of the small domain means are competitive with the EBLUP in situations where the model
is correctly specified. The adaptive nature of the regression weights can improve the robustness of the small domain estimates in situations where the mean model is misspecified particularly when unit-specific sample size have unmodeled association with the unit-specific mean parameters. While enriching the covariates or making the regression model more flexible can reduce the impact of such informative sample size and hence improve the likelihood-based approach, determining the correctness of a model is ultimately an imperfect process and using our approach for constructing regression weights provides an automatic extra layer of robustness.  

As shown in the simulations studies described in Section 5.1 and Section 5.2, the plug-in CBP
typically performs as good or better than the CBP when evaluated by MSPE. We have found that this is generally the case in other simulation studies we have conducted. The discrepancy between CBP and plug-in CBP seems to be largely related to the
differences in performance between EBLUP (REML) and EBLUP (URE). More specifically, EBLUP (REML) generally works better than 
EBLUP (URE) when the model is either correctly specified or nearly correctly specified, and for these cases, 
the plug-in CBP estimates of $\theta_{k}$ will be very close to the corresponding EBLUP (REML) estimates. In contrast, the CBP estimates will more closely resemble the EBLUP (URE) estimates in cases with correct model specification. While EBLUP (URE) can perform better than EBLUP (REML) in many misspecified scenarios, this is irrelevant to the performance of either the CBP or the plug-in CBP as both versions of the CBP will be much closer to the OBP in such cases. 

Our approach provides a more flexible procedure for constructing regression 
weights by using a weight
function that uses a single additional hyperparameter $\alpha$ to determine
the relative weights given to the MLE and BPE regression weights.
While this greater flexibility can offer better performance, there is certainly potential to devise even better weighting schemes by considering more flexible weight functions. The key challenge 
in constructing such more flexible weight functions would be how to tradeoff 
increased flexibility with the 
increased variance that would accompany the use of many additional hyperparameters
to index a class of highly flexible weight functions. 
Constructing an appropriate definition of the ``degrees of freedom" in this context would be one approach for addressing this tradeoff as one could then directly compare weight functions with differing number of hyperparameters by penalizing the additional degrees of freedom appropriately.
A possible related approach that would not require one to directly define an appropriate measure of degrees
of freedom would be to select the weights by minimizing a ``penalized" version of the unbiased estimator $\hat{M}_{K}(\mathbf{w}, \tau)$
where one adds a penalty term that penalizes the complexity of the regression weights $\mathbf{w}$. The exploration of more flexible weighting schemes is beyond the scope of this paper but is certainly an important issue to consider in future research.

We have primarily focused on the use and performance of the CBP for estimating unit-specific mean parameters and did not explore using the CBP framework to generate uncertainty intervals for the mixed effects $\theta_{k}$.
One approach for doing so would be to use the percentiles of the marginal posterior of $\theta_{k}$ under an assumption that the $\theta_{k}$ follow a Gaussian distribution. Rather than computing the percentiles of the marginal posterior by integrating with respect to a specific choice of hyperprior for $\alpha$ and $\tau$, the marginal posterior could be approximated using a bootstrap approach similar to that described in \cite{Laird:1987}.

As is the case in many small domain estimation procedures, the CBP is derived under an assumption that the unit-specific sampling variances $\sigma_{k}^{2}$ are known. Though not explored in this paper, one way of relaxing the assumption of known sampling variances assumption is to introduce a hierarchical model for both the direct estimates and the corresponding estimates of their variances as has been done in, for example, \cite{Dass:2012} and \cite{Sugasawa:2017}. With this approach, one would specify a conditional joint distribution for each direct estimate and its corresponding standard error and specify a distribution for the underlying sampling variances. Using this setup, one could then implement a type of two-stage procedure where one first computes shrunken estimates of the sampling variances based on their posterior means and then uses these shrunken estimates to construct the shrinkage weights $B_{k,\tau}$. Using these alternative shrinkage weights, one could then find the variance component and mixing parameter estimates for the CBP using the unbiased risk estimate described in Section \ref{sec:CBP_describe}. A closely related alternative to this would be to use shrunken values of the standard errors and assume a $t$ distribution for the direct estimates as was suggested in \cite{Lu:2019}.

\section*{Supplementary Material}
An \textbf{R} package entitled \textbf{shrinkcbp} which implements the methods discussed in this paper may be retrieved from \url{https://github.com/nchenderson/shrinkcbp}.

\bibliographystyle{chicago}
\bibliography{sae_bib}

\begin{thebibliography}{}

\bibitem[\protect\citeauthoryear{Battese, Harter, and Fuller}{Battese
  et~al.}{1988}]{Battese:1988}
Battese, G.~E., R.~M. Harter, and W.~A. Fuller (1988).
\newblock An error-components model for prediction of county crop areas using
  survey and satellite data.
\newblock {\em Journal of the American Statistical Association\/}~{\em
  83\/}(401), 28--36.

\bibitem[\protect\citeauthoryear{Bohannon}{Bohannon}{1997}]{Bohannon:1997}
Bohannon, R.~W. (1997).
\newblock Comfortable and maximum walking speed of adults aged 20--79 years:
  {R}eference values and determinants.
\newblock {\em Age and ageing\/}~{\em 26\/}(1), 15--19.

\bibitem[\protect\citeauthoryear{Byrd, Lu, Nocedal, and Zhu}{Byrd
  et~al.}{1995}]{Byrd:1995}
Byrd, R.~H., P.~Lu, J.~Nocedal, and C.~Zhu (1995).
\newblock A limited memory algorithm for bound constrained optimization.
\newblock {\em SIAM Journal on scientific computing\/}~{\em 16\/}(5),
  1190--1208.

\bibitem[\protect\citeauthoryear{Craven and Wahba}{Craven and
  Wahba}{1978}]{Craven1978}
Craven, P. and G.~Wahba (1978).
\newblock Smoothing noisy data with spline functions.
\newblock {\em Numerische mathematik\/}~{\em 31\/}(4), 377--403.

\bibitem[\protect\citeauthoryear{Dass, Maiti, Ren, and Sinha}{Dass
  et~al.}{2012}]{Dass:2012}
Dass, S.~C., T.~Maiti, H.~Ren, and S.~Sinha (2012).
\newblock Confidence interval estimation of small area parameters shrinking
  both means and variances.
\newblock {\em Survey Methodology\/}~{\em 38\/}(2), 173--187.

\bibitem[\protect\citeauthoryear{Datta and Ghosh}{Datta and
  Ghosh}{2012}]{Datta:2012}
Datta, G. and M.~Ghosh (2012).
\newblock Small area shrinkage estimation.
\newblock {\em Statistical Science\/}~{\em 27\/}(1), 95--114.

\bibitem[\protect\citeauthoryear{Donoho and Johnstone}{Donoho and
  Johnstone}{1995}]{Donoho:1995}
Donoho, D.~L. and I.~M. Johnstone (1995).
\newblock Adapting to unknown smoothness via wavelet shrinkage.
\newblock {\em Journal of the American Statistical Association\/}~{\em
  90\/}(432), 1200--1224.

\bibitem[\protect\citeauthoryear{Fay and Herriot}{Fay and
  Herriot}{1979}]{Fay:1979}
Fay, R.~E. and R.~A. Herriot (1979).
\newblock Estimates of income for small places: {A}n application of
  {J}ames-{S}tein procedures to census data.
\newblock {\em Journal of the American Statistical Association\/}~{\em
  74\/}(366), 269--277.

\bibitem[\protect\citeauthoryear{Henderson}{Henderson}{1975}]{Henderson:1975}
Henderson, C.~R. (1975).
\newblock Best linear unbiased estimation and prediction under a selection
  model.
\newblock {\em Biometrics\/}~{\em 31\/}(494), 423--447.

\bibitem[\protect\citeauthoryear{Jiang, Nguyen, and Rao}{Jiang
  et~al.}{2011}]{Jiang:2011}
Jiang, J., T.~Nguyen, and J.~S. Rao (2011).
\newblock Best predictive small area estimation.
\newblock {\em Journal of the American Statistical Association\/}~{\em
  106\/}(494), 732--745.

\bibitem[\protect\citeauthoryear{Jones, Ohlssen, Neuenschwander, Racine, and
  Branson}{Jones et~al.}{2011}]{Jones:2011}
Jones, H.~E., D.~I. Ohlssen, B.~Neuenschwander, A.~Racine, and M.~Branson
  (2011).
\newblock Bayesian models for subgroup analysis in clinical trials.
\newblock {\em Clinical Trials\/}~{\em 8\/}(2), 129--143.

\bibitem[\protect\citeauthoryear{Kou and Yang}{Kou and Yang}{2017}]{Kou:2017}
Kou, S. and J.~J. Yang (2017).
\newblock Optimal shrinkage estimation in heteroscedastic hierarchical linear
  models.
\newblock In {\em Big and Complex Data Analysis}, pp.\  249--284. Springer.

\bibitem[\protect\citeauthoryear{Laird and Louis}{Laird and
  Louis}{1987}]{Laird:1987}
Laird, N.~M. and T.~A. Louis (1987).
\newblock Empirical bayes confidence intervals based on bootstrap samples.
\newblock {\em Journal of the American Statistical Association\/}~{\em
  82\/}(399), 739--750.

\bibitem[\protect\citeauthoryear{Li}{Li}{1985}]{Li:1985}
Li, K.-C. (1985).
\newblock From {S}tein's unbiased risk estimates to the method of generalized
  cross-validation.
\newblock {\em The Annals of Statistics\/}~{\em 13\/}(4), 1352--1377.

\bibitem[\protect\citeauthoryear{Li}{Li}{1986}]{Li:1986}
Li, K.-C. (1986).
\newblock Asymptotic optimality of {$C_L$} and generalized cross-validation in
  ridge regression with application to spline smoothing.
\newblock {\em The Annals of Statistics\/}~{\em 14\/}(3), 1101--1112.

\bibitem[\protect\citeauthoryear{Little}{Little}{2004}]{Little:2004}
Little, R.~J. (2004).
\newblock To model or not to model? {C}ompeting modes of inference for finite
  population sampling.
\newblock {\em Journal of the American Statistical Association\/}~{\em
  99\/}(466), 546--556.

\bibitem[\protect\citeauthoryear{Lu and Stephens}{Lu and
  Stephens}{2019}]{Lu:2019}
Lu, M. and M.~Stephens (2019).
\newblock Empirical {B}ayes estimation of normal means, accounting for
  uncertainty in estimated standard errors.
\newblock {\em arXiv preprint arXiv:1901.10679\/}.

\bibitem[\protect\citeauthoryear{Matloff}{Matloff}{1981}]{Matloff:1981}
Matloff, N.~S. (1981).
\newblock Use of regression functions for improved estimation of means.
\newblock {\em Biometrika\/}~{\em 68\/}(3), 685--689.

\bibitem[\protect\citeauthoryear{Middleton, Fritz, and Lusardi}{Middleton
  et~al.}{2015}]{Middleton:2015}
Middleton, A., S.~L. Fritz, and M.~Lusardi (2015).
\newblock Walking speed: the functional vital sign.
\newblock {\em Journal of aging and physical activity\/}~{\em 23\/}(2),
  314--322.

\bibitem[\protect\citeauthoryear{Normand, Ash, Fienberg, Stukel, Utts, and
  Louis}{Normand et~al.}{2016}]{Normand:2016}
Normand, S.-L.~T., A.~S. Ash, S.~E. Fienberg, T.~A. Stukel, J.~Utts, and T.~A.
  Louis (2016).
\newblock League tables for hospital comparisons.

\bibitem[\protect\citeauthoryear{Rao and Molina}{Rao and
  Molina}{2015}]{Rao:2015}
Rao, J. N.~K. and I.~Molina (2015).
\newblock {\em Small Area Estimation}.
\newblock Hoboken, NJ: Wiley.

\bibitem[\protect\citeauthoryear{Smyth}{Smyth}{2004}]{Smyth2004}
Smyth, G.~K. (2004).
\newblock Linear models and empirical bayes methods for assessing differential
  expression in microarray experiments.
\newblock {\em Statistical Applications in Genetics and Molecular
  Biology\/}~{\em 3\/}(1), 1--25.

\bibitem[\protect\citeauthoryear{Stein}{Stein}{1981}]{Stein:1981}
Stein, C.~M. (1981).
\newblock Estimation of the mean of a multivariate normal distribution.
\newblock {\em The Annals of Statistics\/}~{\em 9\/}(6), 1135--1151.

\bibitem[\protect\citeauthoryear{Studenski, Perera, Patel, Rosano, Faulkner,
  Inzitari, Brach, Chandler, Cawthon, Connor, et~al.}{Studenski
  et~al.}{2011}]{Studenski:2011}
Studenski, S., S.~Perera, K.~Patel, C.~Rosano, K.~Faulkner, M.~Inzitari,
  J.~Brach, J.~Chandler, P.~Cawthon, E.~B. Connor, et~al. (2011).
\newblock Gait speed and survival in older adults.
\newblock {\em {JAMA}\/}~{\em 305\/}(1), 50--58.

\bibitem[\protect\citeauthoryear{Sugasawa, Tamae, and Kubokawa}{Sugasawa
  et~al.}{2017}]{Sugasawa:2017}
Sugasawa, S., H.~Tamae, and T.~Kubokawa (2017).
\newblock Bayesian estimators for small area models shrinking both means and
  variances.
\newblock {\em Scandinavian Journal of Statistics\/}~{\em 44\/}(1), 150--167.

\bibitem[\protect\citeauthoryear{Wakefield}{Wakefield}{2007}]{Wakefield:2007}
Wakefield, J. (2007).
\newblock Disease mapping and spatial regression with count data.
\newblock {\em Biostatistics\/}~{\em 8\/}(2), 158--183.

\bibitem[\protect\citeauthoryear{Xie, Kou, and Brown}{Xie
  et~al.}{2012}]{Xie:2012}
Xie, X., S.~C. Kou, and L.~D. Brown (2012).
\newblock {SURE} estimates for a heteroscedastic hierarchical model.
\newblock {\em Journal of the American Statistical Association\/}~{\em
  107\/}(500), 1465--1479.

\bibitem[\protect\citeauthoryear{Zheng and Little}{Zheng and
  Little}{2005}]{Zheng:2005}
Zheng, H. and R.~J.~A. Little (2005).
\newblock Inference for the population total from
  probability-proportional-to-size samples based on predictions from a
  penalized spline nonparametric model.
\newblock {\em Journal of Official Statistics\/}~{\em 21\/}(1), 1--20.

\end{thebibliography}

\appendix

\section{Proofs of Theorems 1-2}

\noindent
\textit{Proof of Theorem 1.}
First note that we may re-write $\textrm{MSPE}(\mathbf{w}, \tau)$ as
\begin{eqnarray}
\textrm{MSPE}(\mathbf{w},\tau) &=& \bmu^{T}\mathbf{U}_{\mathbf{w}, \tau}^{T}\mathbf{U}_{\mathbf{w}, \tau}\bmu
+ \textrm{tr}\big\{ \mathbf{V}_{Y|\theta}(\mathbf{U}_{\mathbf{w},\tau}^{T}\mathbf{U}_{\mathbf{w},\tau} + \mathbf{U}_{\mathbf{w},\tau} + \mathbf{U}_{\mathbf{w},\tau}^{T} + \mathbf{I}) \big\}
+ \tau_{0}^{2}\textrm{tr}\big\{ \mathbf{U}_{\mathbf{w}, \tau}^{T}\mathbf{U}_{\mathbf{w}, \tau} \big\}  \nonumber \\
&=& \bmu^{T}\mathbf{U}_{\mathbf{w}, \tau}^{T}\mathbf{U}_{\mathbf{w}, \tau}\bmu
+ \textrm{tr}\big\{ \mathbf{U}_{\mathbf{w},\tau}^{T}\mathbf{U}_{\mathbf{w},\tau}\mathbf{V}_{Y|\theta} \big\}  + 2\textrm{tr}\big\{ \mathbf{U}_{\mathbf{w},\tau} \mathbf{V}_{Y|\theta} \big\} + \textrm{tr}\big\{ \mathbf{V}_{Y|\theta} \} 
\nonumber \\
& & + \tau_{0}^{2}\textrm{tr}\big\{ \mathbf{U}_{\mathbf{w}, \tau}^{T}\mathbf{U}_{\mathbf{w}, \tau} \big\}. \nonumber 
\end{eqnarray}
Under model (\ref{eq:mixed_model_true}), it is the case that
\begin{equation}
\mathbf{Y} \sim ( \bmu, \tau_{0}^{2}\mathbf{I} + \mathbf{V}_{Y|\theta} )
\qquad
\textrm{and}
\qquad
\mathbf{Y}|\btheta \sim ( \btheta, \mathbf{V}_{Y|\theta} ),
\label{eq:mean_var_form}
\end{equation}
where the notation $\mathbf{Z} \sim (\mathbf{a}, \mathbf{B})$ means that $\mathbf{Z}$ has mean vector $\mathbf{a}$ and variance-covariance matrix $\mathbf{B}$. Because $\mathbf{Y}|\btheta \sim (\btheta, \mathbf{V}_{Y|\theta})$, it can be directly shown that the conditional MSPE is
\begin{eqnarray}
\textrm{condMSPE}(\mathbf{w}, \tau) &=& 
E\Big[ \big\{ \hatbtheta(\mathbf{w}, \tau) - \btheta \big\}^{T}\big\{\hatbtheta(\mathbf{w}, \tau) - \btheta\big\} \Big| \btheta \Big] \nonumber \\
&=& \btheta^{T}\mathbf{U}_{\mathbf{w}, \tau}^{T}\mathbf{U}_{\mathbf{w},\tau}\btheta
+ 
\textrm{tr}\big\{ \mathbf{U}_{\mathbf{w}, \tau}^{T}\mathbf{U}_{\mathbf{w},\tau} \mathbf{V}_{Y|\theta} \big\} +
2\textrm{tr}\{ \mathbf{U}_{\mathbf{w},\tau}\mathbf{V}_{Y|\theta} \}
+ \textrm{tr}\{ \mathbf{V}_{Y|\theta} \}. \nonumber 
\end{eqnarray}
Now, because of (\ref{eq:mean_var_form}), the marginal and conditional expectations of the quadratic form
$\mathbf{Y}^{T}\mathbf{U}_{\mathbf{w}, \tau}^{T}\mathbf{U}_{\mathbf{w}, \tau}\mathbf{Y}$
are given by
\begin{eqnarray}
E( \mathbf{Y}^{T}\mathbf{U}_{\mathbf{w}, \tau}^{T}\mathbf{U}_{\mathbf{w},\tau}\mathbf{Y})
&=& \bmu^{T}\mathbf{U}_{\mathbf{w}, \tau}^{T}\mathbf{U}_{\mathbf{w},\tau}\bmu
+ \tau_{0}^{2}\textrm{tr}\big\{ \mathbf{U}_{\mathbf{w}, \tau}^{T}\mathbf{U}_{\mathbf{w},\tau} \big\} 
+ 
\textrm{tr}\big\{ \mathbf{U}_{\mathbf{w}, \tau}^{T}\mathbf{U}_{\mathbf{w},\tau} \mathbf{V}_{Y|\theta} \big\}  \label{eq:marginal_qf} \\
E( \mathbf{Y}^{T}\mathbf{U}_{\mathbf{w}, \tau}^{T}\mathbf{U}_{\mathbf{w},\tau}\mathbf{Y} \mid \btheta)
&=& \btheta^{T}\mathbf{U}_{\mathbf{w}, \tau}^{T}\mathbf{U}_{\mathbf{w},\tau}\btheta
+ 
\textrm{tr}\big\{ \mathbf{U}_{\mathbf{w}, \tau}^{T}\mathbf{U}_{\mathbf{w},\tau} \mathbf{V}_{Y|\theta} \big\}. \label{eq:conditional_qf}
\end{eqnarray}
Using (\ref{eq:marginal_qf}), the marginal expectation of $\hat{M}_{K}(\mathbf{w}, \tau)$ is given by
\begin{eqnarray}
E\{ \hat{M}_{K}( \mathbf{w}, \tau) \}
&=& \bmu^{T}\mathbf{U}_{\mathbf{w}, \tau}^{T}\mathbf{U}_{\mathbf{w},\tau}\bmu
+ \tau_{0}^{2}\textrm{tr}\big\{ \mathbf{U}_{\mathbf{w}, \tau}^{T}\mathbf{U}_{\mathbf{w},\tau} \big\} 
+ 
\textrm{tr}\big\{ \mathbf{U}_{\mathbf{w}, \tau}^{T}\mathbf{U}_{\mathbf{w},\tau}\mathbf{V}_{Y|\theta}  \big\} 
\nonumber \\
& & 
+ 2\textrm{tr}\{ \mathbf{U}_{\mathbf{w},\tau}\mathbf{V}_{Y|\theta}\} + \textrm{tr}\{ \mathbf{V}_{Y|\theta} \} = \textrm{MSPE}(\mathbf{w}, \tau), \nonumber
\end{eqnarray}
and from (\ref{eq:conditional_qf}), the conditional expectation is given by
\begin{equation}
E\{ \hat{M}_{K}(\mathbf{w}, \tau)| \btheta \}
= \btheta^{T}\mathbf{U}_{\mathbf{w}, \tau}^{T}\mathbf{U}_{\mathbf{w},\tau}\btheta
+ 
\textrm{tr}\big\{ \mathbf{U}_{\mathbf{w}, \tau}^{T}\mathbf{U}_{\mathbf{w},\tau} \mathbf{V}_{Y|\theta} \big\} +
2\textrm{tr}\{ \mathbf{U}_{\mathbf{w},\tau}\mathbf{V}_{Y|\theta} \}
+ \textrm{tr}\{ \mathbf{V}_{Y|\theta} \}. \nonumber 
\end{equation}

\bigskip

\bigskip

\noindent
\textit{Proof of Theorem 2.} Our aim is to minimize the objective function 
\begin{equation}
F_{\tau}(\alpha) = \hat{M}_{g,0}(\mathbf{w}^{c}(\alpha, \tau), \tau), \nonumber
\end{equation}
where $F_{\tau}(\alpha)$ is defined as
\begin{eqnarray}
&& F_{\tau}(\alpha) 
= \Big(\frac{1}{K}\sum_{k=1}^{K}B_{k,\tau}Y_{k} - \frac{\alpha B_{.\tau}}{K}\sum_{j=1}^{K} w_{j}^{1}(\tau) Y_{j} - \frac{(1 - \alpha) B_{.\tau}}{K}\sum_{j=1}^{K} w_{j}^{0}(\tau)Y_{j} \Big)^{2}  \nonumber \\
&+& \frac{2B_{.\tau}\sigma^{2}\alpha}{K^{2}}\sum_{k=1}^{K} \frac{ w_{k}^{1}(\tau) - w_{k}^{0}(\tau) }{n_{k}} + \frac{2B_{.\tau}\sigma^{2}}{K^{2}}\sum_{k=1}^{K} \frac{ w_{k}^{0}(\tau) }{n_{k}} 
- \frac{2}{K^{2}}\sum_{k=1}^{K} \frac{\sigma^{4}}{n_{k}(\sigma^{2} + n_{k}\tau^{2})}  + \frac{\sigma^{2}}{K \ddot n} \nonumber \\
&=& \alpha^{2}C_{1}(\tau) - 2\alpha C_{2}(\tau) + C_{3}(\tau), \nonumber 
\end{eqnarray}
The terms $C_{1}(\tau)$, $C_{2}(\tau), C_{3}(\tau)$ are defined as 
\begin{eqnarray}
C_{1}(\tau) &=& \frac{1}{K^{2}} \Big( \sum_{j=1}^{K} \{ w_{j}^{1}(\tau) -  w_{j}^{0}(\tau) \} Y_{j}\Big)^{2}B_{.\tau}^{2} \nonumber \\
C_{2}(\tau) &=& \frac{B_{.\tau}}{K^{2}}\Big( \sum_{j=1}^{K} \{ w_{j}^{1}(\tau) -  w_{j}^{0}(\tau) \} Y_{j}\Big)\Big( (\sum_{j=1}^{K} B_{j,\tau} Y_{j}) - B_{.\tau}\sum_{j=1}^{J}w_{j}^{0}(\tau)Y_{j} \Big)  - \frac{B_{.\tau}\sigma^{2}}{K^{2}}\sum_{k=1}^{K} \frac{ w_{k}^{1}(\tau) - w_{k}^{0}(\tau) }{n_{k}}  \nonumber \\
C_{3}(\tau) &=& \frac{1}{K^{2}}\Big( (\sum_{k=1}^{K} B_{k,\tau}Y_{k}) - B_{.\tau}\sum_{k=1}^{K}w_{k}^{0}(\tau)Y_{k}\Big)^{2} + \frac{2B_{.\tau}\sigma^{2}}{K^{2}}\sum_{k=1}^{K} \frac{ w_{k}^{0}(\tau) }{n_{k}} - \frac{2}{K^{2}}\sum_{k=1}^{K} \frac{\sigma^{4}}{n_{k}(\sigma^{2} + n_{k}\tau^{2})}  + \frac{\sigma^{2}}{K \ddot n} \nonumber 
\end{eqnarray}
The derivative of $F_{\tau}(\alpha)$ is then
\begin{equation}
F_{\tau}'(\alpha) =  2\alpha C_{1}(\tau) - 2C_{2}(\tau) \nonumber 
\end{equation}
Because $F_{\tau}''(\alpha) = 2C_{1}(\tau) > 0$ for all $\alpha \in [0,1]$, the optimal $\alpha$ is zero if $F_{\tau}'(0) \geq 0$, is $1$ if $F_{\tau}'(1) \leq 0$, and is the solution of the equation $F_{\tau}'(\alpha) = 0$ if both $F_{\tau}'(0) < 0$ and $F_{\tau}'(1) > 0$. An optimal value of $\alpha \in \{0, 1\}$ can be checked by looking at
\begin{eqnarray}
F_{\tau}'(0) = - 2C_{2}(\tau) \qquad 
F_{\tau}'(1) = 2(C_{1}(\tau) - C_{2}(\tau)) . \nonumber 
\end{eqnarray}
For the case of $F_{\tau}'(0) < 0$ and $F_{\tau}'(1) > 0$, it is clear that the solution of $F_{\tau}'(\alpha) = 0$ is $\alpha_{0}^{*} = C_{2}(\tau)/C_{1}(\tau)$. Thus, the optimal value $\alpha_{opt}(\tau)$ of $\alpha \in [0,1]$ is
\begin{equation}
    \alpha_{opt}(\tau) = 
\begin{cases}
    0,& \text{if } C_{2}(\tau) \leq 0 \\
    1,              & \text{ if } C_{1}(\tau) \leq C_{2}(\tau)  \\
    C_{2}(\tau)/C_{1}(\tau), & \text{ otherwise}
\end{cases}
\nonumber 
\end{equation}

\section{Proofs of Theorems 3 and 4}
\subsection{Lemmas}
\begin{lemma}
Let $w_{K,max}^{c}(\alpha, \tau) = \max\{w_{1}^{c}(\alpha, \tau), \ldots, w_{K}^{c}(\alpha, \tau) \}$ and \\ $w_{K,min}^{c}(\alpha,\tau) = \min\{ w_{1}^{c}(\alpha, \tau), \ldots, w_{K}^{c}(\alpha, \tau) \}$ denote the maximum and minimum values of the compromise weights $w_{k}^{c}(\alpha, \tau)$ for fixed values of $(\alpha,\tau)$. Then, if $\sigma_{max,K}^{2} = \max\{\sigma_{1}^{2}, \ldots, \sigma_{K}^{2} \}$ and $\sigma_{min,K}^{2} = \min\{ \sigma_{1}^{2}, \ldots, \sigma_{K}^{2} \}$ denote the maximum and minimum values of the variances $\sigma_{k}^{2}$ respectively, the following inequality holds for any $K \geq 1$
\begin{equation}
\sup_{\alpha \in [0,1], \tau \geq 0}\Big( \frac{w_{K,max}^{c}(\alpha, \tau) }{w_{K,min}^{c}(\alpha, \tau)} \Big)^{1/2} \leq \frac{2 \sigma_{max,K}^{2}} {\sigma_{min,K}^{2}} + 1 \nonumber
\end{equation}
\end{lemma}

\smallskip

\noindent
\textbf{Proof:} First, recall the formulas for the MLE and BPE weights
\begin{equation}
w_{k}^{MLE}(\tau) = \frac{1}{C_{1}(\tau)(\sigma_{k}^{2} + \tau^{2})}   \qquad \mbox{and} \qquad
w_{k}^{BPE}(\tau) = \frac{\sigma_{k}^{4}}{C_{0}(\tau)(\sigma_{k}^{2} + \tau^{2})^{2}}, \nonumber
\end{equation}
where the terms $C_{1}(\tau)$ and $C_{0}(\tau)$ are defined as
\begin{equation}
C_{1}(\tau) = \sum_{k=1}^{K} \frac{1}{\sigma_{k}^{2} + \tau^{2}} \qquad \mbox{and} \qquad C_{0}( \tau ) = \sum_{k=1}^{K}\frac{\sigma_{k}^{4}}{(\sigma_{k}^{2} + \tau^{2})^{2}} \nonumber
\end{equation}
Let $w_{K,max}^{MLE}(\tau) = \max\{w_{1}^{MLE}(\tau), \ldots, w_{K}^{MLE}(\tau) \}$ and $w_{K,min}^{MLE}(\tau) = \min\{w_{1}^{MLE}(\tau), \ldots, w_{K}^{MLE}(\tau) \}$ and define both $w_{K,max}^{BPE}(\tau)$ and $w_{K,min}^{BPE}(\tau)$ analagously. Because $w_{k}^{MLE}(\tau)$ is a decreasing function of $\sigma_{k}^{2}$ (for fixed $\tau$) and $w_{k}^{BPE}(\tau)$ is an increasing function of $\sigma_{k}^{2}$ (for fixed $\tau$), we have that $w_{K,max}^{MLE}(\tau) = [C_{1}(\tau)(\sigma_{min,K}^{2} + \tau^{2})]^{-1}$, $w_{K,min}^{MLE}(\tau) = [C_{1}(\tau)(\sigma_{max,K}^{2} + \tau^{2})]^{-1}$, \\
$w_{K,max}^{BPE}(\tau) = \sigma_{K,max}^{4}/[C_{0}(\tau)(\sigma_{K,max}^{2} + \tau^{2})^{2}]$, and 
$w_{K,min}^{BPE}(\tau) = \sigma_{K,min}^{4}/[C_{0}(\tau)(\sigma_{K,min}^{2} + \tau^{2})^{2}]$.

\smallskip

\noindent
Now, because $w_{k}^{c}(\alpha, \tau) = \alpha w_{k}^{MLE}(\tau) + (1 - \alpha)w_{k}^{BPE}$ is a convex combination of $w_{k}^{MLE}(\tau)$ and $w_{k}^{BPE}(\tau)$, it is the case that for any $\tau \geq 0$ and $\alpha \in [0,1]$,
\begin{eqnarray}
\frac{w_{K,max}^{c}(\alpha, \tau) }{w_{K,min}^{c}(\alpha, \tau)} &\leq& \frac{\max\{ w_{K,max}^{MLE}(\tau), w_{K,max}^{BPE}(\tau) \}}{ \min\{ w_{K,min}^{MLE}(\tau), w_{K,min}^{BPE}(\tau) \} } \nonumber \\
&\leq& \frac{ w_{K,max}^{MLE}(\tau) }{ w_{K,min}^{MLE}(\tau) } + \frac{ w_{K,max}^{BPE}(\tau) }{ w_{K,min}^{BPE}(\tau) } + \frac{ w_{K,max}^{MLE}(\tau) }{  w_{K,min}^{BPE}(\tau)  } + \frac{ w_{K,max}^{BPE}(\tau) }{ w_{K,min}^{MLE}(\tau) }  \nonumber \\
&=& A_{1K}(\tau) + A_{2K}(\tau) + A_{3K}(\tau) + A_{4K}(\tau). \label{eq:A_inequality}
\end{eqnarray}
Our goal now is to provide upper bounds for each $A_{jK}(\tau)$ which do not depend on $\tau$. We consider each of these terms separately:

\begin{description}
\item[(1)]
First, consider $A_{1K}(\tau)$
\begin{equation}
A_{1K}(\tau) = \frac{ w_{K,max}^{MLE}(\tau) }{ w_{K,min}^{MLE}(\tau) }
= \frac{\sigma_{K,max}^{2} + \tau^{2}}{\sigma_{K,min}^{2} + \tau^{2}}
\leq \frac{\sigma_{K,max}^{2}}{\sigma_{K,min}^{2}}
\leq 1 + \frac{\sigma_{K,max}^{4}}{\sigma_{K,min}^{4}}.
\label{eq:A1_inequality}
\end{equation}
\item[(2)]
Now, consider $A_{2K}(\tau)$
\begin{equation}
A_{2K}(\tau) = \frac{ w_{K,max}^{BPE}(\tau) }{ w_{K,min}^{BPE}(\tau) }
= \frac{\sigma_{K,max}^{4}(\sigma_{K,min}^{2} + \tau^{2})^{2} }{\sigma_{K,min}^{4}(\sigma_{K,max}^{2} + \tau^{2})^{2}}
\leq \frac{\sigma_{K,max}^{4} }{\sigma_{K,min}^{4}}.
\label{eq:A2_inequality}
\end{equation}
\item[(3)]
Now, consider $A_{3K}(\tau)$
\begin{eqnarray}
A_{3K}(\tau) &=& \frac{ w_{K,max}^{MLE}(\tau) }{  w_{K,min}^{BPE}(\tau)  }
= \frac{C_{0}(\tau)(\sigma_{K,min}^{2} + \tau^{2})^{2} }{C_{1}(\tau)\sigma_{K,min}^{4}(\sigma_{K,min}^{2} + \tau^{2})}
= \frac{C_{0}(\tau)(\sigma_{K,min}^{2} + \tau^{2}) }{C_{1}(\tau)\sigma_{K,min}^{4} } \nonumber \\
&\leq& \frac{(\sigma_{K,min}^{2} + \tau^{2})\sum_{k} \sigma_{k}^{4}(\sigma_{k}^{2} + \tau^{2})^{-2} }{\sigma_{K,min}^{4} \sum_{k} (\sigma_{k}^{2} + \tau^{2})^{-1} } 
\leq \frac{\sigma_{K,max}^{4} \sum_{k} \frac{\sigma_{K,min}^{2} + \tau^{2}}{\sigma_{k} + \tau^{2}}(\sigma_{k}^{2} + \tau^{2})^{-1} }{\sigma_{K,min}^{4} \sum_{k} (\sigma_{k}^{2} + \tau^{2})^{-1} } \nonumber \\
&\leq& \frac{\sigma_{K,max}^{4} \sum_{k} (\sigma_{k}^{2} + \tau^{2})^{-1} }{\sigma_{K,min}^{4} \sum_{k} (\sigma_{k}^{2} + \tau^{2})^{-1} } 
= \frac{\sigma_{K,max}^{4}}{\sigma_{K,min}^{4}}. 
\label{eq:A3_inequality}
\end{eqnarray}
Above, the third inequality comes from the fact $(\sigma_{K,min}^{2} + \tau^{2})/(\sigma_{k}^{2} + \tau^{2}) \leq 1$, for each $k$. 
\item[(4)]
Now, consider $A_{4K}(\tau)$
\begin{eqnarray}
A_{4K}(\tau) &=& \frac{ w_{K,max}^{BPE}(\tau) }{ w_{K,min}^{MLE}(\tau) }
= \frac{ C_{1}(\tau)(\sigma_{K,max}^{2} + \tau^{2})\sigma_{K,max}^{4} }{ C_{0}(\tau)(\sigma_{K,max}^{2} + \tau^{2})^{2} }
= \frac{ \sigma_{K,max}^{4} \sum_{k} (\sigma_{k}^{2} + \tau^{2})^{-1} }{ (\sigma_{K,max}^{2} + \tau^{2})\sum_{k} \sigma_{k}^{4}(\sigma_{k}^{2} + \tau^{2})^{-2} }  \nonumber \\
&\leq& \frac{ \sigma_{K,max}^{4} \sum_{k} (\sigma_{k}^{2} + \tau^{2})^{-1} }{ \sigma_{K,min}^{4}(\sigma_{K,max}^{2} + \tau^{2})\sum_{k} (\sigma_{k}^{2} + \tau^{2})^{-2} }
= \frac{ \sigma_{K,max}^{4} \sum_{k} (\sigma_{k}^{2} + \tau^{2})^{-1} }{ \sigma_{K,min}^{4}\sum_{k} \frac{\sigma_{K,max}^{2} + \tau^{2}}{\sigma_{k} + \tau^{2}}(\sigma_{k}^{2} + \tau^{2})^{-1} } \nonumber \\
&\leq& \frac{ \sigma_{K,max}^{4} \sum_{k} (\sigma_{k}^{2} + \tau^{2})^{-1} }{ \sigma_{K,min}^{4}\sum_{k} (\sigma_{k}^{2} + \tau^{2})^{-1} }
= \frac{ \sigma_{K,max}^{4}  }{ \sigma_{K,min}^{4} }.
\label{eq:A4_inequality}
\end{eqnarray}
Above, the second inequality comes from the fact $(\sigma_{K,max}^{2} + \tau^{2})/(\sigma_{k}^{2} + \tau^{2}) \geq 1$, for any $k$. 
\end{description}
Combining (\ref{eq:A_inequality}) with (\ref{eq:A1_inequality}), (\ref{eq:A2_inequality}), (\ref{eq:A3_inequality}), and (\ref{eq:A4_inequality}) allows to conclude that
\begin{eqnarray}
\sup_{\alpha \in [0,1], \tau \geq 0}\frac{w_{K,max}^{c}(\alpha, \tau) }{w_{K,min}^{c}(\alpha, \tau)} 
&\leq& \sup_{\tau \geq 0} |A_{1K}(\tau)| + \sup_{\tau \geq 0} |A_{2K}(\tau)| +  \sup_{\tau \geq 0} |A_{3K}(\tau)| + \sup_{\tau \geq 0} |A_{4K}(\tau)|  \nonumber \\
&\leq& \frac{4\sigma_{K,max}^{4}}{\sigma_{K,min}^{4}} + 1 \nonumber
\end{eqnarray}
The conclusion of the lemma then simply follows from the fact that $\sqrt{4x^{2} + 1} \leq 2x + 1$ for any $x \geq 0$.

\bigskip

\begin{lemma}
If $S_{2K}(\tau)$ is defined as
\begin{equation}
S_{2K}(\tau) = \frac{1}{K}\Big( \mathbf{Y}^{T}(\mathbf{I} - \mathbf{B}_{\tau})\mathbf{Y} - \textrm{tr}\{ (\mathbf{I} - \mathbf{B}_{\tau})\mathbf{V}_{Y|\theta} \} - \btheta^{T}(\mathbf{I} - \mathbf{B}_{\tau})\mathbf{Y} \Big), \nonumber 
\end{equation}
then
\begin{equation}
\lim_{K \longrightarrow \infty} E\Bigg( \sup_{\tau \geq 0}|S_{2K}(\tau)| \Bigg) = 0 \nonumber
\end{equation}
if condition (A3) holds.
\end{lemma}

\smallskip

\noindent
\textbf{Proof:} Note that
\begin{eqnarray}
\sup_{\tau \geq 0}|S_{2K}(\tau)| &=& \frac{1}{K}\sum_{k=1}^{K} \frac{\tau}{\sigma_{k}^{2} + \tau^{2}}(Y_{k}^{2} - \theta_{k}Y_{k} - \sigma_{k}^{2}) \nonumber \\ 
&\leq& \sup_{1 \geq c_{1} \geq \ldots \geq c_{K}}\Big| \frac{1}{K}\sum_{k=1}^{K} c_{k} (Y_{k}^{2} - \theta_{k}Y_{k} - \sigma_{k}^{2}) \Big| \nonumber \\
&=& \max_{1 \leq j \leq K}\Big| \frac{1}{K}\sum_{k=1}^{j} (Y_{k}^{2} - \theta_{k}Y_{k} - \sigma_{k}^{2}) \Big|, \nonumber
\end{eqnarray}
where the second equality follows from Lemma 2.1 of \cite{Li:1986}. Hence,
\begin{equation}
\sup_{\tau \geq 0}|S_{2K}(\tau)| \leq \frac{1}{K} \max_{1 \leq j \leq K} \Big| M_{j} \Big| \leq \frac{1}{K} + \frac{1}{K^{2}}\max_{1 \leq j \leq K} M_{j}^{2}, \nonumber
\end{equation}
where $M_{j} = \sum_{k=1}^{j}(Y_{k}^{2} - Y_{k}\theta_{k} - \sigma_{k}^{2})$ and thus $M_{1}, M_{2}, \ldots$ forms a martingale. 
It follows from the $L^{p}$ maximum inequality for martingales that
\begin{equation}
E\Big( \max_{1 \leq j \leq K} M_{j}^{2} \Big)
\leq 4E(M_{K}^{2}) 
= \sum_{k=1}^{K} E\big[ (\theta_{k}e_{k} + e_{k}^{2} - \sigma_{k}^{2})^{2} \big]
= \sum_{k=1}^{K} \{ (\mu_{k}^{2} + \tau_{0}^{2})\sigma_{k}^{2} + E(e_{k}^{4}) - \sigma_{k}^{4} \}. \nonumber
\end{equation}
Hence,
\begin{equation}
E\Bigg( \sup_{\tau \geq 0}|S_{2K}(\tau)| \Bigg) \leq \frac{1}{K} + \frac{1}{K^{2}}\sum_{k=1}^{K} \{ (\mu_{k}^{2} + \tau_{0}^{2})\sigma_{k}^{2} + E(e_{k}^{4}) - \sigma_{k}^{4} \} \nonumber
\end{equation}
and the result of the lemma then follows from condition (A3).

\begin{lemma}
If we define the $K \times K$ matrix $\mathbf{P}_{\alpha, \tau}$ as $\mathbf{P}_{\alpha, \tau} = \mathbf{X}(\mathbf{X}^{T}\mathbf{W}_{\alpha, \tau}^{c} \mathbf{X})^{-1}\mathbf{X}^{T}\mathbf{W}_{\alpha,\tau}^{c}$, then the following inequality holds
\begin{equation}
\sup_{\alpha \in [0,1], \tau \geq 0} \big| \big| \mathbf{P}_{\alpha, \tau} \mathbf{Y} \big|\big|_{2}^{2}  \leq \Big( \frac{4 \sigma_{max,K}^{4}} {\sigma_{min,K}^{4}} + 1 \Big)|| \mathbf{Y} ||_{2}^{2}. \nonumber
\end{equation}
\end{lemma}

\noindent
\textbf{Proof.} First, note that
\begin{equation}
\mathbf{P}_{\alpha, \tau} = \mathbf{X}(\mathbf{X}^{T}\mathbf{W}_{\alpha, \tau}^{c} \mathbf{X})^{-1}\mathbf{X}^{T}\mathbf{W}_{\alpha,\tau}^{c}
= (\mathbf{W}_{\alpha,\tau}^{c})^{-1/2}\tilde{\mathbf{P}}_{\alpha,\tau}(\mathbf{W}_{\alpha,\tau}^{c})^{1/2}, \nonumber
\end{equation}
where $\tilde{\mathbf{P}}_{\alpha, \tau}$ is the symmetric matrix $\tilde{\mathbf{P}}_{\alpha, \tau} = (\mathbf{W}_{\alpha,\tau}^{c})^{1/2}\mathbf{X}(\mathbf{X}^{T}\mathbf{W}_{\alpha, \tau}^{c} \mathbf{X})^{-1}\mathbf{X}^{T}(\mathbf{W}_{\alpha,\tau}^{c})^{1/2}$. Thus, if we use $|| \mathbf{A} ||_{2,M}$ to denote the spectral norm $|| \mathbf{A} ||_{2,M} = \sup_{||\mathbf{x}||_{2} = 1}|| \mathbf{A}\mathbf{x}||_{2}$, then
\begin{eqnarray}
|| \mathbf{P}_{\alpha, \tau} \mathbf{Y} ||_{2}^{2}
&=& || (\mathbf{W}_{\alpha,\tau}^{c})^{-1/2}\tilde{\mathbf{P}}_{\alpha,\tau}(\mathbf{W}_{\alpha,\tau}^{c})^{1/2}\mathbf{Y} ||_{2}^{2}
\leq || (\mathbf{W}_{\alpha,\tau}^{c})^{-1/2} ||_{2,M}^{2} || \tilde{\mathbf{P}}_{\alpha,\tau} ||_{2,M}^{2} || (\mathbf{W}_{\alpha,\tau}^{c})^{1/2} ||_{2,M}^{2} || \mathbf{Y} ||_{2}^{2} \nonumber \\
&\leq&  \frac{w_{max,K}^{c}(\alpha, \tau) }{w_{min,K}^{c}(\alpha, \tau)} || \mathbf{Y} ||_{2}^{2}
\leq \Big( \frac{4 \sigma_{max,K}^{4}} {\sigma_{min,K}^{4}} + 1 \Big)|| \mathbf{Y} ||_{2}^{2}.
\label{eq:py_inequality}
\end{eqnarray}
Above, the first inequality follows from the fact that $|| \mathbf{A}\mathbf{x} ||_{2} \leq || \mathbf{A} ||_{2,M}|| \mathbf{x} ||_{2}$; the second inequality follows from the fact that $|| \tilde{\mathbf{P}}_{\alpha,\tau}||_{2,M} = 1$ for any $(\alpha,\tau)$; and the third inequality follows from Lemma 1.

\begin{lemma}
Suppose that conditions (A2) and (A4) hold. Then,
\begin{equation}
\lim_{K \longrightarrow \infty}\frac{1}{K^{\delta}}E\Bigg\{ \max_{1 \leq j \leq K} \Big( || \mathbf{P}_{X}[\mathbf{Y} - \btheta]_{j:K} ||_{2}^{2} \Big) \Bigg\} = 0, \nonumber 
\end{equation}
where $[\mathbf{Y} - \btheta]_{j:K}$ is the $K \times 1$ vector whose first $(j-1)$ elements are zero and whose $k^{th}$ element equals $Y_{k} - \theta_{k}$ for $k \geq j$.
\end{lemma}

\noindent
\textbf{Proof.} First, the following inequality follows directly from the proof of Theorem 1 in \cite{Kou:2017}
\begin{equation}
E\Bigg\{ \max_{1 \leq j \leq K} \Big( || \mathbf{P}_{X}[\mathbf{Y} - \btheta]_{j:K} ||_{2}^{2} \Big) \Bigg\}
\leq 4\textrm{tr}( \mathbf{P}_{X}\mathbf{V}_{Y|\theta} ). \nonumber
\end{equation}
Let $P_{X}^{(k,k)}$ denote the $k^{th}$ diagonal element of $\mathbf{P}_{X}$. Then, because $\mathbf{V}_{Y|\theta}$ is diagonal, we have
\begin{equation}
\frac{1}{K^{\delta}}E\Bigg\{ \max_{1 \leq j \leq K} \Big( || \mathbf{P}_{X}[\mathbf{Y} - \btheta]_{j:K} ||_{2}^{2} \Big) \Bigg\}
\leq \frac{4}{K^{\delta}}\sum_{k=1}^{K} P_{X}^{(k,k)}\sigma_{k}^{2}
\leq 4\Big( K^{1-\delta/2} D_{max}(\mathbf{P}_{X}) \Big)\Big(\frac{1}{K^{1 + \delta/2}}\sum_{k=1}^{K} \sigma_{k}^{2} \Big), \nonumber
\end{equation}
where $D_{max}(\mathbf{P}_{X}) = \max\{P_{X}^{(1,1)}, \ldots, P_{X}^{(K,K)}\}$. The conclusion of the lemma now follows from conditions (A2) and (A4).

\subsection{Proof of Theorem 3}
For this proof, let $\mathbf{U}_{\alpha, \tau} = \mathbf{U}_{\mathbf{w}^{c}(\alpha,\tau), \tau}$ where $\mathbf{U}_{\mathbf{w},\tau}$ is as defined in Section 2.1 and let $\checkbtheta_{\alpha,\tau} = \hatbtheta\big(\mathbf{w}^{c}(\alpha,\tau), \tau\big)$.
Now, note that $\hat{M}_{K}^{c}(\alpha, \tau)$ may be written as
\begin{equation}
\hat{M}_{K}^{c}(\alpha, \tau) 
=\mathbf{Y}^{T}\mathbf{U}_{\alpha,\tau}^{T}\mathbf{U}_{\alpha,\tau}\mathbf{Y} + 2\textrm{tr}\{ (\mathbf{U}_{\alpha,\tau} + \mathbf{I})\mathbf{V}_{Y|\theta} \}
- \textrm{tr}\{ \mathbf{V}_{Y|\theta} \}.
\label{eq:Mexpression}
\end{equation}
Because $\checkbtheta_{\alpha,\tau} = (\mathbf{U}_{\alpha, \tau} + \mathbf{I}) \mathbf{Y}$, we also have
\begin{eqnarray}
K \mathcal{L}_{K}(\btheta, \checkbtheta_{\alpha, \tau}) &=& (\checkbtheta_{\alpha,\tau} - \btheta)^{T}(\checkbtheta_{\alpha,\tau} - \btheta) \nonumber \\
&=& \mathbf{Y}^{T}(\mathbf{U}_{\alpha,\tau}^{T} + \mathbf{I})(\mathbf{U}_{\alpha,\tau} + \mathbf{I})\mathbf{Y} - 2\btheta^{T}(\mathbf{U}_{\alpha,\tau} + \mathbf{I})\mathbf{Y} + \btheta^{T}\btheta \nonumber \\
&=& \mathbf{Y}^{T}\mathbf{U}_{\alpha,\tau}^{T}\mathbf{U}_{\alpha,\tau}\mathbf{Y} - \mathbf{Y}^{T}\mathbf{Y} + 2(\mathbf{Y} - \btheta)^{T}(\mathbf{U}_{\alpha,\tau} + \mathbf{I})\mathbf{Y}  + \btheta^{T}\btheta. \label{eq:Lexpression}
\end{eqnarray}
Combining (\ref{eq:Mexpression}) and (\ref{eq:Lexpression}) gives
\begin{eqnarray}
& & \hat{M}_{K}^{c}(\alpha, \tau)/K - \mathcal{L}_{K}(\checkbtheta_{\alpha, \tau}, \btheta) \nonumber \\
&=& \frac{1}{K}\Bigg( 2\textrm{tr}\{ (\mathbf{U}_{\alpha,\tau} + \mathbf{I})\mathbf{V}_{Y|\theta} \}
- \textrm{tr}\{ \mathbf{V}_{Y|\theta} \} + \mathbf{Y}^{T}\mathbf{Y} - 2(\mathbf{Y} - \btheta)^{T}(\mathbf{U}_{\alpha,\tau} + \mathbf{I})\mathbf{Y} - \btheta^{T}\btheta \Bigg) \nonumber \\
&=& \frac{1}{K}\Big( \mathbf{Y}^{T}\mathbf{Y} - \textrm{tr}\{ \mathbf{V}_{Y|\theta} \} - \btheta^{T}\btheta \Big) 
- \frac{2}{K}\Big( (\mathbf{Y} - \btheta)^{T}(\mathbf{I} - \mathbf{B}_{\tau})\mathbf{Y} - \textrm{tr}\{ (\mathbf{I} - \mathbf{B}_{\tau})\mathbf{V}_{Y|\theta} \}  \Big) \nonumber \\
& &  - \frac{2}{K}\Big( (\mathbf{Y} - \btheta)^{T}(\mathbf{U}_{\alpha,\tau} + \mathbf{B}_{\tau})\mathbf{Y}   \Big)  + \frac{2}{K}\textrm{tr}\{ (\mathbf{U}_{\alpha,\tau} + \mathbf{B}_{\tau})\mathbf{V}_{Y|\theta} \} \nonumber \\
&=& S_{1K} - 2S_{2K}(\tau) - 2S_{3K}(\alpha,\tau) + 2S_{4K}(\alpha,\tau). \label{eq:Sdecomposition}
\end{eqnarray}
First consider $S_{1K}$. Note that $E(S_{1K}) = 0$ and that
\begin{equation}
E( S_{1K}^{2} ) = \frac{1}{K^{2}}\Big( \sum_{k=1}^{K} 4\mu_{k}^{2}\sigma_{k}^{2} + 4\tau_{0}^{2}\sum_{k=1}^{K} \sigma_{k}^{2} + 4\sum_{k=1}^{K}\mu_{k}E(e_{k}^{3}) + \sum_{k=1}^{K} E(e_{k}^{4}) - \sum_{k=1}^{K} \sigma_{k}^{4} \Big). \nonumber 
\end{equation}
Condition (A3) then guarantees that $\lim_{K \longrightarrow \infty} E( S_{1K}^{2}) = 0$. Hence, $\lim_{K \longrightarrow \infty}E(|S_{1K}|) = 0$.  
\medskip

\noindent
Next, we turn to $S_{2K}( \tau)$. Assuming condition (A3) holds, it follows from Lemma 2 that $ \sup_{\tau \geq 0}|S_{2K}(\tau)| \longrightarrow 0$ in $L^{1}$. 

\medskip

\noindent
Now, consider $S_{3K}(\alpha, \tau)$. 
Note that $\mathbf{U}_{\alpha,\tau} + \mathbf{B}_{\tau} = \mathbf{B}_{\tau}\mathbf{P}_{\alpha, \tau}$, where $\mathbf{P}_{\alpha,\tau}$ is the $K \times K$ matrix
$\mathbf{P}_{\alpha, \tau} = \mathbf{X}(\mathbf{X}^{T}\mathbf{W}_{\alpha,\tau}^{c}\mathbf{X})^{-1}\mathbf{X}^{T}\mathbf{W}_{\alpha,\tau}^{c}$. Hence, if we let $\hat{\mathbf{Y}}_{\alpha, \tau}$ denote the vector $\hat{\mathbf{Y}}_{\alpha,\tau} = \mathbf{P}_{\alpha, \tau}\mathbf{Y} = (\hat{Y}_{1}(\alpha, \tau), \ldots, \hat{Y}_{K}(\alpha,\tau))^{T}$, we have
$S_{3K}(\alpha, \tau) = \frac{1}{K}\sum_{k=1}^{K}B_{k,\tau}\Big( (Y_{k} - \theta_{k})\hat{Y}_{k}(\alpha, \tau)  \Big)$.
We can assume without loss of generality here that $\sigma_{1}^{2} \geq \ldots \geq \sigma_{K}^{2}$  so that $1 \geq B_{1,\tau} \geq \ldots B_{K,\tau} \geq 0$. When this is the case, we have
\begin{eqnarray}
\sup_{\alpha \in [0,1], \tau \geq 0}|S_{3K}(\alpha, \tau)|
 &\leq& \sup_{\alpha \in [0,1], \tau \geq 0}\sup_{1 \geq c_{1} \geq \cdots \geq c_{K} \geq 0}\Bigg| \frac{1}{K}\sum_{k=1}^{K} c_{k} (Y_{k} - \theta_{k})\hat{Y}_{k}(\alpha, \tau)\Bigg| \nonumber \\
&=& \sup_{\alpha \in [0,1], \tau \geq 0}\max_{1 \leq j \leq K} \Bigg| \frac{1}{K}\sum_{k=j}^{K} (Y_{k} - \theta_{k})\hat{Y}_{k}(\alpha, \tau)\Bigg| \nonumber \\
&=& \frac{1}{K}\sup_{\alpha \in [0,1], \tau \geq 0}\max_{1 \leq j \leq K} \Bigg| \mathbf{Y}^{T}\mathbf{P}_{\alpha,\tau}^{T}[\mathbf{Y} - \btheta]_{j:K} \Bigg|,
\label{eq:S3_inequality} 
\end{eqnarray}
where the first equality above follows from Lemma 2.1 in \cite{Li:1986} and 
where $[\mathbf{Y} - \btheta]_{j:K}$ is as defined in Lemma 4.
Now, building on (\ref{eq:S3_inequality}) and using the fact that $\mathbf{P}_{\alpha, \tau}^{T} = \mathbf{P}_{\alpha,\tau}^{T}\mathbf{P}_{X}^{T}$, we have that
\begin{eqnarray}
\sup_{\alpha \in [0,1], \tau \geq 0}|S_{3K}(\alpha, \tau)|
&\leq& \frac{1}{K}\sup_{\alpha \in [0,1], \tau \geq 0}\max_{1 \leq j \leq K} \Bigg| \mathbf{Y}^{T}\mathbf{P}_{\alpha,\tau}^{T}\mathbf{P}_{X}^{T}[\mathbf{Y} - \btheta]_{j:K} \Bigg| \nonumber \\
&\leq& \frac{1}{K}\Big( \sup_{\alpha \in [0,1], \tau \geq 0} || \mathbf{P}_{\alpha,\tau} \mathbf{Y} ||_{2} \Big)\max_{1 \leq j \leq K} \Big( || \mathbf{P}_{X}[\mathbf{Y} - \btheta]_{j:K} ||_{2} \Big) \nonumber \\
&\leq& \frac{1}{K}|| \mathbf{Y} ||_{2} \Big( \frac{4\sigma_{max,K}^{4}}{\sigma_{min,K}^{4}} + 1 \Big)^{1/2} \max_{1 \leq j \leq K} \Big( || \mathbf{P}_{X}[\mathbf{Y} - \btheta]_{j:K} ||_{2} \Big) \nonumber
\end{eqnarray}
where the second inequality follows Cauchy-Schwarz and the third inequality follows from Lemma 3. Another application of Cauchy-Schwarz yields
\begin{eqnarray}
& & E\Bigg( \sup_{\alpha \in [0,1], \tau \geq 0}|S_{3K}(\alpha, \tau)| \Bigg) \nonumber \\
&\leq& \frac{1}{K}\sqrt{E(|| \mathbf{Y} ||_{2})} \Big( \frac{4\sigma_{max,K}^{4}}{\sigma_{min,K}^{4}} + 1 \Big)^{1/2} \sqrt{E\Big\{ \max_{1 \leq j \leq K} \Big( || \mathbf{P}_{X}[\mathbf{Y} - \btheta]_{j:K} ||_{2}^{2} \Big) \Big\}} \nonumber \\
&=& \sqrt{\frac{1}{K^{2-\delta}}\sum_{k=1}^{K}(\mu_{k}^{2} + \tau_{0}^{2} + \sigma_{k}^{2}) \Big( \frac{4\sigma_{max,K}^{4}}{\sigma_{min,K}^{4}} + 1 \Big)} \sqrt{\frac{1}{K^{\delta}}E\Big\{ \max_{1 \leq j \leq K} \Big( || \mathbf{P}_{X}[\mathbf{Y} - \btheta]_{j:K} ||_{2}^{2} \Big) \Big\}} \nonumber 
\end{eqnarray}
Thus, it now follows from the above inequality,  conditions (A1) and (A2), and Lemma 4 (which also assumes (A4)) that $E\Big( \sup_{\alpha\in [0,1], \tau \geq 0} | S_{3K}(\alpha,\tau) | \Big) \longrightarrow 0$. 

\bigskip

\noindent
Finally, we turn to $S_{4K}(\alpha, \tau)$. If we let $\tilde{P}_{\alpha,\tau}^{(k,k)}$ denote the $(k,k)$ element of the projection matrix
$\tilde{\mathbf{P}}_{\alpha, \tau} = (\mathbf{W}_{\alpha,\tau}^{c})^{1/2}\mathbf{P}_{\alpha,\tau}(\mathbf{W}_{\alpha,\tau}^{c})^{-1/2}$ and recall that $\mathbf{U}_{\alpha,\tau} + \mathbf{B}_{\tau} = \mathbf{B}_{\tau}\mathbf{P}_{\alpha,\tau}$, then
\begin{eqnarray}
| S_{4K}(\alpha, \tau) | &=& \frac{1}{K}\Big| \textrm{tr}\Big\{ \mathbf{B}_{\tau}\mathbf{P}_{\alpha, \tau} \mathbf{V}_{Y|\theta} \Big\} \Big|
= \frac{1}{K}\Big| \textrm{tr}\{ \mathbf{B}_{\tau}(\mathbf{W}_{\alpha,\tau}^{c})^{-1/2}\tilde{\mathbf{P}}_{\alpha, \tau}(\mathbf{W}_{\alpha,\tau}^{c})^{1/2} \mathbf{V}_{Y|\theta} \} \Big| \nonumber \\
&=& \frac{1}{K}\Big| \textrm{tr}\Big\{ \tilde{\mathbf{P}}_{\alpha, \tau}(\mathbf{W}_{\alpha,\tau}^{c})^{1/2} \mathbf{V}_{Y|\theta}\mathbf{B}_{\tau}(\mathbf{W}_{\alpha,\tau}^{c})^{-1/2} \Big\} \Big|
= \frac{1}{K}\Big| \sum_{k=1}^{K} \tilde{P}_{\alpha,\tau}^{(k,k)}B_{k,\tau}\sigma_{k}^{2}  \Big|, \nonumber
\end{eqnarray}
where the last equality follows from the fact that $\mathbf{W}_{\alpha,\tau}^{c}$, $\mathbf{V}_{Y|\theta}$, and $\mathbf{B}_{\tau}$ are all diagonal matrices. Now, by Cauchy-Schwarz, we have that
\begin{eqnarray}
| S_{4K}(\alpha, \tau) | &\leq& \Big( \frac{1}{K^{2}} \sum_{k=1}^{K} B_{k,\tau}^{2}\sigma_{k}^{4} \Big)^{1/2}\Big( \sum_{k=1}^{K} ( \tilde{P}_{\alpha, \tau}^{(k,k)})^{2} ) \Big)^{1/2}
\leq \Big( \frac{1}{K^{2}}\sum_{k=1}^{K} \sigma_{k}^{4} \Big)^{1/2}\Big( \sum_{k=1}^{K} \tilde{P}_{\alpha, \tau}^{(k,k)}  \Big)^{1/2}  \nonumber \\
&=& \Big( \frac{1}{K^{2}}\sum_{k=1}^{K} \sigma_{k}^{4} \Big)^{1/2}\Big( \textrm{tr}\big( \tilde{\mathbf{P}}_{\alpha, \tau} \big) \Big)^{1/2}
= \Big( \frac{p}{K^{2}}\sum_{k=1}^{K} \sigma_{k}^{4} \Big)^{1/2}, \label{eq:S4_inequality}
\end{eqnarray}
where the second inequality comes from the fact that both $0 \leq B_{k,\tau} \leq 1$ and $0 \leq \tilde{P}_{\alpha, \tau}^{(k,k)} \leq 1$ ($0 \leq \tilde{P}_{\alpha,\tau}^{(k,k)} \leq 1$ follows from the fact that $\tilde{\mathbf{P}}_{\alpha,\tau}$ is both symmetric and idempotent).
Now, it follows from (\ref{eq:S4_inequality}) and condition (A3) that $\lim_{K \longrightarrow \infty} E\big( \sup_{\alpha \in [0,1], \tau \geq 0} |S_{4K}(\alpha, \tau)|  \big) = 0$.

\medskip

\noindent
So, we have now established that $| S_{1K} |$, $\sup_{\tau \geq 0} |S_{2K}(\tau)|$, $\sup_{\alpha \in [0,1], \tau \geq 0} | S_{3K}(\alpha, \tau) |$, and $\sup_{\alpha \in [0,1], \tau \geq 0} | S_{4K}(\alpha, \tau) |$ all converge to zero in $L^{1}$. Thus, from (\ref{eq:Sdecomposition}), we may conclude that $\sup_{\alpha \in [0,1], \tau \geq 0}\Big|\hat{M}_{K}^{c}(\alpha, \tau)/K - \mathcal{L}_{K}(\btheta, \checkbtheta_{\alpha, \tau})\Big|$
goes to zero in $L^{1}$.

\subsection{Proof of Theorem 4}
Let $(\alpha^{*}, \tau^{*})$ be as defined in Section 2.2. Namely, 
\begin{equation}
(\alpha^{*}, \tau^{*}) = \argmin_{\alpha \in [0,1], \tau \geq 0}\hat{M}_{K}^{c}(\alpha, \tau) \nonumber
\end{equation}
Now, note that 
\begin{eqnarray}
& & \mathcal{L}_{K}\{ \btheta, \checkbtheta^{CBP} \} - \mathcal{L}_{K}\{ \btheta, \hatbtheta^{OR} \}  \nonumber \\
&=& \big[\mathcal{L}_{K}\{ \btheta, \hatbtheta(\mathbf{w}^{c}(\alpha^{*}, \tau^{*}), \tau^{*} ) \} - \hat{M}_{K}^{c}(\alpha^{*}, \tau^{*})/K \big] + \big[ \hat{M}_{K}^{c}(\alpha^{*}, \tau^{*})/K - \hat{M}_{K}^{c}(\alpha^{OR}, \tau^{OR})/K \big] \nonumber \\
& & + \big[ \hat{M}_{K}^{c}(\alpha^{OR}, \tau^{OR})/K - \mathcal{L}_{K}\{ \btheta, \hatbtheta(\mathbf{w}^{c}(\alpha^{OR}, \tau^{OR}), \tau^{OR} ) \} \big] \nonumber \\
&\leq& \big[\mathcal{L}_{K}\{ \btheta, \hatbtheta(\mathbf{w}^{c}(\alpha^{*}, \tau^{*}), \tau^{*} ) \} - \frac{1}{K}\hat{M}_{K}^{c}(\alpha^{*}, \tau^{*}) \big] 
 + \big[ \frac{1}{K}\hat{M}_{K}^{c}(\alpha^{OR}, \tau^{OR}) - \mathcal{L}_{K}\{ \btheta, \hatbtheta(\mathbf{w}^{c}(\alpha^{OR}, \tau^{OR}), \tau^{OR} ) \} \big] \nonumber \\
&\leq& 2 \Bigg( \sup_{\alpha \in [0,1], \tau \geq 0}\Big| \mathcal{L}_{K}\big\{ \btheta, \hatbtheta\big( \mathbf{w}^{c}(\alpha, \tau), \tau \big)  \big\} - \hat{M}_{K}^{c}(\alpha, \tau)/K \Big| \Bigg), \label{eq:loss_inequality}
\end{eqnarray}
where the first inequality follows from the fact that $(\alpha^{*},\tau^{*})$ minimizes $\hat{M}_{K}^{c}(\alpha, \tau)$. It then follows from (\ref{eq:loss_inequality}) that
\begin{equation}
0 \leq \mathcal{R}_{K}(\checkbtheta^{CBP} ) - \mathcal{R}_{K}( \hatbtheta^{OR} )
\leq 2 E \Bigg( \sup_{\alpha \in [0,1], \tau \geq 0}\Big| \mathcal{L}_{K}\big\{ \btheta, \hatbtheta\big( \mathbf{w}^{c}(\alpha, \tau), \tau \big)  \big\} - \hat{M}_{K}^{c}(\alpha, \tau)/K \Big| \Bigg), \nonumber
\end{equation}
and thus the desired result follows from Theorem 3.

\section{The Nested-Error Regression Model }

In the \textit{nested-error regression model} (NER model), one assumes that 
in the $k^{th}$ ``cluster" or ``population" we sample $n_{k}$ values $y_{kj}$ while the 
remaining $N_{k} - n_{k}$ values are not sampled. 

In our formulation of the nested-error regression model, we consider the following superpopulation model for the $k^{th}$ population
\begin{equation}
Y_{kj}^{P} = \mu_{kj}^{P} + v_{k} + e_{kj}^{P}, \qquad k = 1,\ldots, N_{k}, \label{eq:superpop_model}
\end{equation}
and we assume there are $K$ such populations, i.e., $k = 1,\ldots,K$. 
In model (\ref{eq:superpop_model}), it is assumed that $v_{1}, \ldots, v_{K}$ are independent with $E(v_{k}) = 0$ 
and $E( v_{k}^{2} ) = \tau_{0}^{2}$, $e_{kj}^{P}$ are all independent with $E(e_{kj}^{P}) = 0$ and $E\{ (e_{kj}^{P})^{2} \} = \sigma_{kj}^{2}$, 
and the $v_{k}^{P}$ and $e_{jk}^{P}$ are independent.

If we let $\mathcal{S}_{k}$ denote the set of indices of the $n_{k}$ sampled cases in population $k$, we can express the sampled values
$y_{kj}$ in the $k^{th}$ population as:
\begin{equation}
y_{kj} = \mu_{ij} + v_{i} + e_{ij}, \qquad \textrm{ for } j \in \mathcal{S}_{k} \nonumber 
\end{equation}
where $y_{kj} = Y_{kj}^{P}$ if the $j^{th}$ case in population $k$ is sampled, i.e., $j \in \mathcal{S}_{k}$. 
We will consider a version of the NER model where the design is such that the number of elements in $\mathcal{S}_{k}$ may not be exactly
equal to $n_{k}$, but rather, $\mathcal{S}_{k}$ is constructed from $N_{k}$ independent Bernoulli trials
with success probability $n_{k}/N_{k}$. Namely,
\begin{equation}
\mathcal{S}_{k} = \{ j: S_{kj} = 1\}, \quad \textrm{ where } S_{kj} \sim \textrm{Bernoulli}(n_{k}/N_{k}). 
\label{eq:sampling_design}
\end{equation}
This sampling design (\ref{eq:sampling_design}) can be thought of as a close version of simple random sampling where a subset of exactly $n_{k}$
units would be drawn from the $N_{k}$ units in population $k$.

For the $k^{th}$ population, we are primarily interested in the terms $\theta_{1}, \ldots, \theta_{K}$ defined as
\begin{equation}
\theta_{k} = \frac{1}{N_{k}}\sum_{k=1}^{N_{k}} \mu_{kj}^{P} + v_{k}
= \mathbf{a}_{k}^{T}\bmu + v_{k}, \nonumber 
\end{equation}
where $\bmu^{P} = (\mu_{11}^{P}, \ldots, \mu_{KN_{k}}^{P})$ and $\mathbf{a}_{k}$ is the vector whose
$h^{th}$ component $a_{kh}$ is given by $a_{kh} = 1/N_{k}$ if $\sum_{j=1}^{k-1} N_{j} < h \leq \sum_{j=1}^{k} N_{j}$ and is equal to $0$ otherwise.  
Thus, if we let $\mathbf{A}$ denote the $K \times N$ matrix (where $N = \sum_{k=1}^{K} N_{k}$) whose $k^{th}$ row
is $\mathbf{a}_{k}^{T}$, we can express the vector $\btheta = (\theta_{1}, \ldots, \theta_{K})$ as
\begin{equation}
\btheta = \mathbf{A}\bmu^{P} + \mathbf{v}, \label{eq:theta_ner_def}
\end{equation}
where $\mathbf{v} = (v_{1}, \ldots, v_{K})$. 

Note that, for the $N \times 1$ vector of superpopulation
responses $\mathbf{Y}^{P} = (Y_{11}^{P}, Y_{12}^{P}, ...., Y_{21}^{P}, ...., Y_{KN_{k}}^{P})$, we can express the superpopulation model 
in vector form as
\begin{equation}
\mathbf{Y}^{P} = \bmu^{P} + \mathbf{D}\mathbf{A}^{T}\mathbf{v} + \mathbf{e}^{P}, \nonumber 
\end{equation}
where $\mathbf{D}$ is the $N \times N$ diagonal matrix
$\mathbf{D} = \textrm{diag}\{ N_{1}, N_{1}, ..., N_{2}, ...., N_{K} \}$ and where $\mathbf{e}^{P} = (e_{11}^{P}, e_{12}^{P}, ...., e_{21}^{P}, ...., e_{KN_{k}}^{P})$. 
Note also that $\mathbf{A}\mathbf{D}\mathbf{A}^{T} = \mathbf{I}$, where $\mathbf{I}$ here denotes the 
$K \times K$ identity matrix. The variance-covariance matrix $\mathbf{V}_{Y}$ of $\mathbf{Y}^{P}$ is
then given by $\mathbf{V}_{Y} = \mathbf{D}\mathbf{A}^{T}\mathbf{V}_{v}\mathbf{A}\mathbf{D}^{T} + \mathbf{V}_{e}$,
where $\mathbf{V}_{v}$ is the $K \times K$ matrix $\mathbf{V}_{v} = \textrm{Var}(\mathbf{v})$ and $\mathbf{V}_{e}$ is the $N \times N$ matrix 
$\mathbf{V}_{e} = \textrm{Var}( \mathbf{e}^{P} )$.

The ``direct" estimate  of $\theta_{k}$ is the sample mean from population $k$, and we denote this direct estimate
by $Y_{k}$, i.e.
\begin{equation}
    Y_{k} = \frac{1}{n_{k}} \sum_{j=1}^{N_{k}} Y_{kj}^{P}S_{kj}, \nonumber  
\end{equation}
where the $S_{kj}$ are as defined in (\ref{eq:sampling_design}).

Let $\mathbf{Y}$ be the $K \times 1$ vector $\mathbf{Y} = (Y_{1}, \ldots, Y_{K})$.
We can express $\mathbf{Y}$ in terms of the superpopulation vector $\mathbf{Y}^{P}$ as 
$\mathbf{Y} = \mathbf{S}\mathbf{Y}^{P}$ where $\mathbf{S}$ is $K \times N$ matrix
whose $k^{th}$ row is $\mathbf{s}_{k}$ and 
where the components $\mathbf{s}_{k} = (s_{k1}, s_{k2}, \ldots, s_{kN})$ of $\mathbf{s}_{k}$ are given by
\begin{equation}
s_{kh} = 
\begin{cases}
1/n_{k} & \text{ if } h - \sum_{l=1}^{k-1} N_{l} \in \mathcal{S}_{k} \nonumber \\
0 & \text{ otherwise } \nonumber 
\end{cases}
\end{equation}
For our estimator of the MSPE associated with a predictor in an NER model, we will assume the independent Bernoulli sampling design
described in (\ref{eq:sampling_design}) so that all of the $s_{kh}$ are independent with $P(s_{kh} > 0) = n_{k}/N_{k}$. Note that this implies that $E( \mathbf{S}) = \mathbf{A}$.

We will consider an estimator/predictor $\hatbtheta(\alpha, \tau)$ of the vector $\btheta = (\theta_{1}, \ldots, \theta_{K})$ that can be expressed as 
\begin{equation}
\hatbtheta(\alpha, \tau) = \mathbf{U}_{\alpha, \tau}\mathbf{Y} + \mathbf{Y}, 
\label{eq:predictor_form}
\end{equation}
where the $K \times K$ matrix $\mathbf{U}_{\alpha, \tau}$ is a matrix that can be expressed solely as a function of two hyperparameters $\tau$ and $\alpha$.

To develop an unbiased estimator of the mean-squared prediction error (MSPE) associated
with a predictor of the form (\ref{eq:predictor_form}), we also need to work with the 
within-population average second moment parameters $\kappa_{2k}$ which are defined as
\begin{equation}
\kappa_{2k} = \frac{1}{N_{k}}\sum_{j=1}^{N_{k}} E\{ (Y_{kj}^{P})^{2} \} = \textrm{Var}(v_{k}) + \frac{1}{N_{k}}\sum_{j=1}^{N_{k}} \{ (\mu_{kj}^{P})^{2} + \textrm{Var}(e_{kj}^{P}) \}. \label{eq:kappa2}
\end{equation}
An unbiased estimate of $\kappa_{2k}$ is
\begin{equation}
\hat{\kappa}_{2k} = \frac{1}{n_{k}}\sum_{j \in \mathcal{S}_{k} }  (Y_{kj}^{P})^{2} = \sum_{h=1}^{N}  (Y_{kj}^{P})^{2}s_{kh}. \nonumber  
\end{equation}

\medskip

The following theorem defines an estimator $\hat{M}_{K}^{NER}(\alpha, \tau)$ and establishes that it is 
an unbiased estimator of the MSPE associated with a predictor of the form (\ref{eq:predictor_form}).
\begin{theorem}
In the nested-error regression model with the sampling conducted as described in (\ref{eq:sampling_design}), 
the following estimator
\begin{equation}
\hat{M}_{K}^{NER}(\alpha, \tau) = \mathbf{Y}^{T}\mathbf{U}_{\alpha, \tau}^{T}\mathbf{U}_{\alpha, \tau}\mathbf{Y} - 2\textrm{tr}\Big( (\mathbf{U}_{\alpha,\tau} + \tfrac{1}{2}\mathbf{I})\mathbf{A}\mathbf{V}_{e}\mathbf{A}^{T} \Big) 
- 2\sum_{k=1}^{K} \frac{\hat{\kappa}_{2k}(N_{k} - n_{k})( U_{\alpha,\tau, kk} + 1/2)}{N_{k}n_{k}},  \nonumber
\end{equation}
where $U_{\alpha,\tau, kk}$ is the $k^{th}$ diagonal element of $\mathbf{U}_{\alpha, \tau}$,
is an unbiased estimator of the MSPE associated with $\hatbtheta(\alpha, \tau) = \mathbf{U}_{\alpha, \tau}\mathbf{Y} + \mathbf{Y}$. That is,
\begin{equation}
E\{ \hat{M}_{K}^{NER}(\alpha, \tau) \} = E\Big( \{ \hatbtheta(\alpha, \tau) - \btheta\}^{T}\{ \hatbtheta(\alpha, \tau) - \btheta\} \Big). \nonumber
\end{equation}
\label{thm:unbiased_estimate_ner}
\end{theorem}

\vspace{-1cm}

\textbf{Proof of Theorem \ref{thm:unbiased_estimate_ner}.} First, note that $\hatbtheta(\alpha, \tau) = \mathbf{U}_{\alpha, \tau}\mathbf{S}\mathbf{Y}^{P} + \mathbf{S}\mathbf{Y}^{P}$ so that we can write
\begin{eqnarray}
\hatbtheta(\alpha, \tau) - \btheta &=& (\mathbf{U}_{\alpha, \tau} + \mathbf{I})\mathbf{S}(\bmu^{P} + \mathbf{D}\mathbf{A}^{T}\mathbf{v} + \mathbf{e}^{P}) - \mathbf{A}\bmu^{P} - \mathbf{v}.  \nonumber 
\end{eqnarray}
This then implies that
\begin{eqnarray}
\{ \hatbtheta(\alpha, \tau) - \btheta\}^{T}\{ \hatbtheta(\alpha, \tau) - \btheta\}
&=& (\bmu^{P} + \mathbf{D}\mathbf{A}^{T}\mathbf{v} + \mathbf{e}^{P})^{T}\mathbf{S}^{T}(\mathbf{U}_{\alpha, \tau}^{T} + \mathbf{I})(\mathbf{U}_{\alpha, \tau} + \mathbf{I})\mathbf{S}(\bmu^{P} + \mathbf{D}\mathbf{A}^{T}\mathbf{v} + \mathbf{e}^{P}) \nonumber \\
&-& 2(\bmu^{P} + \mathbf{D}\mathbf{A}^{T}\mathbf{v} + \mathbf{e}^{P})^{T}\mathbf{S}^{T}(\mathbf{U}_{\alpha, \tau}^{T} + \mathbf{I})(\mathbf{A}\bmu^{P} + \mathbf{v}) \nonumber \\
&+& ((\bmu^{P})^{T}\mathbf{A}^{T} + \mathbf{v}^{T})(\mathbf{A}\bmu^{P} + \mathbf{v}). \nonumber 
\end{eqnarray}
Hence,
\begin{eqnarray}
&& E\Big( \{ \hatbtheta(\alpha, \tau) - \btheta\}^{T}\{ \hatbtheta(\alpha, \tau) - \btheta\} \Big) \nonumber \\
&=& E\Big\{ (\bmu^{P} + \mathbf{D}\mathbf{A}^{T}\mathbf{v} + \mathbf{e}^{P})^{T}\mathbf{S}^{T}(\mathbf{U}_{\alpha, \tau}^{T} + \mathbf{I})(\mathbf{U}_{\alpha, \tau} + \mathbf{I})\mathbf{S}(\bmu^{P} + \mathbf{D}\mathbf{A}^{T}\mathbf{v} + \mathbf{e}^{P}) \Big\} \nonumber \\
&-& 2E\Big( (\bmu^{P})^{T}\mathbf{S}^{T}(\mathbf{U}_{\alpha, \tau}^{T} + \mathbf{I})\mathbf{A}\bmu^{P} \Big) 
- 2E\Big( \mathbf{v}^{T}\mathbf{A}\mathbf{D}\mathbf{S}^{T}(\mathbf{U}_{\alpha, \tau}^{T} + \mathbf{I})\mathbf{v} \Big) 
+ E\Big( (\bmu^{P})^{T}\mathbf{A}^{T}\mathbf{A}\bmu^{P} \Big) + E\Big( \mathbf{v}^{T}\mathbf{v} \Big) \nonumber \\
&=& E\Big\{ \mathbf{Y}^{T}\mathbf{U}_{\alpha, \tau}^{T}\mathbf{U}_{\alpha, \tau}\mathbf{Y} \Big\} 
+ E\Big\{ \mathbf{Y}^{T}\mathbf{U}_{\alpha, \tau}\mathbf{Y} \Big\} 
+ E\Big\{ \mathbf{Y}^{T}\mathbf{U}_{\alpha, \tau}^{T}\mathbf{Y} \Big\} 
+ E\Big\{ \mathbf{Y}^{T}\mathbf{Y} \Big\} \nonumber \\
&-& 2E\Big( (\bmu^{P})^{T}\mathbf{A}^{T}\mathbf{U}_{\alpha, \tau}^{T}\mathbf{A}\bmu^{P} \Big) 
- 2E\Big( \mathbf{v}^{T}\mathbf{A}\mathbf{D}\mathbf{A}^{T}\mathbf{U}_{\alpha, \tau}^{T}\mathbf{v} \Big)  - E\Big( (\bmu^{P})^{T}\mathbf{A}^{T}\mathbf{A}\bmu^{P} \Big) - E\Big( \mathbf{v}^{T}\mathbf{v} \Big) \nonumber \\
&=& E\Big\{ \mathbf{Y}^{T}\mathbf{U}_{\alpha, \tau}^{T}\mathbf{U}_{\alpha, \tau}\mathbf{Y} \Big\} 
+ 2E\Big\{ \mathbf{Y}^{T}\mathbf{U}_{\alpha, \tau}\mathbf{Y} \Big\} 
+ E\Big\{ \mathbf{Y}^{T}\mathbf{Y} \Big\} \nonumber \\
&-& 2 (\bmu^{P})^{T}\mathbf{A}^{T}\mathbf{U}_{\alpha, \tau}^{T}\mathbf{A}\bmu^{P}
- 2\textrm{tr}\Big( \mathbf{A}\mathbf{D}\mathbf{A}^{T}\mathbf{U}_{\alpha, \tau}^{T}\mathbf{V}_{v} \Big)  - (\bmu^{P})^{T}\mathbf{A}^{T}\mathbf{A}\bmu^{P}  - \textrm{tr}\Big( \mathbf{V}_{v} \Big), \label{eq:mspe_simplify}
\end{eqnarray}
where $\mathbf{V}_{v} = \textrm{Var}( \mathbf{v} )$. From Lemma \ref{lem:expectation_quad_form}, we know that, 
\begin{eqnarray}
E\Big\{ \mathbf{Y}^{T}\mathbf{U}_{\alpha, \tau}\mathbf{Y} \Big\} 
&=& (\bmu^{P})^{T}\mathbf{A}^{T}\mathbf{U}_{\alpha, \tau}\mathbf{A}\bmu^{P} + \sum_{k=1}^{K} \frac{(N_{k} - n_{k})}{N_{k}n_{k}} U_{\alpha,\tau, kk}\kappa_{2k} + \textrm{tr}\Big( \mathbf{U}_{\alpha, \tau}(\mathbf{V}_{v} + \mathbf{A}\mathbf{V}_{e}\mathbf{A}^{T}) \Big), \nonumber \\
E\Big\{ \mathbf{Y}^{T}\mathbf{Y} \Big\} 
&=& (\bmu^{P})^{T}\mathbf{A}^{T}\mathbf{A}\bmu^{P} + \sum_{k=1}^{K} \frac{(N_{k} - n_{k})}{N_{k}n_{k}} \kappa_{2k} + \textrm{tr}\Big( \mathbf{V}_{v} + \mathbf{A}\mathbf{V}_{e}\mathbf{A}^{T} \Big). \label{eq:quadform_ner} 
\end{eqnarray}
Plugging (\ref{eq:quadform_ner}) into (\ref{eq:mspe_simplify}) and using the fact that $\mathbf{A}\mathbf{D}\mathbf{A}^{T} = \mathbf{I}$, we now have that
\begin{eqnarray}
&& E\Big( \{ \hatbtheta(\alpha, \tau) - \btheta\}^{T}\{ \hatbtheta(\alpha, \tau) - \btheta\} \Big) \nonumber \\
&=& E\Big\{ \mathbf{Y}^{T}\mathbf{U}_{\alpha, \tau}^{T}\mathbf{U}_{\alpha, \tau}\mathbf{Y} \Big\} 
+ 2(\bmu^{P})^{T}\mathbf{A}^{T}\mathbf{U}_{\alpha, \tau}\mathbf{A}\bmu^{P}
+ 2\sum_{k=1}^{K} \frac{(N_{k} - n_{k})}{N_{k}n_{k}} U_{\alpha,\tau, kk}\kappa_{2k} \nonumber \\ 
&+& 2\textrm{tr}\Big( \mathbf{U}_{\alpha,\tau}(\mathbf{V}_{v} + \mathbf{A}\mathbf{V}_{e}\mathbf{A}^{T}) \Big)  
+ (\bmu^{P})^{T}\mathbf{A}^{T}\mathbf{A}\bmu^{P} + \sum_{k=1}^{K} \frac{(N_{k} - n_{k})}{N_{k}n_{k}} \kappa_{2k} + \textrm{tr}\Big( \mathbf{V}_{v} + \mathbf{A}\mathbf{V}_{e}\mathbf{A}^{T} \Big) \nonumber \\
&-& 2 (\bmu^{P})^{T}\mathbf{A}^{T}\mathbf{U}_{\alpha, \tau}^{T}\mathbf{A}\bmu^{P} 
- 2\textrm{tr}\Big( \mathbf{U}_{\alpha, \tau}^{T}\mathbf{V}_{v} \Big)  - (\bmu^{P})^{T}\mathbf{A}^{T}\mathbf{A}\bmu^{P}  - \textrm{tr}\Big( \mathbf{V}_{v} \Big) \nonumber \\
&=& E\Big\{ \mathbf{Y}^{T}\mathbf{U}_{\alpha, \tau}^{T}\mathbf{U}_{\alpha, \tau}\mathbf{Y} \Big\} 
+ 2\sum_{k=1}^{K} \frac{(N_{k} - n_{k})}{N_{k}n_{k}} U_{\alpha,\tau, kk}\kappa_{2k} 
+ 2\textrm{tr}\Big( \mathbf{U}_{\alpha,\tau}\mathbf{A}\mathbf{V}_{e}\mathbf{A}^{T} \Big)  \nonumber \\
&+& \sum_{k=1}^{K} \frac{(N_{k} - n_{k})}{N_{k}n_{k}} \kappa_{2k} 
+ \textrm{tr}\Big( \mathbf{A}\mathbf{V}_{e}\mathbf{A}^{T} \Big) \nonumber \\
&=& E\Big\{ \mathbf{Y}^{T}\mathbf{U}_{\alpha, \tau}^{T}\mathbf{U}_{\alpha, \tau}\mathbf{Y} \Big\} 
+ 2\textrm{tr}\Big( (\mathbf{U}_{\alpha,\tau} + \tfrac{1}{2}\mathbf{I})\mathbf{A}\mathbf{V}_{e}\mathbf{A}^{T} \Big) + 2\sum_{k=1}^{K} \frac{\kappa_{2k}(N_{k} - n_{k})( U_{\alpha,\tau, kk} + 1/2)}{N_{k}n_{k}}. \nonumber 
\end{eqnarray}
The result then follows from the fact that $E( \hat{\kappa}_{2k} ) = \kappa_{2k}$.

\vspace{1.5cm}

\begin{lemma}
For any $N \times N$ matrix $\mathbf{F}$, we have that
\begin{equation}
E\Big\{ \mathbf{S}\mathbf{F}\mathbf{S}^{T} \Big\} = \mathbf{D}_{F} + \mathbf{A}\mathbf{F}\mathbf{A}^{T}, \nonumber 
\end{equation}
where $\mathbf{D}_{F}$ is the $K \times K$ matrix whose $i^{th}$ diagonal element is given by
\begin{equation}
[\mathbf{D}_{F}]_{i,i} = \sum_{k=1}^{N} a_{ik}^{2}F_{kk}\Big( \frac{N_{i}}{n_{i}} - 1 \Big), \nonumber 
\end{equation}
where $a_{ik}$ is the $(i,k)$ component of $\mathbf{A}$ and $F_{kk}$ is the $(k,k)$ component of $\mathbf{F}$.
\label{lem:S_expectation}
\end{lemma}

\textbf{Proof.} First, note that the $(i,j)$ component of $\mathbf{S}\mathbf{F}\mathbf{S}^{T}$ is given by
\begin{equation}
\Big[ \mathbf{S}\mathbf{F}\mathbf{S}^{T} \Big]_{i,j} = \sum_{k=1}^{N}\sum_{h=1}^{N} s_{ik}s_{jh}F_{kh}, \nonumber 
\end{equation}
where $F_{k,h}$ is the $(k,h)$ component of $\mathbf{F}$. Because the $s_{ij}$ are assumed to be mutually independent
we have, for $i \neq j$, that
\begin{equation}
\Big[ E(\mathbf{S}\mathbf{F}\mathbf{S}^{T}) \Big]_{i,j} = \sum_{k=1}^{N}\sum_{h=1}^{N} E(s_{ik})E(s_{jh})F_{kh}
= \sum_{k=1}^{N}\sum_{h=1}^{N} a_{ik} a_{jh} F_{kh}, \nonumber 
\end{equation}
where $a_{ik}$ is the $(i,k)$ component of $\mathbf{A}$. For $i = j$, we have
\begin{eqnarray}
\Big[ E(\mathbf{S}\mathbf{F}\mathbf{S}^{T}) \Big]_{i,i} &=& \sum_{k=1}^{N}\sum_{h=1}^{N} E(s_{ik} s_{ih})F_{kh}
= \sum_{k=h} E(s_{ik} s_{ih})F_{kh} + \sum_{k \neq h} E(s_{ik} s_{ih})F_{kh} \nonumber \\
&=& \sum_{k=1}^{N} E(s_{ik}^{2})F_{kk} + \sum_{k \neq h} E(s_{ik}) E(s_{ih})F_{kh} \nonumber \\
&=& \sum_{k=1}^{N} \frac{N_{i}}{n_{i}}a_{ik}^{2} F_{kk} + \sum_{k \neq h} a_{ik} a_{ih} F_{kh} \nonumber \\
&=& \sum_{k=1}^{N} a_{ik}^{2}F_{kk}\Big( \frac{N_{i}}{n_{i}} - 1 \Big) + \sum_{k=1}^{N}\sum_{h=1}^{N} a_{ik} a_{ih} F_{kh}. \nonumber
\end{eqnarray}

\bigskip

\begin{lemma}
For any $K \times K$ matrix $\mathbf{F}$, we have that
\begin{eqnarray}
E\Big\{ \mathbf{Y}^{T}\mathbf{F}\mathbf{Y} \Big\} 
&=& (\bmu^{P})^{T}\mathbf{A}^{T}\mathbf{F}\mathbf{A}\bmu^{P} + \sum_{k=1}^{K} \frac{(N_{k} - n_{k})}{N_{k}n_{k}} F_{kk}\kappa_{2k} + \textrm{tr}\Big( \mathbf{F}(\mathbf{V}_{v} + \mathbf{A}\mathbf{V}_{e}\mathbf{A}^{T}) \Big), \nonumber 
\end{eqnarray}
where $F_{kk}$ is the $k^{th}$ diagonal element of $\mathbf{F}$.
\label{lem:expectation_quad_form}
\end{lemma}

\textbf{Proof.} First, note that since $\mathbf{S}$ and $\mathbf{Y}^{P}$ are independent, 
$E(\mathbf{S}\mathbf{Y}^{P}) = E(\mathbf{S})E(\mathbf{Y}^{P}) = \mathbf{A}\bmu^{P}$. 
Second, note that
\begin{eqnarray}
\textrm{Var}\Big( E(\mathbf{S}\mathbf{Y}^{P}|\mathbf{S}) \Big)
&=& \textrm{Var}\Big( \mathbf{S}\bmu^{P} \Big) = E\Big\{ (\mathbf{S}\bmu^{P} - \mathbf{A}\bmu^{P})(\mathbf{S}\bmu^{P} - \mathbf{A}\bmu^{P})^{T} \Big\} \nonumber \\
&=& E\Big\{ \mathbf{S}\bmu^{P}(\bmu^{P})^{T}\mathbf{S}^{T} \Big\} -
E\Big\{ \mathbf{A}\bmu^{P}(\bmu^{P})^{T}\mathbf{S}^{T} \Big\} 
- E\Big\{ \mathbf{S}\bmu^{P}(\bmu^{P})^{T}\mathbf{A}^{T} \Big\}
+ \mathbf{A}\bmu^{P}(\bmu^{P})^{T}\mathbf{A}^{T}  \nonumber \\
&=& E\Big\{ \mathbf{S}\bmu^{P}(\bmu^{P})^{T}\mathbf{S}^{T} \Big\} - \mathbf{A}\bmu^{P}(\bmu^{P})^{T}\mathbf{A}^{T} \nonumber \\
&=& \mathbf{D}_{\mu^{P}(\mu^{P})^{T}} + \mathbf{A}\bmu^{P}(\bmu^{P})^{T}\mathbf{A}^{T} - \mathbf{A}\bmu^{P}(\bmu^{P})^{T}\mathbf{A}^{T} \nonumber \\
&=& \mathbf{D}_{\mu^{P}(\mu^{P})^{T}}, \label{eq:var_condexp_ner}
\end{eqnarray}
with the second-to-last equality following from Lemma \ref{lem:S_expectation}.
Third, note that
\begin{eqnarray}
E\Big( \textrm{Var}(\mathbf{S}\mathbf{Y}^{P}|\mathbf{S}) \Big) = E\Big\{ \mathbf{S}\mathbf{V}_{Y}\mathbf{S}^{T} \Big\}
= \mathbf{D}_{V_{Y}} + \mathbf{A}\mathbf{V}_{Y}\mathbf{A}^{T}, \label{eq:exp_condvar_ner}
\end{eqnarray}
with the second equality also following from Lemma \ref{lem:S_expectation} and where 
$\mathbf{V}_{Y} = \textrm{Var}(\mathbf{Y}^{P}) = \mathbf{D}\mathbf{A}^{T}\mathbf{V}_{v}\mathbf{A}\mathbf{D}^{T} + \mathbf{V}_{e}$.
So, we may conclude from  (\ref{eq:var_condexp_ner}) and (\ref{eq:exp_condvar_ner}) that
\begin{eqnarray}
\textrm{Var}( \mathbf{S}\mathbf{Y}^{P} ) &=& \mathbf{D}_{\mu^{P}(\mu^{P})^{T}} + \mathbf{D}_{V_{Y}} + \mathbf{A}\mathbf{V}_{Y}\mathbf{A}^{T} \nonumber \\
&=& \mathbf{D}_{\mu^{P}(\mu^{P})^{T}} + \mathbf{D}_{V_{Y}} + \mathbf{A}(\mathbf{D}\mathbf{A}^{T}\mathbf{V}_{v}\mathbf{A}\mathbf{D}^{T} + \mathbf{V}_{e})\mathbf{A}^{T} \nonumber \\
&=&  \mathbf{D}_{\mu^{P}(\mu^{P})^{T}} + \mathbf{D}_{V_{Y}} + \mathbf{V}_{v} + \mathbf{A}\mathbf{V}_{e}\mathbf{A}^{T}. \nonumber 
\end{eqnarray}
Thus, 
\begin{eqnarray}
E\Big\{ \mathbf{Y}^{T}\mathbf{F}\mathbf{Y} \Big\} 
&=& E\Big\{ (\mathbf{S}\mathbf{Y}^{P})^{T}\mathbf{F} (\mathbf{S}\mathbf{Y}^{P}) \Big\} \nonumber \\
&=& E(\mathbf{S}\mathbf{Y}^{P})^{T}\mathbf{F}E(\mathbf{S}\mathbf{Y}^{P}) + \textrm{tr}\Big( \mathbf{F}\textrm{Var}( \mathbf{S}\mathbf{Y}^{P} ) \Big) \nonumber \\
&=& (\bmu^{P})^{T}\mathbf{A}^{T}\mathbf{F}\mathbf{A}\bmu^{P} + \textrm{tr}\Big( \mathbf{F}(\mathbf{D}_{\mu^{P}(\mu^{P})^{T}} + \mathbf{D}_{V_{Y}} + \mathbf{V}_{v} + \mathbf{A}\mathbf{V}_{e}\mathbf{A}^{T}) \Big). \nonumber 
\end{eqnarray}
Now, note that the $i^{th}$ diagonal element of the matrix $\mathbf{D}_{\mu^{P}(\mu^{P})^{T}} + \mathbf{D}_{V_{Y}}$ is
\begin{eqnarray}
[\mathbf{D}_{\mu^{P}(\mu^{P})^{T}} + \mathbf{D}_{V_{Y}}]_{i,i} &=& \sum_{h=1}^{N} a_{ih}^{2}\Big( \frac{N_{i}}{n_{i}} - 1 \Big)\{ (\mu_{i(h)j(h)}^{P})^{2} + \tau_{0}^{2} + \sigma_{i(h)j(h)}^{2})\} \nonumber \\
&=& \frac{1}{N_{i}^{2}}\Big( \frac{N_{i}}{n_{i}} - 1 \Big)\sum_{h=\sum_{j=1}^{i-1} N_{j} }^{\sum_{j=1}^{i} N_{j} } \{ (\mu_{i(h)j(h)}^{P})^{2} + \tau_{0}^{2} + \sigma_{i(h)j(h)}^{2})\} \nonumber \\
&=& \frac{1}{N_{i}^{2}}\Big( \frac{N_{i}}{n_{i}} - 1 \Big)\sum_{ j=1 }^{ N_{i} } \{ (\mu_{ij}^{P})^{2} + \tau_{0}^{2} + \sigma_{ij}^{2})\} \nonumber \\
&=& \frac{1}{N_{i}}\Big( \frac{N_{i} - n_{i}}{n_{i}} \Big)\kappa_{2i}, \nonumber 
\end{eqnarray}
where $j(h) = h - i(h)$ and $i(h) = 1$, for $h = 1, ..., N_{1}$; $i(h) = 2$, for $h = N_{1} + 1, ...., N_{1} + N_{2}$; .... 

Hence,
\begin{equation}
\textrm{tr}\Big( \mathbf{F}(\mathbf{D}_{\mu^{P}(\mu^{P})^{T}} + \mathbf{D}_{V_{Y}}) \Big)
= \sum_{k=1}^{K}  \frac{1}{N_{k}}\Big( \frac{N_{k} - n_{k}}{n_{k}} \Big) F_{kk}\kappa_{2k}. \nonumber 
\end{equation}

\end{document}